\documentclass[11pt,twoside]{article}

\usepackage[utf8]{inputenc}
\usepackage{geometry}
\usepackage{setspace}
\usepackage{graphicx}
\usepackage{lscape}
\usepackage{float}
\floatstyle{plaintop}
\restylefloat{table}

\usepackage{booktabs}
\usepackage{multirow}
\usepackage{tabularx}
\usepackage{threeparttable}
\usepackage{array}
\usepackage{dcolumn}
\newcolumntype{d}[1]{D{.}{.}{#1}}
\newcolumntype{C}[1]{>{\centering\arraybackslash}p{#1}}

\usepackage{amsmath,amssymb,amsfonts,bm}

\usepackage{enumitem}
\newlist{steps}{enumerate}{1}
\setlist[steps,1]{wide=0pt,leftmargin=1cm,label=Step \arabic*:,font=\scshape}

\usepackage[dvipsnames]{xcolor}
\definecolor{nblue}{HTML}{000660}

\usepackage[
  hypertexnames=false,
  colorlinks=true,
  urlcolor=nblue,
  linkcolor=nblue,
  citecolor=nblue
]{hyperref}

\usepackage{natbib}
\usepackage[hang,flushmargin]{footmisc}
\usepackage{caption}
\usepackage{subcaption}
\usepackage{fancyhdr}
\usepackage{titlesec}

\titleformat{\section}
  {\bfseries\sffamily\Large\singlespacing}
  {\thesection. }{0em}{}

\titleformat{\subsection}
  {\bfseries\sffamily\large\singlespacing}
  {\thesubsection. }{0em}{}

\titleformat{\subsubsection}
  {\large\singlespacing}
  {}{0em}{\itshape}

\titlespacing*{\section}{0pt}{2.0ex plus .5ex minus .2ex}{1.0ex}
\titlespacing*{\subsection}{0pt}{1.6ex plus .4ex minus .2ex}{0.8ex}
\titlespacing*{\subsubsection}{0pt}{1.4ex plus .3ex minus .2ex}{0.6ex}

\usepackage{cleveref}
\crefname{section}{Section}{Sections}
\crefname{subsection}{Section}{Sections}
\crefname{subsubsection}{Section}{Sections}
\crefname{figure}{Figure}{Figures}
\crefname{table}{Table}{Tables}
\crefname{equation}{Eq.}{Eqs.}
\crefname{appendix}{Appendix}{Appendices}
\Crefname{section}{Section}{Sections}
\Crefname{subsection}{Subsection}{Subsections}
\Crefname{subsubsection}{Subsection}{Subsections}
\Crefname{figure}{Figure}{Figures}
\Crefname{table}{Table}{Tables}
\Crefname{equation}{Eq.}{Eqs.}
\Crefname{appendix}{Appendix}{Appendices}
\creflabelformat{equation}{(#2#1#3)}
\let\autoref\cref

\usepackage[title,titletoc]{appendix}
\makeatletter

\renewenvironment{appendices}{%
  \begin{oldappendices}%
  \renewcommand{\thefigure}{\thesection.\arabic{figure}}%
  \@addtoreset{figure}{section}%
  \@addtoreset{table}{section}}
{\end{oldappendices}}
\makeatother

\usepackage{placeins}
\usepackage{verbatim}
\usepackage{authblk}
\usepackage{ifsym}

\newcommand{\diag}{\text{diag}}
\newcommand{\bi}{\begin{itemize}}
\newcommand{\ei}{\end{itemize}}
\newcommand{\be}{\begin{equation}}
\newcommand{\ee}{\end{equation}}

\def\titletext{Beyond Aggregate VARs: A Bayesian Benchmark for HANK Models}

\title{\sffamily\huge{\textbf{\titletext}}}
\author{}
\date{}

\begin{document}

\maketitle
\vspace*{-6em}
\begin{center}

\end{center}
\begin{center}
\begin{minipage}{.32\textwidth}
  \centering\normalsize Florian \MakeUppercase{Huber}\\[0.25em]
  \small \textit{University of Salzburg}
\end{minipage}
\begin{minipage}{.32\textwidth}
  \centering\normalsize Gary \MakeUppercase{Koop}\\[0.25em]
  \small \textit{University of Strathclyde}
\end{minipage}
\begin{minipage}{.32\textwidth}
  \centering\normalsize Christian \MakeUppercase{Matthes}\\[0.25em]
  \small \textit{University of Notre Dame}
\end{minipage}
\end{center}

\begin{center}
\small This version: \today
\end{center}

\begin{abstract}
\noindent
Heterogeneous-agent New Keynesian (HANK) models characterize how entire cross-sectional distributions respond to structural shocks. Traditional representative-agent models are routinely disciplined by impulse responses from aggregate vector autoregressions (VARs). HANK models have no comparable established empirical benchmark because they make predictions not only about aggregates, but also about distributions of micro-level data. We propose a Bayesian benchmark that jointly models macroeconomic aggregates and several marginal distributions from repeated cross sections, including distributions observed in different surveys. Our approach can use both standard structural VAR identification approaches on macroeconomic aggregates and identification restrictions imposed on micro-level data. The model delivers a joint posterior of the distributional effects of shocks, without the need for household panel data or a separate first-stage density estimate.
\end{abstract}

\vspace*{1em}
\begin{center}
\begin{minipage}{0.85\textwidth}
\end{minipage}
\end{center}

\vfill\noindent{\footnotesize\textit{Acknowledgements}: We would like to thank Ludwig Straub for very useful comments.}

\thispagestyle{empty}\renewcommand{\footnotelayout}{\setstretch{1}}%
\onehalfspacing\normalsize\renewcommand{\thepage}{\arabic{page}}
\newpage

\section{Introduction}\label{sec:introduction}
Heterogeneous-agent New Keynesian (HANK) models predict that aggregate variables and household distributions respond together to structural shocks. Building on the incomplete-markets economies of \citet{krusell1998income}, \citet{kaplan2018monetary} show that household heterogeneity changes how monetary policy works. Most of the consumption response operates through general-equilibrium effects on labor income rather than through intertemporal substitution. Heterogeneity also weakens forward guidance \citep{mckay2016power}, turns redistribution into a transmission channel \citep{auclert2019monetary}, and makes fiscal multipliers depend on the distribution of marginal propensities to consume \citep{auclert2024intertemporal}. \citet{kaplan2018microeconomic} survey this research program.

Vector autoregressions (VARs) are the standard benchmark for aggregate predictions of dynamic stochastic general equilibrium (DSGE) models, but they do not recover the distributional responses that distinguish HANK models from their representative-agent counterparts. This paper delivers such a benchmark, while using both the Bayesian toolkit and VAR methods that are familiar to macroeconomists \citep{canova2007methods}.

We keep the VAR's transparent identification and add the distributions the theory is designed to explain. The Joint Aggregate--Micro Mixture VAR (JAMM-VAR) augments a Bayesian structural VAR (SVAR) with several marginal distributions constructed from repeated cross sections. Because the JAMM-VAR models marginal rather than joint distributions, the cross sections may come from separate surveys: earnings can come from the Current Population Survey (CPS) and consumption from the Consumer Expenditure Survey (CEX) even though the surveys neither follow nor contain the same households. We represent each marginal distribution as a Gaussian mixture, allowing a small number of normal components to approximate fat tails, skewness, and multimodality \citep{marron1992exact,FS_book}. Estimation requires neither household panels nor a preliminary density estimate.

The JAMM-VAR accommodates standard SVAR identification schemes for shocks in the aggregate block, including recursive orderings, sign and narrative restrictions, and instruments included in the system \citep{plagborg2021local}. It also identifies an aggregate shock represented by an innovation to a common latent factor, using restrictions on the factor's distributional and aggregate effects. We call it a common micro shock because cross-sectional information helps identify it, not because it is an idiosyncratic household shock. The joint posterior therefore delivers both the distributional responses to shocks identified within the aggregate SVAR and the aggregate responses to the common micro shock.

The JAMM-VAR links the aggregate and distributional blocks in both directions. Contemporaneous and lagged aggregate variables, together with common latent factors, determine the mixture weights and thus the shape of each distribution. The latent factors capture movements shared across distributions that observable aggregates do not explain. Conversely, lagged cross-sectional quantiles enter the macro block, allowing distributional conditions to shape subsequent aggregate dynamics. This feedback allows nonlinear aggregate dynamics; setting the quantile-feedback coefficients to zero yields a linear aggregate SVAR.

The posterior bands for impulse responses account for uncertainty from estimating the cross-sectional distributions. We draw the mixture components, weights, factors, and VAR coefficients from a joint posterior rather than estimating densities first and treating them as data. Two-step procedures can omit this source of uncertainty.

Our approach uses familiar Bayesian and VAR tools. Shrinkage priors regularize the high-dimensional weight and coefficient blocks, truncated priors impose the identifying restrictions, and information criteria compare specifications. The Gibbs sampler is modular: missing survey waves and stochastic volatility require additional blocks but leave the core algorithm unchanged. \autoref{sec:extensions} develops both extensions. Because the aggregate block remains an SVAR, the model also supports variance decompositions, historical decompositions, and conditional forecasts.

We evaluate the method in two simulation exercises. The first uses the model itself as the data-generating process. We verify that the sampler recovers the weights, densities, and impulse response functions (IRFs) when the model nests the data-generating process. The second exercise is more challenging, but also much more relevant for macroeconomists. We estimate our model on data simulated from a HANK economy that it does not strictly nest. Despite this deliberate misspecification, and conditional on matching the policy-variable impact, the model recovers the broad propagation of a monetary policy shock across aggregates, earnings quantiles, and consumption quantiles in a long simulated sample.

Finally, we apply our approach to U.S. data. We combine a standard policy VAR with CPS earnings and CEX consumption to estimate posterior response targets for identified fiscal and monetary policy shocks and for a common micro shock. We use the monetary-policy targets to evaluate the estimated HANK model of \citet{bayer2024shocks}. The model reproduces the cross-quantile pattern and the magnitude of the earnings responses on impact. Its aggregate responses are too large, and both its aggregate and earnings responses fade too quickly.

Our paper continues a long tradition of using VAR evidence to discipline equilibrium models. \citet{gali1999technology} uses the estimated responses to a long-run-identified technology shock to discriminate between competing business-cycle mechanisms. \citet{christiano2005nominal} choose the structural parameters of a New Keynesian model to match the VAR responses to an identified monetary policy shock, and \citet{altig2011firm} extend this matching strategy to technology and investment-specific shocks. \citet{delnegro2004priors} turn the mapping around and use an equilibrium model as a prior for a VAR. \citet{delnegro2007fit} use the resulting hybrid to measure how far New Keynesian models fall short of VAR fit. \citet{canova2011business} evaluate what sign-restricted VARs can recover from equilibrium models, and \citet{loria2022economic} use a VAR as a common empirical framework for confronting several structural theories with the data. Throughout this tradition the evidence is aggregate. We extend it to distributions.

The closest work is the functional-VAR literature, which jointly models aggregates and cross-sectional distributions. \citet{chang2024heterogeneity} estimate a period-specific sieve representation of each distribution and embed the resulting coefficients in a Bayesian state-space model. They also use functional-VAR responses to discipline heterogeneous-agent models. \citet{chang2024monetary} extend this framework to several marginal distributions drawn from separate data sources. The JAMM-VAR differs in two respects. First, a joint likelihood combines the micro cross sections with the aggregate dynamics, so no estimated functional coefficients are passed to a second stage and the posterior retains the uncertainty from fitting the distributions. Second, the same posterior contains both distributional responses to aggregate shocks identified with standard SVAR restrictions and aggregate responses to a common shock identified through restrictions on micro data.

In related work, \citet{nagasaka2026identifying} uses a two-step procedure. The first step approximates the time series of cross-sectional densities with a finite set of functional principal components, following \citet{chang2024functional}. The second places the resulting functional principal-component loadings and aggregate variables in a mixed autoregression and uses direct effects estimated from microeconometric research designs as prior restrictions to identify aggregate shocks. \citet{nagasaka2026identifying} builds on \citet{matthes2025missing}, who identify aggregate shocks through units' heterogeneous exposure but require a long panel. The density-based approach of \citet{nagasaka2026identifying} works with the repeated cross sections available at the household level. Unlike panel and pseudo-panel approaches, including \citet{baumeister2026havar} and \citet{koop2026pseudo}, we neither model individual transitions nor require stable household groups. Naturally, if panel data were available, one could still use our approach by neglecting the panel dimension of the data. \citet{ettmeier2024functional} discuss this distinction between distributional and household-level questions.

Other work estimates heterogeneous-agent models directly. \citet{liu2023full} develop full-information Bayesian inference that combines aggregate time series with repeated micro cross sections, while \citet{parraalvarez2023estimation} construct the cross-sectional likelihood from the model's Fokker--Planck equation. \citet{auclert2021using} use sequence-space Jacobians to make large heterogeneous-agent models tractable to estimate, and \citet{bayer2024shocks} estimate a HANK model with state-space methods. These approaches estimate structural parameters under the full restrictions of a particular equilibrium model. The JAMM-VAR is complementary: it summarizes joint aggregate--distributional evidence without imposing any one model's structure. The resulting posterior response objects can therefore be used to evaluate several structural models under common measurement and identification choices.

The paper proceeds as follows. \autoref{sec:hank_targets} describes how the method fits into the workflow of a researcher using heterogeneous-agent models. It presents the posterior targets it delivers and a step-by-step protocol for using them. \autoref{sec:econometrics} develops the econometric framework. It covers the mixture representation of the cross sections, the macro block, identification, priors, and the posterior sampler. \autoref{sec:validation} assesses the method on simulated data, first with the model itself as the data-generating process and then with a HANK economy that the model does not nest. \autoref{sec:empirical} presents the U.S. application, namely the distributional effects of identified fiscal and monetary policy shocks and the aggregate effects of a common micro shock. \autoref{sec:hankbench} uses the empirical estimates to benchmark an estimated HANK model. \autoref{sec:extensions} develops two extensions, cross sections observed at only some dates and stochastic volatility. \autoref{sec:conclusions} concludes.

\section{Our Method Delivers Targets for HANK Models}\label{sec:hank_targets}
This section explains how researchers can use the posterior targets of the JAMM-VAR to evaluate heterogeneous-agent models. We focus on HANK models, but the protocol applies more broadly whenever a model generates predictions for distributions observed in repeated cross sections. It applies, for example, to models of firm heterogeneity that predict how the distributions of productivity, investment, or employment respond to aggregate shocks \citep[e.g.,][]{liu2023full,winberry2018method,ottonello2020financial,marcellino2025firm}.

The JAMM-VAR supplies aggregate and distributional response targets for evaluating a HANK model, much as an aggregate SVAR supplies targets for a representative-agent DSGE model. The additional evidence consists of responses of the marginal distributions HANK models are designed to explain. Unlike direct structural estimation, the JAMM-VAR neither imposes the structure of a particular equilibrium model nor estimates household policy rules or deep parameters, although structural models can inform its priors. The targets below are not exhaustive; researchers can add others suited to their application.

\subsection{Posterior Targets}

The posterior delivers at least three sets of targets:
\begin{enumerate}[label=(\roman*), nosep]
\item aggregate responses to shocks identified with the specified SVAR restrictions;
\item responses of marginal distributions, including their quantiles and inequality measures, to the same shocks;
\item aggregate and distributional responses to any shock identified using restrictions on micro data.
\end{enumerate}
Density and quantile IRFs summarize the same underlying distributional response and therefore contain overlapping information. We report both because density IRFs show how probability mass shifts across the support, whereas responses of selected quantiles make magnitudes and cross-quantile patterns easy to compare across horizons and with structural models.

A candidate HANK model should be evaluated against the joint posterior of these response objects, not against point estimates assembled from separate procedures. The joint posterior preserves estimation uncertainty and dependence across the targets. Separate procedures may also use different samples, information sets, or identifying assumptions, producing targets that need not be mutually coherent.

To give some examples where our approach can be useful, in \citet{kaplan2018monetary}, intertemporal substitution accounts for a small share of the consumption response to a monetary policy shock and general-equilibrium labor-income effects account for the rest. The two transmission regimes have different implications for our targets, because direct transmission front-loads the consumption-quantile responses, which jump on impact and decay, while indirect transmission makes them inherit the hump shape and peak timing of the earnings-quantile responses. One diagnostic could then compare the peak horizon and peak size of each consumption-quantile response with those of the corresponding earnings quantiles, with credible bands from one posterior. Because the surveys are separate repeated cross sections, corresponding earnings and consumption quantiles need not contain the same households, so such comparisons are diagnostics to be read through a structural model rather than mechanisms identified by the quantile responses alone. The earnings-heterogeneity channel of \citet{auclert2019monetary} operates through unequal incidence of aggregate labor-income movements, and one implication is the cross-quantile ordering of the earnings responses. After a contractionary shock, the bottom quantiles fall by more than the top, so the P90--P10 spread (the 90th minus the 10th percentile) rises. In \citet{auclert2024intertemporal}, intertemporal marginal propensities to consume (MPCs) determine fiscal multipliers, and high-MPC households spend when income arrives rather than when news arrives. Because the fiscal shock in our application is a news shock, the announcement-versus-realization timing of the consumption-quantile responses provides a timing diagnostic related to the model's intertemporal-MPC profile. Finally, \citet{bayer2024shocks} estimate impulse responses of income and consumption inequality to structural shocks inside a HANK model. Such model responses can be compared with our posterior inequality bands horizon by horizon.

The benchmark does not identify household-level transitions. Repeated cross sections reveal how a marginal distribution changes, not which households move across its regions. Similarly, combining CPS earnings with CEX consumption does not identify the household-level joint distribution of earnings and consumption. The JAMM-VAR instead models the dynamics and comovement of these marginal distributions. The framework could be extended to model joint distributions, but datasets containing the required joint outcomes are relatively rare.

\subsection{A Protocol for HANK Researchers}
We now describe a prototypical workflow for disciplining a heterogeneous-agent model.

\begin{steps}
\item \textbf{Choose the empirical objects.} Select the aggregate variables and the marginal distributions of earnings, consumption, wealth, expectations, or other outcomes that the HANK model seeks to explain.
\item \textbf{Construct repeated cross sections.} Choose a common sample period for the aggregate and cross-sectional data. The cross sections may come from separate surveys and need not contain the same households. \autoref{sub:ext_missing} relaxes the common-sample requirement when cross sections are observed in only some aggregate periods.
\item \textbf{Identify the shocks.} If the comparison uses impulse responses, apply a standard SVAR identification scheme for the aggregate shocks. If desired, state separately the restrictions that identify a shock using the micro data.
\item \textbf{Estimate the joint posterior.} Use posterior draws to construct the selected targets, including impulse responses for aggregates, densities, quantiles, and inequality measures when shocks are identified.
\item \textbf{Make the HANK objects comparable.} Simulate the candidate model using the same variable definitions, transformations, survey observation rules, shock size, and response horizons as in the empirical benchmark.
\item \textbf{Diagnose the model.} Compare the model-implied targets with their posterior distributions and identify the aggregate or distributional margins on which they differ. The comparison need not be limited to responses to identified shocks; it can also use second and higher moments of the aggregate time series or cross-sectional distributions. Existing methods for comparing calibrated-model predictions with statistical estimates can guide this step \citep{smith1993estimating,canova1994statistical,dejong1996bayesian}.
\end{steps}

\section{Econometric Framework}\label{sec:econometrics}
We jointly model $M$ macroeconomic aggregates $\bm{Q}_t$ and $S$ marginal cross-sectional distributions. At date $t$, cross section $s$ contains
\[
\bm{y}_{t,s} = (y_{1t,s}, \dots, y_{n_{t,s}t,s})',
\]
where $n_{t,s}$ is the sample size and $p_s(y_{it,s} \mid \bm{\vartheta}_s)$ is the density for observation $i$. Sample sizes may vary across dates and distributions. The data need not track the same units over time, and different distributions may come from separate surveys. Conditional on the common variables and distribution-specific parameters, we treat observations as independent within each cross section. Cross section $(t,s)$ therefore contributes $\prod_{i=1}^{n_{t,s}}p_s(y_{it,s} \mid \bm{\vartheta}_s)$ to the joint likelihood.

We now provide a stylized description of the most common approach in the literature. We add this description here to make clear how our approach differs. One could approximate each cross-sectional distribution by a univariate Gaussian, $p_s \approx \mathcal{N}(\mu_{t,s}, \sigma^2_{t,s})$, estimated separately at each date. Then stack the macro aggregates $\bm{Q}_t$ with the cross-section-specific means and log variances into
\[
\bm{z}_t = (\bm{Q}'_t, \mu_{t,1}, \log\sigma^2_{t,1}, \dots, \mu_{t,S}, \log\sigma^2_{t,S})'.
\]
A natural statistical model would then be a VAR for $\bm{z}_t$ that links $\bm{Q}_t$ to all cross-sectional distributions:
\begin{equation}
    \bm{z}_t = \bm{A}_1 \bm{z}_{t-1} + \dots + \bm{A}_P \bm{z}_{t-P}
    + \bm{\varepsilon}_t, \label{eq:fVAR_simple}
\end{equation}
where $\bm{A}_j$ is a $(M + 2S) \times (M + 2S)$ coefficient matrix and $\bm{\varepsilon}_t \sim \mathcal{N}(\bm{0}, \bm{\Sigma}_z)$. Responses of $\bm{z}_t$ to shocks driving $\bm{Q}_t$ trace the effects on the first two moments of each distribution. A Gaussian approximation, however, cannot capture the pronounced asymmetry and long right tails of income and wealth distributions. Adding a small number of higher moments relaxes the Gaussian approximation, but it still compresses each distribution into a few chosen statistics: different shifts in probability mass can produce the same responses of those statistics. Adding many moments or quantiles increases the dimension of the VAR, and separately modeled quantiles need not remain ordered.

\citet{chang2024heterogeneity} address these limitations with a flexible spline representation. In their comparisons, the functional approach yields substantially tighter response bands than a VAR containing selected percentiles. A VAR containing inequality measures also produces implausible long-horizon responses. They approximate each time-varying cross-sectional density with spline basis functions and include the resulting coefficients in a VAR similar to \autoref{eq:fVAR_simple}. The JAMM-VAR instead estimates a parametric mixture representation jointly with the aggregate dynamics using the underlying micro observations. The mixture yields densities and quantiles directly and accommodates several marginal distributions without adding every basis coefficient to the VAR state.

\subsection{Marginal Distributions from Repeated Cross Sections}
We represent each cross-sectional distribution with a finite Gaussian mixture \citep{FS_book}. The fully parametric mixture requires no preliminary kernel-density estimate and is estimated jointly with the aggregate dynamics. The functional form is common across distributions $s \in \{1, \dots, S\}$, but the component parameters and number of components may differ. We therefore present the specification for a generic distribution $s$.

Rather than approximating $p_s$ by a single Gaussian, we use a mixture of $G_s$ Gaussians:
\begin{equation}
    p_s(y_{it,s} \mid \bm{\vartheta}_s) \approx \sum_{g=1}^{G_s}
    w_{tg,s}  \mathcal{N}(y_{it,s} \mid \mu_{g,s}, \sigma^2_{g,s}),
    \quad i = 1, \dots, n_{t,s}, \label{eq: mixture_1}
\end{equation}
where $w_{tg,s}$, $\mu_{g,s}$, and $\sigma^2_{g,s}$ are the component weights, means, and variances for cross section $s$. We allow the number of components $G_s$ to differ across cross sections. Only the weights vary over time. Component locations and scales are fixed. The weights follow a multinomial logistic specification:
\begin{equation}
    w_{tg,s} = \frac{e^{\eta_{tg,s}}}{\sum_{j=1}^{G_s - 1}
    e^{\eta_{tj,s}} + 1}, \quad g = 1, \dots, G_s - 1, \label{eq:softmax}
\end{equation}
with component $G_s$ as the reference, meaning $\eta_{tG_s,s}=0$, so $w_{tG_s,s} = \big(\sum_{j=1}^{G_s-1} e^{\eta_{tj,s}} + 1\big)^{-1}$. The unnormalized log weight $\eta_{tg,s}$ is a linear function of contemporaneous and lagged macro aggregates and common latent factors:
\begin{equation}
    \eta_{tg,s} = \beta_{0g,s} + \bm{\beta}'_{g,s} \bm{x}_t
    + \bm{\lambda}'_{g,s} \bm{f}_t.
    \label{eq:log_weights}
\end{equation}
Here $\bm{x}_t = (\bm{Q}'_t, \dots, \bm{Q}'_{t-P})'$, $\beta_{0g,s}$ is a component- and cross-section-specific intercept, and $\bm{\beta}_{g,s}$ is a $(P+1)M$-dimensional vector of macro loadings for component $g$ in cross section $s$. These loadings vary freely across $g$ and $s$, allowing each distribution to respond differently to the same macroeconomic conditions.

The $R$ static factors $\bm{f}_t \sim \mathcal{N}(\bm{0}, \bm{I}_R)$ are common to all $S$ distributions, with component- and distribution-specific loadings $\bm{\lambda}_{g,s}$. They induce comovement by shifting the mixture weights of each distribution on which they load. Absent additional restrictions, the factors are statistical objects without a structural interpretation. \autoref{sub:identification} describes the restrictions on their loadings and realizations that identify selected factor innovations as structural shocks.

The mixture likelihood is invariant to permutations of the component labels within each distribution, creating the standard label-switching problem. We resolve it by imposing the ordering $\mu_{1,s} < \dots < \mu_{G_s,s}$ for each $s$ throughout estimation.

\paragraph{A two-component example.}
To make the mixture mechanism concrete, consider one distribution with two components ($G=2$) and one macro aggregate $Q_t$. We suppress the distribution subscript $s$. The density is
\begin{equation*}
p(y_{it}\mid\bm \vartheta)
= w_{t1}\mathcal{N}(y_{it}\mid\mu_1,\sigma_1^2)
+ (1-w_{t1})\mathcal{N}(y_{it}\mid\mu_2,\sigma_2^2).
\end{equation*}
The ordering restriction $\mu_1 < \mu_2$ means that component $1$ is centered at lower outcomes and component $2$ at higher outcomes. Because the component means and variances are fixed, changes in $w_{t1}$ generate all time variation in the distribution.

To see how $Q_t$ changes the distribution through $w_{t1}$, consider a logit specification without latent factors:
\begin{equation*}
w_{t1} = \frac{\exp(\eta_{t1})}{1+\exp(\eta_{t1})},
\qquad
\eta_{t1} = \alpha_1 + \beta_1 Q_t.
\end{equation*}
Suppose $\beta_1<0$. Then
\begin{equation*}
\frac{\partial w_{t1}}{\partial Q_t}
= \beta_1 w_{t1}(1-w_{t1}) < 0.
\end{equation*}
An increase in $Q_t$ therefore lowers $w_{t1}$ and shifts probability mass toward the higher-outcome component. The shift is largest when the two components have equal weight ($w_{t1}=0.5$) and approaches zero as either component's weight approaches one.

The implied cross-sectional mean is
\begin{equation*}
\mathbb{E}_t(y_{it})
= w_{t1}\mu_1 + (1-w_{t1})\mu_2.
\end{equation*}
Differentiating with respect to $Q_t$ gives
\begin{equation*}
\frac{\partial \mathbb{E}_t(y_{it})}{\partial Q_t}
= (\mu_1-\mu_2)\beta_1 w_{t1}(1-w_{t1}) > 0.
\end{equation*}
The derivative is positive because $\mu_1-\mu_2$ and $\beta_1$ are both negative. The shift toward the higher-outcome component therefore raises the cross-sectional mean. The cross-sectional variance also depends on the mixture weight:
\begin{equation*}
\mathrm{Var}_t(y_{it})
= w_{t1}\sigma_1^2 + (1-w_{t1})\sigma_2^2
+ w_{t1}(1-w_{t1})(\mu_1-\mu_2)^2.
\end{equation*}
Thus changes in $w_{t1}$ can alter both the location and dispersion of the cross-sectional distribution even though the component means and variances remain fixed.

\subsection{The Aggregate Time Series Model}
The macro block is a structural VAR($P$) for $\bm Q_t$ augmented by two links to the cross-sectional distributions: lagged quantiles and the common latent factors. The lagged quantiles allow the distributions to affect subsequent aggregate dynamics, while the factors capture contemporaneous comovement between the macro and mixture blocks. Let $\bm A_0$ denote the $M \times M$ contemporaneous coefficient matrix. The structural form is
\begin{equation}
  \bm A_0 \bm Q_t = \bm c + \bm A_1 \bm Q_{t-1} + \dots + \bm A_P \bm Q_{t-P} + \sum_{s=1}^{S} \sum_{r \in \mathcal{R}} \bm \alpha_{r,s} \mathcal{Q}_r(\bm y_{t-1,s}) + \bm \Lambda_q \bm f_t + \bm u_t, \quad \bm u_t \sim \mathcal{N}(\bm 0, \bm D). \label{eq: VAR}
\end{equation}
We normalize the diagonal of $\bm A_0$ to one and parameterize it as $\bm A_0 = \bm I_M - \bm W$, where $\bm W$ has a zero diagonal and contains the free contemporaneous coefficients. The structural-shock covariance matrix is $\bm D = \diag(d_1,\dots,d_M)$. Because $\bm D$ is diagonal, the elements of $\bm u_t$ are mutually uncorrelated. Premultiplying the structural system by $\bm A_0^{-1}$ yields the reduced form
\begin{align}
  \bm Q_t
  &=\bm A_0^{-1}\bm c
  +\sum_{j=1}^{P}\bm A_0^{-1}\bm A_j\bm Q_{t-j}
  +\bm A_0^{-1}\sum_{s=1}^{S}\sum_{r\in\mathcal{R}}
  \bm\alpha_{r,s}\mathcal{Q}_r(\bm y_{t-1,s}) \notag\\
  &\quad+\bm A_0^{-1}\bm\Lambda_q\bm f_t
  +\bm\varepsilon_t,
  \qquad
  \bm\varepsilon_t=\bm A_0^{-1}\bm u_t,
  \qquad
  \operatorname{Var}(\bm\varepsilon_t)=\bm A_0^{-1}\bm D\bm A_0^{-\prime}.
  \label{eq:VAR_reduced}
\end{align}
The reduced-form innovations $\bm\varepsilon_t$ are correlated across equations. The data pin down only their covariance $\bm\Sigma=\bm A_0^{-1}\bm D\bm A_0^{-\prime}$. The restrictions in \autoref{sub:identification} resolve the rotation from $\bm\varepsilon_t$ to the orthogonal structural shocks $\bm u_t$, exactly as in a standard SVAR. The vector $\bm c$ collects the intercepts, the matrices $\bm A_1,\dots,\bm A_P$ are the $M \times M$ structural lag coefficients, $\mathcal{R}$ is a finite set of $n_q$ quantile levels in $(0,1)$, and $\mathcal{Q}_r(\bm y_{t-1,s})$ is the $r$-quantile of lagged cross section $s$.

The static factors enter both blocks, with macro loadings $\bm \Lambda_q$. Absent additional restrictions, a factor has no structural interpretation. It links the macro series to the micro distributions. In the empirical application, restrictions on $\bm A_0$ (equivalently $\bm W$) identify the aggregate shocks, and restrictions on the loadings $\bm \Lambda_q$ and $\bm \lambda_{g,s}$ label the common micro shock. In our implementation, we think of the aggregate structural shocks as being elements of $\bm u_t$.

The lagged quantiles make the joint system nonlinear. The log-weight indices in \autoref{eq:log_weights} are linear in the macro variables and factors, but the multinomial-logit map converts these indices into mixture weights nonlinearly. The quantiles implied by those weights then feed back into $\bm Q_t$ through \autoref{eq: VAR}. Impulse responses can therefore depend on a shock's sign and size and on the initial state. Setting $\bm\alpha_{r,s}=\bm 0$ for every $r$ and $s$ and $\bm \Lambda_q = \bm 0$ removes this feedback and nests a linear structural VAR for the aggregate block. If we fix  $\bm\alpha_{r,s}=\bm 0$ while allowing the quantiles to enter the VAR, the aggregate block is a linear factor-augmented SVAR. The priors in \autoref{sub:priors} center the quantile loadings at zero, so the posterior departs from the linear submodel only when the likelihood supports distributional feedback.

Unlike functional VARs that include basis coefficients or functional principal-component loadings among the dependent variables, our aggregate block keeps selected quantiles and factors on the right-hand side and therefore retains $M$ equations. Excluding contemporaneous coefficients, each aggregate equation contains $MP$ lag coefficients and an intercept. The quantiles and factors add $Sn_q+R$ regressors per equation, or $M(Sn_q+R)$ coefficients to the system, while $\bm A_0$ and $\bm D$ are unchanged.

In the empirical application, $M=4$, $P=4$, $S=2$, $n_q=5$, and $R=1$. The aggregate-lag block alone has $17$ regressors per equation, including the intercept. The ten quantiles and one factor add $11$ regressors per equation, or $44$ coefficients across the system. Excluding intercepts, the aggregate block therefore contains $64$ lag coefficients, $40$ quantile loadings, and $4$ factor loadings, for a total of $108$.

A functional VAR that stacks $K$ basis coefficients per distribution among the dependent variables has $M+SK$ equations and $(M+SK)^2P$ lag coefficients. At $K=10$, it becomes a $24$-variable VAR with $2{,}304$ lag coefficients. The $108$ count pertains only to our aggregate block; the JAMM-VAR also estimates the log-weight equations and the mixture-component parameters. The comparison therefore concerns the size of the VAR block rather than the total number of model parameters. Shrinkage priors regularize the quantile, factor, and log-weight coefficients (\autoref{sub:priors}).

\subsection{Structural Identification}\label{sub:identification}

We identify aggregate shocks and the common micro shock separately. The aggregate block supports the same identification schemes as a standard SVAR, including recursive, sign, magnitude, narrative, and instrument-based restrictions. These restrictions identify selected elements of $\bm u_t$ as aggregate shocks. Because the mixture weights depend on contemporaneous and lagged aggregates, the mixture block maps each identified aggregate shock into responses of the marginal distributions.

We identify the common micro shock as an innovation to a selected factor in $\bm f_t$. The loadings $\bm\lambda_{g,s}$ map this innovation into changes in mixture weights across marginal distributions, while $\bm\Lambda_q$ maps it into the aggregate equations. The normalization $\bm f_t\sim\mathcal{N}(\bm 0,\bm I_R)$ fixes the shock's scale. Sign and zero restrictions on both sets of loadings, together with narrative restrictions on selected factor realizations, give the shock an economic interpretation. Because these restrictions impose some impact responses, the empirical analysis distinguishes the imposed impact signs from the subsequent responses learned from the posterior.

For both types of shock, we report posterior responses of the aggregate variables and of each marginal distribution.

\subsection{Prior Distributions}\label{sub:priors}
The model contains many coefficients relative to the short macroeconomic sample, so the priors serve two roles: they regularize estimation and encode the identifying restrictions. We use an ordered Normal--inverse-Gamma prior for the mixture-component means and variances, horseshoe shrinkage for the log-weight and VAR coefficients, inverse-Gamma priors for the structural variances, and Gaussian priors for the free contemporaneous coefficients. Sign and magnitude restrictions truncate the relevant prior support, zero restrictions fix selected coefficients, and narrative restrictions truncate selected factor realizations. Each restriction therefore holds at every posterior draw \citep{baumeister2015sign}.

\subsubsection{Priors on the Mixture Components and Weights}
For each component $g$ of cross section $s$, we use a conjugate prior that links its location and scale parameters:
\begin{equation*}
  p(\mu_{g, s}, \sigma_{g, s}^2) \propto p(\mu_{g, s}| \sigma_{g, s}^2) ~ p(\sigma_{g,s}^2)
\end{equation*}
where the marginal prior for $\sigma^2_{g,s}$ and the conditional prior for $\mu_{g,s}$ are
\begin{equation}
  \sigma^2_{g,s} \sim \mathcal{IG}(a_0,b_0),\qquad
  \mu_{g,s} | \sigma^2_{g,s},\kappa_{0,s}\sim
  \mathcal{N}(m_{0,s},\tfrac{\sigma^2_{g,s}}{\kappa_{0,s}}) ~ \mathbb{I}(\mu_{1,s}<\dots<\mu_{G_s,s}),
  \label{eq:prior_mu}
\end{equation}
where $\mathcal{IG}$ denotes the inverse-Gamma distribution. The ordering constraint resolves the label-switching problem discussed in \autoref{sec:econometrics}. We set $a_0=b_0=0.01$ and center the component means for cross section $s$ on its pooled sample mean,
\begin{equation*}
m_{0,s}
=
\left(\sum_{t=1}^{T}n_{t,s}\right)^{-1}
\sum_{t=1}^{T}\sum_{i=1}^{n_{t,s}}y_{it,s}.
\end{equation*}
The precision $\kappa_{0,s}$ controls how tightly the component means cluster around $m_{0,s}$: larger values pull them toward the pooled mean, while smaller values permit greater separation across component means. We place a heavy-tailed hyperprior on $\kappa_{0,s}$ and update it within the sampler. Posterior draws can place component means close together or assign negligible weight to some components. Thus $G_s$ is an upper bound on the effective number of distinct components.

For each nonreference component $g=1,\dots,G_s-1$ in cross section $s$, collect the intercept, macro loadings, and factor loadings in
$\bm b_{g,s}=(\beta_{0g,s},\bm\beta_{g,s}',\bm\lambda_{g,s}')'$. Each cross section contains one such log-weight equation for every nonreference component, and the coefficients vary freely across components and cross sections. We regularize this high-dimensional block with a horseshoe prior \citep{carvalho2010}:
\begin{equation}
  b_{l,g,s} | \varphi_{l,g,s},\tau_{g,s} \sim
  \mathcal{N}(0,\varphi_{l,g,s}^2\tau_{g,s}^2),\qquad
  \varphi_{l,g,s}\sim \mathcal{C}^+(0,1),\qquad
  \tau_{g,s}\sim \mathcal{C}^+(0,1),
  \label{eq:prior_B}
\end{equation}
where $l$ indexes the elements of $\bm b_{g,s}$, $\tau_{g,s}$ is a component-specific global scale, $\varphi_{l,g,s}$ is a coefficient-specific local scale, and $\mathcal{C}^+$ denotes the half-Cauchy distribution. We sample the scale parameters using the inverse-Gamma augmentation of \citet{makalic2016}. The horseshoe strongly shrinks small coefficients toward zero while leaving large coefficients weakly penalized.

Some elements of $\bm b_{g,s}$ carry identifying restrictions. Restrictions on the contemporaneous macro loadings in $\bm\beta_{g,s}$ govern the distributional impact of shocks identified in the aggregate block, while restrictions on the factor loadings in $\bm\lambda_{g,s}$ help identify the common micro shock. We impose these sign and magnitude restrictions by multiplying the joint horseshoe prior for $\bm b_{g,s}$ and its scales by an indicator for the admissible region and normalizing the joint density once. Conditional on an admissible coefficient draw, the indicator and joint normalizer are constant with respect to the scales, so their inverse-Gamma full conditionals are unchanged. In the augmented sampler, the coefficient full conditional is Gaussian truncated to the admissible region. Every posterior draw therefore satisfies the restrictions without a scale-dependent normalizing-constant correction. We apply the same joint-truncation convention to inequality-restricted coefficients in the macro block, treating $d_i$ as part of the joint prior; zero restrictions instead fix coefficients at zero. The empirical application specifies the exact restrictions.

\subsubsection{Priors on the Macro Block}
The structural macro block in \autoref{eq: VAR} contains three parameter groups: the regression coefficients in $\bm\Phi$, the structural variances in $\bm D=\diag(d_1,\dots,d_M)$, and the free contemporaneous coefficients in $\bm W$, where $\bm A_0=\bm I_M-\bm W$. Let $\bm m_t$ collect the $P$ lags of $\bm Q_t$, the lagged cross-sectional quantiles $\mathcal{Q}_r(\bm y_{t-1,s})$, the common factors $\bm f_t$, and a constant. Stacking the corresponding coefficients in $\bm\Phi$ gives
$\bm A_0\bm Q_t=\bm\Phi'\bm m_t+\bm u_t$.
The $i$th column $\bm\phi_i$ is the coefficient vector for equation $i$; it contains the coefficients on aggregate lags, lagged quantiles, factors, and the intercept. Conditional on its structural variance $d_i$, we assign the Gaussian prior
\begin{equation}
  \bm\phi_i\mid d_i\sim\mathcal{N}\big(\underline{\bm\phi}_i,
  d_i\underline{\bm V}_i\big),
  \label{eq:prior_phi}
\end{equation}
where the prior mean $\underline{\bm\phi}_i$ follows the Minnesota convention \citep{litterman1986}. The own first lag is centered at $\delta$; cross-lags, higher-order own lags, quantile and factor loadings, and the intercept are centered at zero. We set $\delta=0.8$ in the empirical application. The prior precision is
\[
\underline{\bm V}_i^{-1}
=
\diag\left(\frac{1}{\tau_i^2\varphi_{l,i}^2}\right),
\]
where $\tau_i$ is an equation-specific global scale and $\varphi_{l,i}$ is a coefficient-specific local scale, both following the horseshoe hierarchy in \autoref{eq:prior_B}. This prior centers each equation on a persistent univariate process and shrinks the distributional-feedback and factor loadings toward zero.

Identifying restrictions apply to the factor loadings in $\bm\Phi$. Following the joint-truncation convention above, sign restrictions truncate the corresponding multivariate Gaussian conditional. A zero restriction instead fixes the selected factor loading at zero and draws the remaining coefficients from their Gaussian conditional given that restriction.

For the structural variances, we use the conditionally conjugate prior
\begin{equation}
  d_i\sim\mathcal{IG}\Big(\tfrac{M+2}{2},\tfrac{\hat\sigma^2_i}{2}\Big),
  \label{eq:prior_D}
\end{equation}
where $\hat\sigma^2_i$ denotes the residual variance obtained by estimating the reduced-form of the model on an equation-by-equation basis using a unit ridge penalty. The prior is weakly informative, with mean
\[
  \mathbb{E}[d_i]=\frac{\hat\sigma^2_i}{M},
\]
and, for $M>2$, variance
\[
  \operatorname{Var}(d_i)
  =
  \frac{2(\hat\sigma^2_i)^2}{M^2(M-2)}.
\]
The free off-diagonal elements of $\bm W$ follow independent Gaussian priors
\begin{equation*}
  W_{ij}\sim\mathcal{N}(0,\underline{l}_W),
\end{equation*}
restricted to intervals $[\underline{w}_{ij},\overline{w}_{ij}]$ that encode the identifying restrictions on $\bm A_0$. A sign restriction sets one endpoint to zero; a magnitude restriction sets a nonzero bound calibrated using the corresponding prior in \citet{baumeister2018inference}; and an exclusion restriction sets both endpoints to zero. We set $\underline{l}_W=2$. We normalize the shared factors as $\bm f_t\sim\mathcal{N}(\bm 0,\bm I_R)$, as stated in \autoref{sec:econometrics}. Narrative restrictions impose sign constraints on selected factor realizations.

\subsection{Posterior Simulation}\label{sub:posterior}
In the unaugmented posterior, three features rule out direct conjugate updates: the multinomial-logit likelihood for the log-weight coefficients, the determinant term in the conditional for $\bm A_0$, and the shared factors' entry in every component index. P\'olya--Gamma augmentation \citep{polson2013} yields conditionally Gaussian Gibbs updates for the log-weight coefficients. We update $\bm A_0$ and the shared factors with Metropolis--Hastings steps that use Gaussian proposals and target their exact conditional distributions; all remaining blocks are sampled by Gibbs. The steps below describe one MCMC sweep. \autoref{app:technical} derives the full conditionals and Metropolis--Hastings acceptance ratios.

\begin{steps}
\item \textbf{Component allocation.} For each micro observation $y_{it,s}$, draw a label $z_{it,s}\in\{1,\dots,G_s\}$ from its multinomial full conditional, with probabilities proportional to
\[
w_{tg,s}\mathcal{N}(y_{it,s}\mid\mu_{g,s},\sigma^2_{g,s}),
\]
using the current weights from \autoref{eq:softmax}.

\item \textbf{Component parameters and mean-shrinkage precision.} Conditional on the allocations, the observations assigned to each component form a Gaussian subsample. For each component, draw $\sigma^2_{g,s}$ from its inverse-Gamma conditional given the current mean and then $\mu_{g,s}$ from its Gaussian conditional truncated to the interval between the neighboring means, which preserves $\mu_{1,s}<\dots<\mu_{G_s,s}$. The two draws form an exact Gibbs update of the ordered Normal--inverse-Gamma conditional implied by \autoref{eq:prior_mu}. Then draw the cross-section-specific precision $\kappa_{0,s}$ from its generalized-inverse-Gaussian ($\mathcal{GIG}$) full conditional.

\item \textbf{Log-weight coefficients.} For each cross section $s$ and nonreference component $g=1,\dots,G_s-1$, update $\bm b_{g,s}$ using the one-vs-rest representation of the multinomial-logit likelihood. Introducing $\omega_{tg,s}\sim\mathrm{PG}(n_{t,s},\psi_{tg,s})$ makes the full conditional for $\bm b_{g,s}$ Gaussian \citep{polson2013}, where $n_{t,s}$ is the sample size of cross section $s$ at date $t$ and $\psi_{tg,s}$ is the current one-vs-rest log odds. Draw each $\omega_{tg,s}$ using the saddlepoint approximation of \citet{windle2014}. Then draw $\bm b_{g,s}$ from its Gaussian full conditional, truncated to the admissible region when the coefficients carry sign or magnitude restrictions. Finally, update the local and global horseshoe scales using the inverse-Gamma augmentation of \citet{makalic2016}. Under the joint-truncation convention in \autoref{sub:priors}, these restrictions leave the scale full conditionals unchanged. \autoref{app:technical} gives the full conditional formulas.

\item \textbf{Macro coefficients.} Conditional on $\bm A_0$, $\bm D$, the shared factors, and the lagged quantiles, the structural system separates by equation. For each equation $i$, combining the likelihood with the Gaussian prior in \autoref{eq:prior_phi} yields a Gaussian full conditional for $\bm\phi_i$. Draw unrestricted coefficient vectors directly from this conditional and coefficient vectors with sign-restricted factor loadings from the corresponding truncated multivariate Gaussian. If a factor loading is fixed at zero, set it to zero and draw the remaining coefficients from their Gaussian conditional given that restriction. Finally, update the local and global horseshoe scales using the same inverse-Gamma augmentation as in the log-weight block.

\item \textbf{Structural variances.} Conditional on $\bm\Phi$, $\bm A_0$, the shared factors, and the lagged quantiles, draw each $d_i$ from its inverse-Gamma full conditional. The update combines the prior in \autoref{eq:prior_D}, the structural residual sum of squares for equation $i$, and the quadratic term from the $d_i$-scaled coefficient prior in \autoref{eq:prior_phi}. \autoref{app:technical} gives the exact parameters.

\item \textbf{Contemporaneous matrix.} The full conditional for each row of $\bm A_0=\bm I_M-\bm W$ contains the Jacobian $|\det \bm A_0|^{T}$ and is therefore non-Gaussian. Because $\det \bm A_0$ is linear in a single row, a second-order Laplace expansion adds a rank-one term to the Gaussian precision, which the Sherman--Morrison identity evaluates in $O(M^2)$. Use the resulting approximate Gaussian, truncated to the sign, magnitude, and zero restrictions on $\bm A_0$, as the proposal in a Metropolis--Hastings step targeting the exact conditional. The sign restrictions on the impact responses in $\bm A_0^{-1}$ involve all rows of $\bm A_0$ at once. They enter the target of every row update as an indicator, so a proposed row is rejected whenever the implied impact responses violate them.

\item \textbf{Common factors.} Draw the factor history on a $t$ by $t$ basis. The P\'olya--Gamma representation is valid for the log-weight coefficients because it holds the competing component indices fixed while $\bm b_{g,s}$ is updated. The factors are shared. They enter \emph{every} component index, so moving $\bm f_t$ also moves the log-sum-exp offset of each one-vs-rest representation, and the resulting kernel is no longer Gaussian in $\bm f_t$. We therefore build a Gaussian proposal that pools information from the macro shocks, through $\bm\Lambda_q$ and $\bm D$, with information from the log-weight equations, through $\bm\lambda_{g,s}$ and {the conditional means of the P\'olya--Gamma weights evaluated at the value the proposal is centered on}. We accept the proposal with a Metropolis--Hastings probability computed from the exact multinomial logistic likelihood{, with the proposal density evaluated in both directions}. This leaves the target posterior invariant. Acceptance rates are between $0.92$ and {$0.95$} across chains in the empirical application. At dates with a narrative restriction, the proposal is drawn from the corresponding truncated Gaussian, so the restricted sign of (elements of) $\bm f_t$ is imposed exactly. \autoref{app:technical} gives the exact conditional and the acceptance ratio.
\end{steps}

We discard burn-in draws and retain every $k$th remaining draw. Thinning reduces storage and the cost of the simulation-based impulse responses. We run several independent chains in parallel and pool the thinned draws to compute impulse responses for each identified shock. The reported MCMC budget for each exercise includes burn-in, retained draws, and thinning. For each retained draw, we propagate the estimated system forward from the sample mean of the state to construct the structural impulse responses. The individual mixture parameters mix more slowly than the fitted distributions and impulse responses. We discuss the convergence diagnostics in \autoref{app:mcmc_diagnostics}.

The sampler is modular. The two extensions developed in \autoref{sec:extensions}---allowing cross sections to be observed at only some dates and introducing stochastic volatility---each add or replace one sampling block; the remaining updates are unchanged or modified only through known weights. Because each retained draw contains the full structural system, the posterior can also be used to construct variance decompositions, historical decompositions, and conditional forecasts in addition to the reported impulse responses.

Before turning to the empirical applications, a brief word on computation times is in order. The estimation times are based on running a single chain with an Apple M4 Max processor and include the simulation of the impulse responses. For the HANK exercise, with $T=500$ dates, twelve aggregate series, and two cross sections of $1{,}000$ {simulated} households per date summarized by $500$ quantile points, a chain of $10{,}000$ draws ($5{,}000$ burn-in and $5{,}000$ retained) takes about $35$ minutes. For the controlled DGP of \autoref{sub:rf_dgp}, a chain of $2{,}500$ draws takes about five minutes. For the empirical application of \autoref{sec:empirical}, with $T=71$ dates, a chain of $15{,}000$ draws ($5{,}000$ burn-in and $10{,}000$ retained) takes about $15$ minutes. Additional chains run in parallel on separate CPU cores at little extra cost in elapsed time.

\section{Assessing Our Approach Using Simulated Data}\label{sec:validation}
We assess how well our approach does in two exercises. The first uses a data-generating process (DGP) that the JAMM-VAR nests. We think of this as a low bar to pass, but nonetheless a bar we want to check. The second exercise uses a HANK DGP outside the model class and asks whether the statistical model still recovers dynamics, {cross sections}, and impulse responses. The HANK economy's known shocks and responses provide a controlled laboratory for assessing the JAMM-VAR's identification and specification choices.

\subsection{Our Model as the Data-Generating Process}\label{sub:rf_dgp}
The controlled DGP combines a two-variable VAR(1), two cross sections, and one common standard-normal factor. The VAR includes the lagged median of each cross section and the factor as regressors. At each of $T=250$ dates, each cross section contains $500$ observations drawn from a three-component Gaussian mixture whose weights depend on the contemporaneous and lagged macro state and on the factor. With only two cross sections, the factor cannot be separated from the observed macro state without additional structure. We identify it using sign restrictions of the same form as those in the empirical application. The design contains two structural channels: an innovation to the macro block and an innovation to the common factor. \autoref{app:validation} reports the full parameterization.

Posterior median weights track the true weights almost exactly, and the fitted densities are nearly indistinguishable from the true densities (\autoref{fig:rf_weights,fig:rf_densities} in the appendix). For both structural shocks, the posterior median aggregate, density, and quantile responses track the truth closely. The 90\% credible bands contain the true quantile responses at every percentile and horizon, the true density responses at every point of the support and horizon shown except a small part of the support at impact for the factor shock, and the true aggregate responses at every horizon except the impact response of $Q_{2t}$ to the factor shock (\autoref{fig:rf_dist_quant_macro,fig:rf_dist_quant_micro} in the appendix). We next turn to a HANK model that is not nested in the JAMM-VAR model.

\subsection{HANK Data-Generating Process}

We use the one-asset HANK model of \citet{auclert2021using} as the DGP. Households face incomplete markets, trade a single liquid asset subject to a zero borrowing limit, and supply labor with idiosyncratic productivity that follows an AR(1) in logs. Firms produce with a linear technology, and Rotemberg price adjustment yields a New Keynesian Phillips curve. Monetary policy follows a Taylor rule. The government keeps the supply of real bonds fixed and levies lump-sum taxes to finance interest payments. \autoref{app:hank_calibration} reports the full calibration and discretization.

We solve the model with the sequence-space Jacobian method of \citet{auclert2021using}, using a modified version of their public toolkit.\footnote{\url{https://github.com/shade-econ/sequence-jacobian}.} Two aggregate AR(1) disturbances drive the economy. A shock to the Taylor-rule intercept has persistence $0.61$ and an innovation standard deviation of $25$ basis points; a TFP shock has persistence $0.8$ and an innovation standard deviation of $1$ percent. We discard the first $2{,}000$ quarters of the simulation and retain the next $T=500$ quarters.

The statistical model observes twelve aggregate series: the exogenous policy-rate shock (MP) and the model's output, consumption, labor demand, real interest rate, real wage, inflation, aggregate assets, hours, effective labor, dividends, and taxes. Because MP is the observed Taylor-rule disturbance, we identify the monetary policy shock as the innovation to its equation. TFP is not among the observables, so one of the two aggregate shocks is hidden from the statistical model. This is a deliberate source of misspecification beyond the mixture approximation itself. Aggregate assets equal the fixed bond supply and are therefore constant.

We estimate the model with $G=(6,6)$ components and $R=2$ latent factors. With two latent factors and unrestricted loadings, the factor configuration is identified only up to rotation. We therefore anchor each factor to one equation in the spirit of \citet{geweke1996}. The first factor is excluded from the equation of the policy instrument, and its loading on output is restricted to be negative. The second factor is excluded from the output equation, and its loading on the policy instrument is restricted to be positive. Both factors are excluded from the asset equation, which is constant in this simulation because bonds are in fixed supply. These restrictions are labeling conventions, analogous to the ordering of the mixture means, and match the sparse loading pattern an unrestricted run recovers.

At each date, we draw separate cross sections of consumption and gross labor earnings, each containing $1{,}000$ households. As in the empirical application, each simulated cross section is summarized by $500$ equally weighted quantile points before estimation. Gross labor earnings equal the product of the real wage, individual productivity, and individual hours. We obtain both cross sections by inverse-CDF sampling from the model-implied distribution over productivity and assets. We use one simulated history and fixed random seeds throughout. We compare the estimates with the HANK model's true impulse responses to a one-time $25$-basis-point monetary policy shock.

Because the HANK DGP is not a finite mixture, it has no true mixture components or weights to recover. The estimated components and weights are approximation devices. We therefore evaluate the cross-sectional distributions implied jointly by the components and weights. The fitted six-component mixtures reproduce the broad support, modes, skewness, and tails of consumption and earnings throughout the sample, although the fit is not exact in every bin (\autoref{fig:hank_weights,fig:hank_densities} in the appendix). The impulse-response analysis therefore focuses on changes in the fitted distributions rather than in individual components or weights.

\subsubsection{Responses to the Monetary Policy Shock}
\begin{figure}[htbp]
  \centering
  \begin{subfigure}[b]{0.80\textwidth}
    \centering
    \caption{Macro IRFs}
    \label{fig:hank_macro}
    \includegraphics[width=\textwidth, trim=0 0 0 10bp, clip]{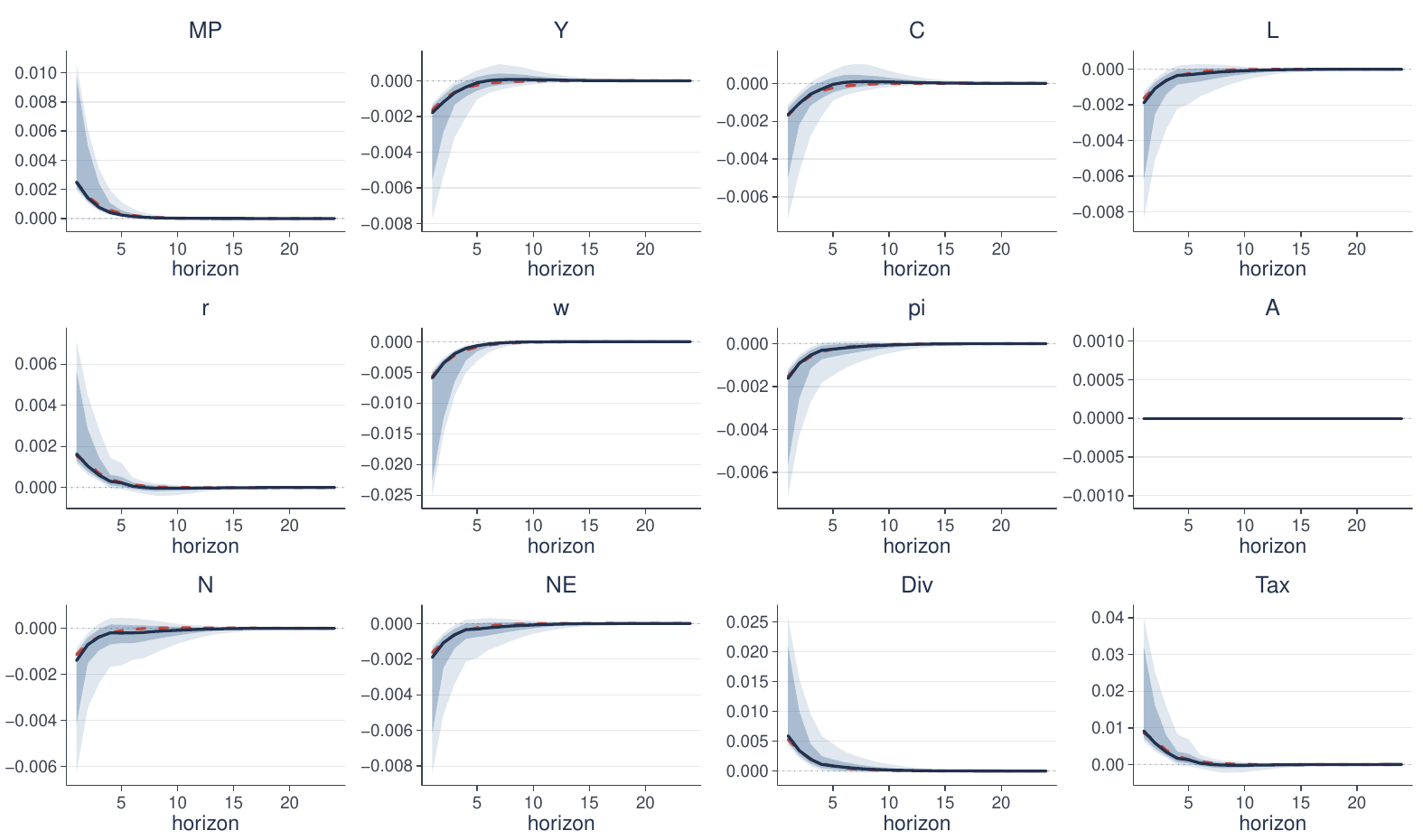}
  \end{subfigure}\\[0.4em]
  \begin{subfigure}[b]{0.80\textwidth}
    \centering
    \caption{Quantile IRFs: consumption}
    \label{fig:hank_quant_cons}
    \includegraphics[width=\textwidth, trim=0 0 0 10bp, clip]{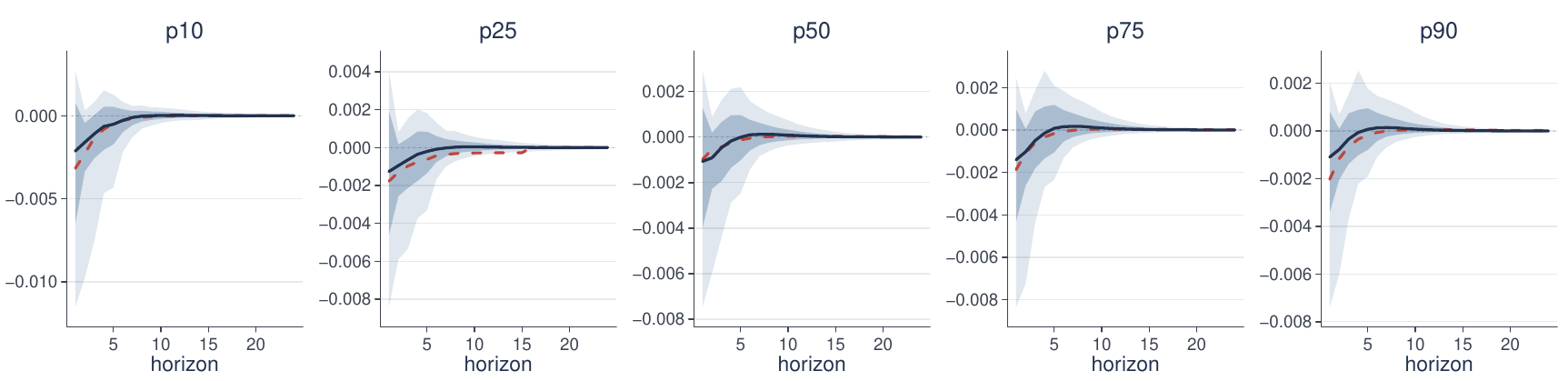}
  \end{subfigure}\\[0.4em]
  \begin{subfigure}[b]{0.80\textwidth}
    \centering
    \caption{Quantile IRFs: earnings}
    \label{fig:hank_quant_earn}
    \includegraphics[width=\textwidth, trim=0 0 0 10bp, clip]{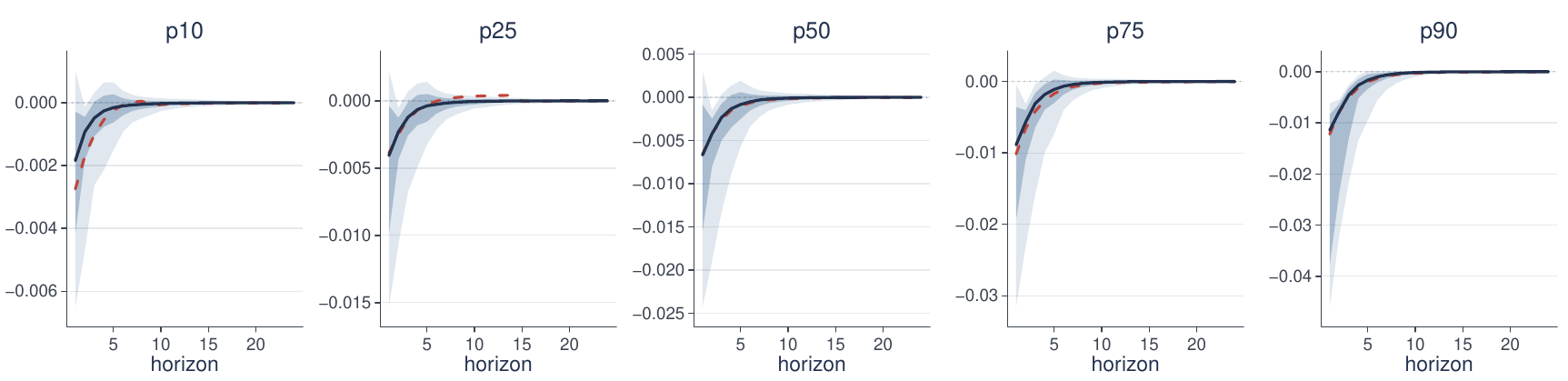}
  \end{subfigure}
  \caption*{\footnotesize \textbf{Notes}: Panel (a) reports responses of the twelve macro variables to a monetary policy shock on a $3\times4$ grid. Aggregate assets ($A$) equal the fixed bond supply and appear as a flat line on a {narrow} symmetric scale{ ($\pm 10^{-3}$)}. Panels (b) and (c) report responses of the 10th, 25th, 50th, 75th, and 90th percentiles of consumption and earnings. Navy lines are posterior medians, shaded areas are 68\% and 90\% credible bands, and dashed red lines are the true HANK responses. IRFs are normalized so that the estimated impact response of the policy variable equals the truth at horizon~1.}
  \caption{HANK simulation: macro and quantile IRFs}
  \label{fig:hank_irfs}
\end{figure}

\autoref{fig:hank_irfs} compares the estimated responses to a monetary policy shock with the true HANK responses. Panel~(a) reports responses for the twelve aggregate variables, while panels~(b) and~(c) report responses for the 10th, 25th, 50th, 75th, and 90th percentiles of consumption and earnings.

After normalizing the estimated impact response of the policy variable to match the truth, the aggregate responses in panel~(a) reproduce the signs and decay of the HANK responses. Posterior medians closely track the truth for output, consumption, the real rate, inflation, wages, dividends, and taxes. The labor-market responses are slightly attenuated on impact, but the discrepancies are concentrated at short horizons and the posterior bands contain the true paths.

For consumption (see panel~(b)), the model recovers the negative impact response across the distribution, its decay, and the larger responses in the tails than at the median. Posterior medians nevertheless attenuate the impact contraction at every percentile except the median and return to zero too quickly. Earnings responses, shown in panel~(c), are recovered more closely. The medians reproduce the negative impact response and its increasing absolute magnitude toward the top of the distribution. The main localized error is a small positive overshoot in the true 25th-percentile response at intermediate horizons. The 90\% credible bands, which are wide at short horizons, generally contain the true consumption and earnings paths.

Overall, our approach recovers the broad aggregate and distributional propagation of the HANK economy. The discrepancies discussed above concern the point estimates, summarized by the posterior medians. Once we account for posterior uncertainty, the credible bands generally contain the true aggregate and distributional response paths.

\section{Structural Evidence from U.S. Repeated Cross Sections}\label{sec:empirical}
We combine a four-variable aggregate VAR with CPS earnings and CEX consumption distributions. Earnings and consumption enter as separate marginal distributions linked only through the aggregate state and the latent factor. The CPS and CEX samples therefore need not contain the same households, and we do not link individual records across surveys. We report distributional responses to identified fiscal and monetary policy shocks and aggregate responses to the common micro shock. The monetary policy responses provide the targets for the HANK model comparison in \autoref{sec:hankbench}; the responses to the fiscal and common micro shocks are additional empirical results.

\subsection{Data and Identification}\label{sub:emp_data_spec}
We follow the small-scale macroeconomic VAR in \citet{baumeister2018inference}, adding only the Ben Zeev--Pappa fiscal-news series \citep{benzeev2017}, denoted $\text{BZP}_t$. The aggregate vector contains $M=4$ endogenous variables:
\begin{equation*}
  \bm Q_t = (\text{BZP}_t, OG_t, \pi_t, R_t)'.
\end{equation*}
Here $OG_t = 100 \cdot \log(\text{GDPC1}_t / \text{GDPPOT}_t)$ is the output gap based on Congressional Budget Office potential output, $\pi_t$ is annual log-difference inflation in personal consumption expenditures (\texttt{PCECTPI}), and $R_t$ is the federal funds rate (FFR). All series are quarterly, and the sample runs from $1990$Q2 through $2007$Q4. Availability of the BZP series determines the end date, which also keeps the zero lower bound and pandemic periods outside the sample.

The BZP series measures news about future defense spending. \citet{benzeev2017} identify the corresponding shock by requiring it to be orthogonal to current defense spending and to best explain subsequent movements in defense spending. We obtain the series from the harmonized structural-shock dataset maintained by Jonathan Adams\footnote{\url{https://github.com/jonathanjadams/structuralshocks}.} and documented in \citet{adams2025empirical} and \citet{adams2026ricardian}.

We construct the earnings and consumption cross sections as in \citet{chang2024heterogeneity} and \citet{chang2024monetary}. For earnings, we combine the monthly CPS weekly-earnings variable, annualized to a yearly rate, with the CPS employment indicator. Following \citet{chang2024heterogeneity}, we define relative earnings as
\begin{equation*}
  z_{i,t} = \frac{\text{annual earnings}_{i,t}}{\tfrac{2}{3}\cdot\text{nominal per-capita GDP}_t},
\end{equation*}
where $2/3$ approximates labor's share of gross domestic product (GDP). The mixture model uses the inverse hyperbolic sine (IHS) transformation,
\begin{equation*}
  x_{i,t} = \operatorname{asinh}(z_{i,t}) = \log\left(z_{i,t}+\sqrt{z_{i,t}^{2}+1}\right),
\end{equation*}
which is approximately linear near zero and behaves like $\log z_{i,t}+\log 2$ for large positive values of $z_{i,t}$. The transformation compresses the right tail of the CPS earnings distribution. It is the same transformation used in \citet[eq.~(30)]{chang2024heterogeneity} and \citet[eq.~(26)]{chang2024monetary}. The earnings cross sections consist of employed individuals with reported weekly earnings. They contain no mass at zero, so the earnings distribution describes the intensive margin among the employed. The quantiles of our processed earnings data closely match those of \citet{chang2024heterogeneity}.

For consumption, we use quarterly household expenditures from the CEX. Following \citet{chang2024monetary}, we normalize household consumption by nominal National Income and Product Accounts (NIPA) consumption per person aged 16 or older:
\[
  z^{c}_{i,t} = \frac{\text{consumption}_{i,t}}{C^{\text{NIPA}}_t / \text{pop}^{16+}_t}.
\]
The mixture model uses $z^{c}_{i,t}$ directly. Consumption has a less extreme right tail than earnings, so we do not apply the IHS transformation. The quarterly cross sections contain about $12{,}000$ to $16{,}000$ individual observations for earnings and $4{,}500$ to $8{,}200$ households for consumption. We represent each quarterly cross section using $500$ equally weighted quantile points, which closely preserve the shape of the empirical distribution. The common grid prevents differences in survey sample sizes across dates and outcomes from mechanically changing their relative influence on the estimates. The empirical results are conditional on this fixed-grid representation, and the micro likelihood treats the grid points as the cross-sectional observations. Such an approximation helps to lower the influence of extreme outliers in the micro data. Alternatively, one could clean the data from outliers, i.e. trim or winsorize the raw microdata, as in related applications using earnings and consumption data \citep[e.g.,][]{heathcote2010macro,moffitt2012trends,coibion2021consumption}. In our Monte Carlo exercises, we instead directly use simulated micro data.  

Both micro variables normalize household outcomes by aggregate quantities: earnings by labor-share-adjusted nominal GDP per capita and consumption by nominal NIPA consumption per person aged 16 or older. We report density, quantile, and inequality responses in these normalized units. An earnings quantile response of $+0.05$ means that the corresponding percentile of $x_{i,t}$ rises by $0.05$. Since $\operatorname{asinh}'(1)=1/\sqrt{2}$, a $0.05$ change in $x_{i,t}$ corresponds locally to a change of about $0.07$ in $z_{i,t}$ when $z_{i,t}=1$, or roughly a 7\% increase in earnings relative to two-thirds of nominal GDP per capita. Consumption is not transformed, so a response of $0.05$ raises the ratio to per-capita NIPA consumption by five percentage points. We do not convert these responses into raw earnings or consumption levels because doing so would also require the response of each aggregate normalizer and, for earnings, inversion of the IHS transformation.

The multinomial logistic weights depend on a constant, contemporaneous $\bm Q_t$, and its first four lags, matching the VAR lag order $P=4$. A single latent factor ($R=1$), denoted $f_t$, captures distributional movements not explained by the macro aggregates and enters the aggregate VAR through the loading vector $\bm\Lambda_q$. We set its loading in the BZP equation to zero, so $f_t$ does not enter that equation contemporaneously, and restrict its loadings in the output-gap and inflation equations to be positive. For each cross section, we order the component means to resolve label switching.

We choose the number of components $G_s$ using the Widely Applicable Information Criterion (WAIC) \citep{watanabe2010,gelman2014waic}, computed from posterior draws for $G_s\in\{4,5,6,8,10\}$. \autoref{app:model_fit_selection} reports the comparison. The fitted six-component mixtures closely track the observed consumption distributions and reproduce the location, spread, and skewness of the earnings distributions at the dates shown in \autoref{fig:emp_densities_ot}. The estimated mixture weights vary substantially over time (\autoref{fig:emp_weights}). As in the HANK exercise, the components are approximation devices, so we interpret the density and quantile responses implied jointly by all components rather than any individual weight path.

We combine seven sets of restrictions. They identify four shocks in the VAR block and the common micro shock. The four VAR shocks are supply, demand, monetary policy, and fiscal policy shocks. We jointly identify them using the sign pattern of \citet{baumeister2018inference}, the exogeneity restriction on BZP, and impact sign restrictions. The common micro shock is the latent-factor innovation. Restrictions on its aggregate effects and on its values at selected policy events identify its direction. \autoref{tab:restrictions} summarizes these restrictions, and \autoref{app:technical} provides details.
\begin{table}[htbp]
  \centering
  \caption{Identifying restrictions in the U.S. application}
  \label{tab:restrictions}
  \small
  \begin{tabular}{@{}p{0.30\textwidth}p{0.46\textwidth}p{0.16\textwidth}@{}}
    \toprule
    Object restricted & Restriction & Type \\
    \midrule
    Contemporaneous $(OG,\pi,R)$ block of $\bm A_0$ & Sign and magnitude bounds calibrated using the priors in \citet{baumeister2018inference} (\autoref{tab:Wbounds}) & Sign, magnitude \\
    First row of $\bm W$ (BZP equation) & BZP does not respond contemporaneously to the other VAR variables (exogeneity) & Zero \\
    Impact responses to fiscal and monetary shocks & The fiscal shock raises the output gap; the contractionary monetary shock raises the federal funds rate and lowers the output gap and inflation & Sign \\
    Latent factor $f_t$ at six policy events & The sign of $f_t$ is restricted at the 1993 Omnibus Budget Reconciliation Act, the minimum-wage increases to \$4.75, \$5.15, and \$5.85, the Bush rebates, and the 2003 Jobs and Growth Tax Relief Reconciliation Act & Narrative \\
    Aggregate loadings of $f_t$ in $\bm\Lambda_q$ & The BZP loading is set to zero; the output-gap and inflation loadings are restricted to be positive & Zero, sign \\
    Macro loadings in the log-weight equations & A higher output gap or inflation shifts mass toward higher-mean components. A higher policy rate shifts mass toward lower-mean components. The coefficient floors are $0.02$ and $0.08$. & Sign, magnitude \\
    Factor loadings in the log-weight equations & A positive factor innovation moves mass from both tails toward the middle components. The magnitude floors are $0.05$ for the tails and $0.08$ for the middle. & Sign, magnitude \\
    \bottomrule
  \end{tabular}
\end{table}

\subsection{Distributional Effects of Monetary and Fiscal Policy Shocks}\label{sub:emp_macro_shocks}
We focus on monetary and fiscal policy shocks. The joint identification also includes supply and demand shocks, but we do not report their responses. {We scale each reported shock to three standard deviations.} For each shock, one figure reports aggregate responses and another reports density and quantile responses; a separate figure collects the inequality responses for both shocks.

\subsubsection{Responses to a Fiscal Policy Shock}
\begin{figure}[!htb]
  \centering
  \includegraphics[width=\textwidth, trim=0 3bp 0 10bp, clip]{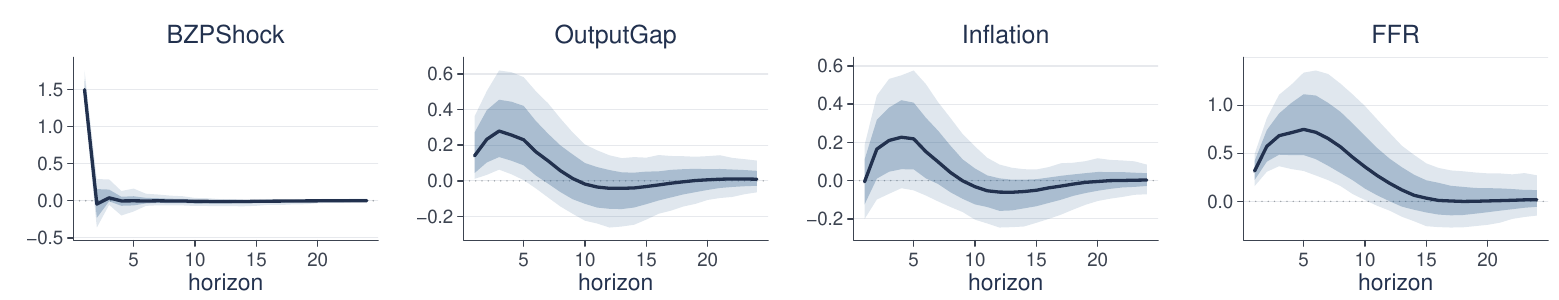}
  \caption*{\footnotesize \textbf{Notes}: Aggregate responses (BZP, the output gap, inflation, and the federal funds rate) to a three-standard-deviation fiscal shock. Navy lines are posterior medians, and shaded areas are 68\% and 90\% credible bands.}
  \caption{Macro IRFs to the fiscal shock}
  \label{fig:emp_fiscal_macro}
\end{figure}

The fiscal-shock identification imposes an increase in the output gap on impact. The response paths at later horizons are learned from the posterior. The median output-gap response is hump-shaped between roughly quarters two and five, while the federal funds rate and inflation also rise (\autoref{fig:emp_fiscal_macro}). The inflation response is less precisely estimated.

The posterior median output gap peaks at about $0.3$ percent around quarter three, while inflation rises by about $0.23$ percentage points over the same interval. The federal funds rate response peaks near $0.7$ percentage points around quarter five and then declines slowly. The joint increases in activity, inflation, and the policy rate are consistent with the central bank leaning against a demand expansion.

\begin{figure}[!t]
  \centering
  \begin{subfigure}[b]{\textwidth}
    \centering
    \caption{Density IRF: earnings}
    \label{fig:emp_fiscal_dens_earn}
    \includegraphics[width=\textwidth, trim=0 0 0 10bp, clip]{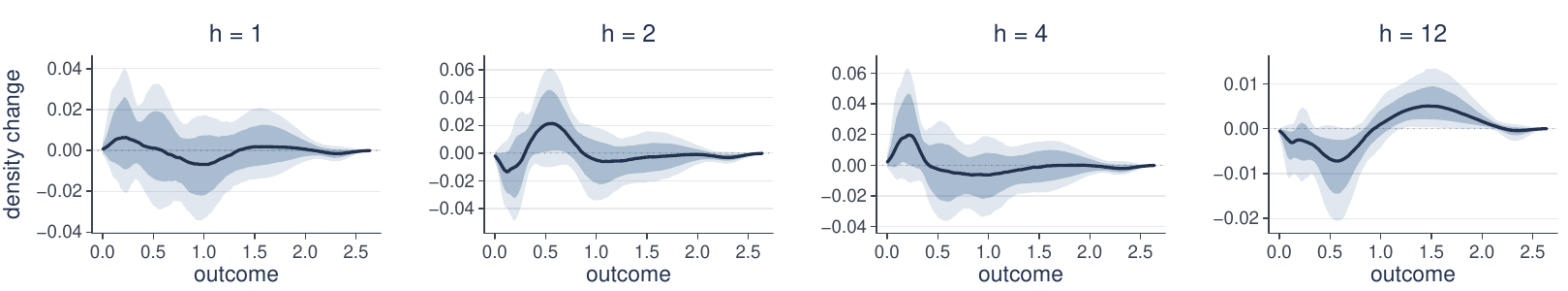}
  \end{subfigure}\\[0.2em]
  \begin{subfigure}[b]{\textwidth}
    \centering
    \caption{Density IRF: consumption}
    \label{fig:emp_fiscal_dens_cons}
    \includegraphics[width=\textwidth, trim=0 0 0 10bp, clip]{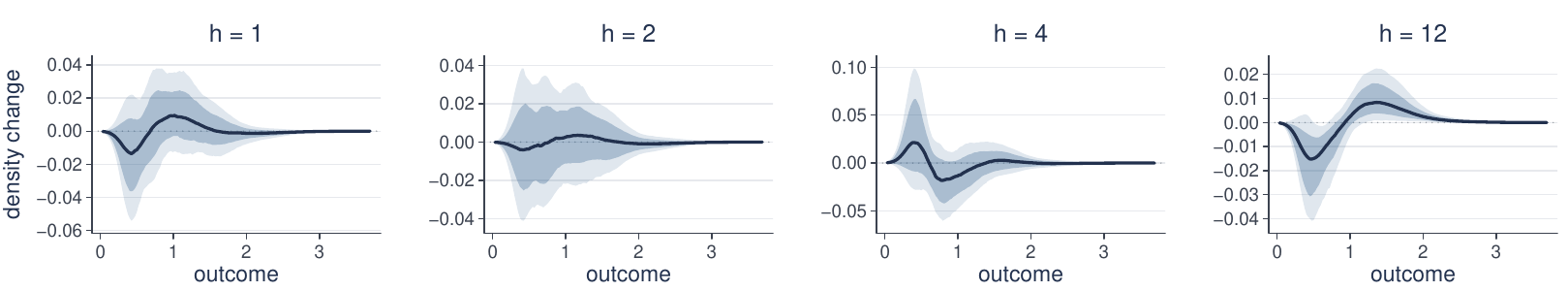}
  \end{subfigure}\\[0.2em]
  \begin{subfigure}[b]{\textwidth}
    \centering
    \caption{Quantile IRF: earnings}
    \label{fig:emp_fiscal_quant_earn}
    \includegraphics[width=\textwidth, trim=0 0 0 10bp, clip]{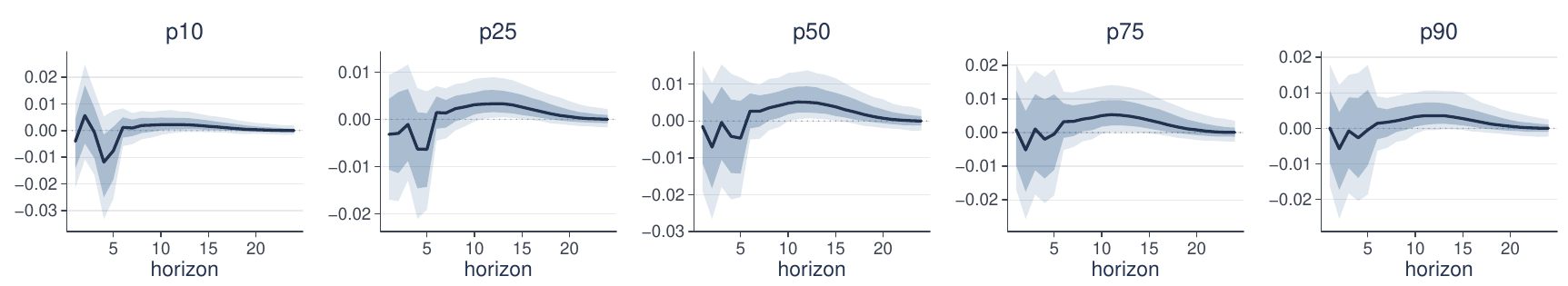}
  \end{subfigure}\\[0.2em]
  \begin{subfigure}[b]{\textwidth}
    \centering
    \caption{Quantile IRF: consumption}
    \label{fig:emp_fiscal_quant_cons}
    \includegraphics[width=\textwidth, trim=0 0 0 10bp, clip]{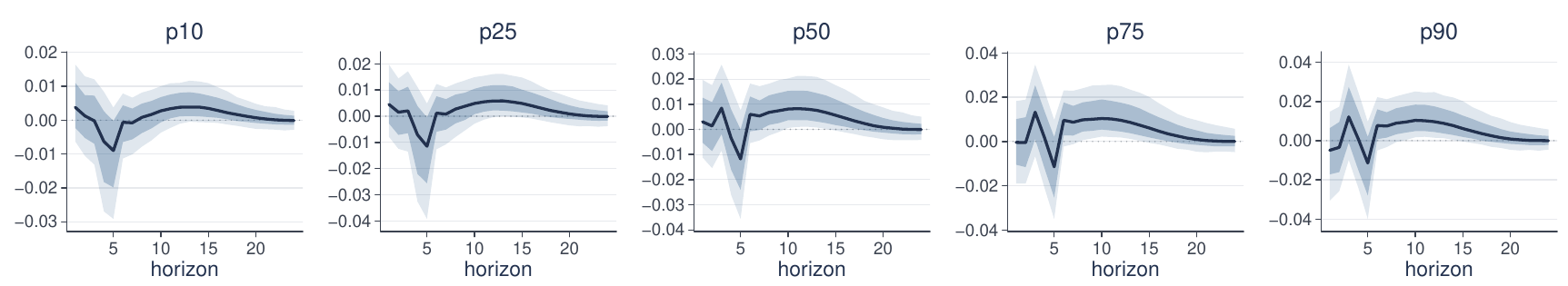}
  \end{subfigure}
  \caption*{\footnotesize \textbf{Notes}: Panels (a) and (b) report the density change $\Delta f(x;h)=f_{\mathrm{shocked}}(x;h)-f_{\mathrm{baseline}}(x;h)$ at selected horizons in response to a three-standard-deviation fiscal shock, and panels (c) and (d) report the 10th, 25th, 50th, 75th, and 90th percentiles of earnings and consumption. Navy lines are posterior medians, and shaded areas are 68\% and 90\% credible bands.}
  \caption{Density and quantile IRFs to the fiscal shock}
  \label{fig:emp_fiscal_dist}
\end{figure}

The posterior median density responses in panels (a) and (b) of \autoref{fig:emp_fiscal_dist} trace how the fiscal shock redistributes probability mass. For earnings, mass initially shifts toward below-average values. At longer horizons, it moves from the lower to the upper part of the distribution, consistent with positive responses across the earnings quantiles. For consumption, mass accumulates at low values around quarter four, but this shift reverses at later horizons.

The posterior median earnings-quantile responses in panel (c) of \autoref{fig:emp_fiscal_dist} oscillate at short horizons. They dip around quarter four, coinciding with the peak in the federal funds rate, and are positive at every reported percentile from quarter eight onward. At their peak, they reach about $0.005$ in the IHS-transformed earnings ratio, which near $z_{i,t}=1$ corresponds to roughly a $0.7$ percent increase in earnings relative to two-thirds of nominal GDP per capita. The posterior median consumption-quantile responses in panel (d) fall around quarter five by about two percentage points in the normalized consumption ratio at the median and higher percentiles, then exhibit a small positive hump. The 90\% posterior intervals contain zero at most horizons. Taken together, the posterior medians indicate an eventual rightward shift in the earnings distribution and a temporary consumption decline concentrated in its upper half.

\subsubsection{Responses to a Monetary Policy Shock}

\begin{figure}[!htb]
  \centering
  \includegraphics[width=\textwidth, trim=0 3bp 0 10bp, clip]{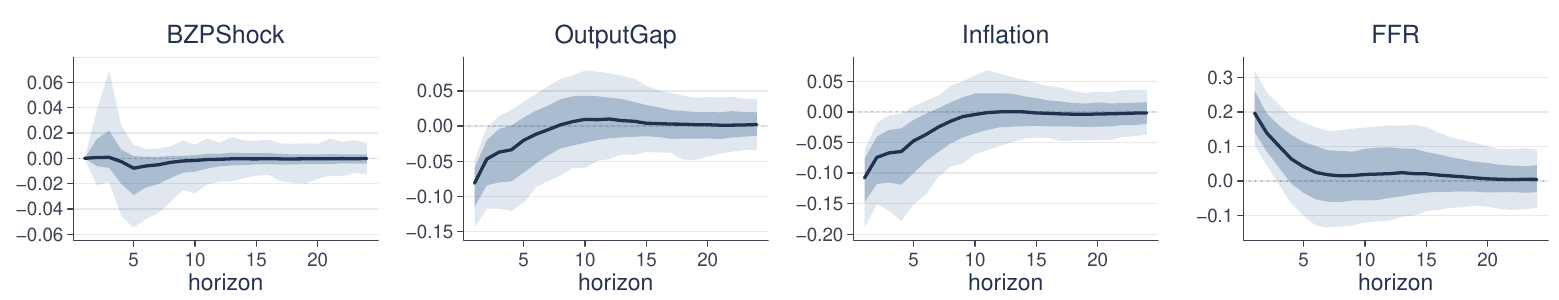}
  \caption*{\footnotesize \textbf{Notes}: Aggregate responses (BZP, the output gap, inflation, and the federal funds rate) to a three-standard-deviation monetary policy shock. Navy lines are posterior medians, and shaded areas are 68\% and 90\% credible bands.}
  \caption{Macro IRFs to the monetary policy shock}
  \label{fig:emp_mp_macro}
\end{figure}

The sign restrictions define a contractionary monetary policy shock as one that raises the federal funds rate and lowers the output gap and inflation on impact. The response paths beyond impact are learned from the posterior. The median policy-rate response rises by about $0.20$ percentage points on impact and {declines substantially within roughly five quarters}. The median output-gap and inflation responses fall by about $0.08$ percent and $0.11$ percentage points, respectively, and {recover most of the decline} within roughly five to seven quarters. All median aggregate responses are close to zero {by around quarters ten to twelve} (\autoref{fig:emp_mp_macro}).

\begin{figure}[!t]
  \centering
  \begin{subfigure}[b]{\textwidth}
    \centering
    \caption{Density IRF: earnings}
    \label{fig:emp_mp_dens_earn}
    \includegraphics[width=\textwidth, trim=0 0 0 10bp, clip]{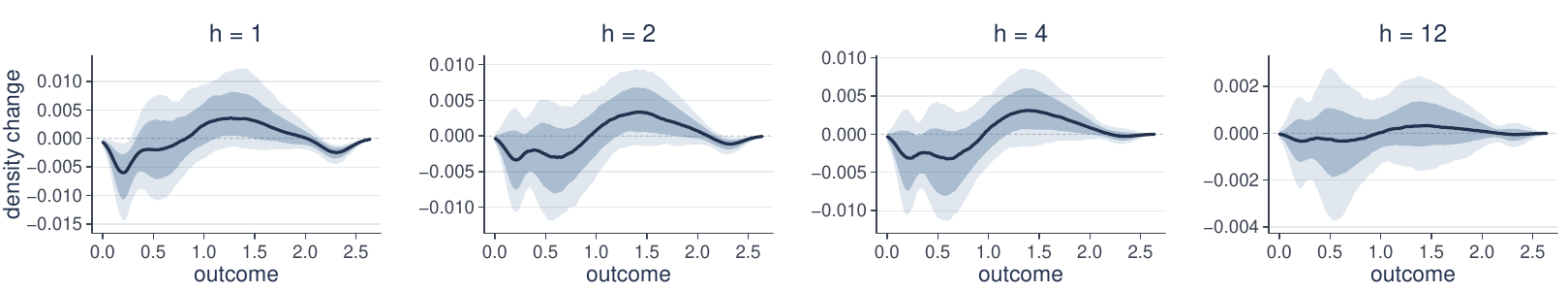}
  \end{subfigure}\\[0.2em]
  \begin{subfigure}[b]{\textwidth}
    \centering
    \caption{Density IRF: consumption}
    \label{fig:emp_mp_dens_cons}
    \includegraphics[width=\textwidth, trim=0 0 0 10bp, clip]{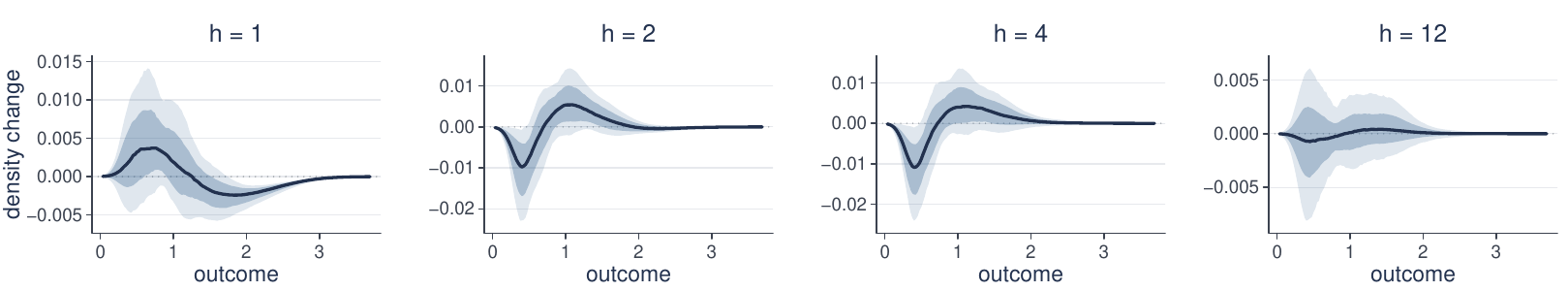}
  \end{subfigure}\\[0.2em]
  \begin{subfigure}[b]{\textwidth}
    \centering
    \caption{Quantile IRF: earnings}
    \label{fig:emp_mp_quant_earn}
    \includegraphics[width=\textwidth, trim=0 0 0 10bp, clip]{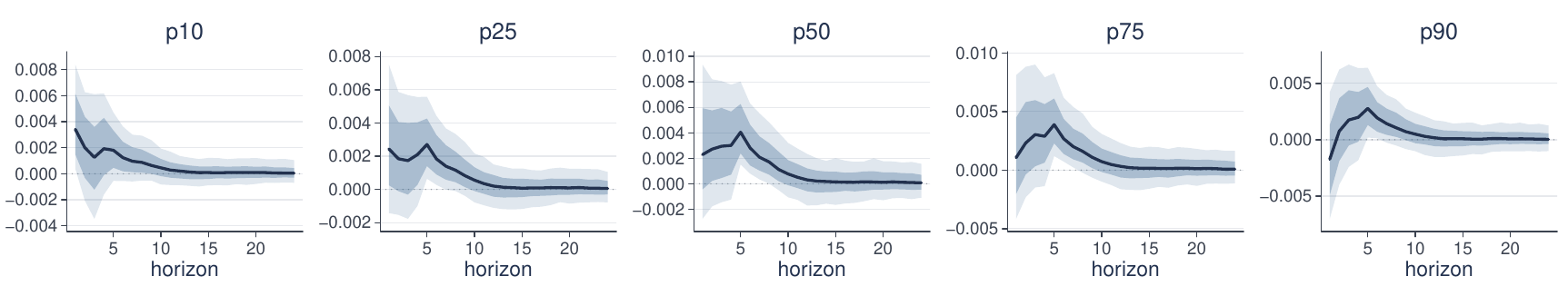}
  \end{subfigure}\\[0.2em]
  \begin{subfigure}[b]{\textwidth}
    \centering
    \caption{Quantile IRF: consumption}
    \label{fig:emp_mp_quant_cons}
    \includegraphics[width=\textwidth, trim=0 0 0 10bp, clip]{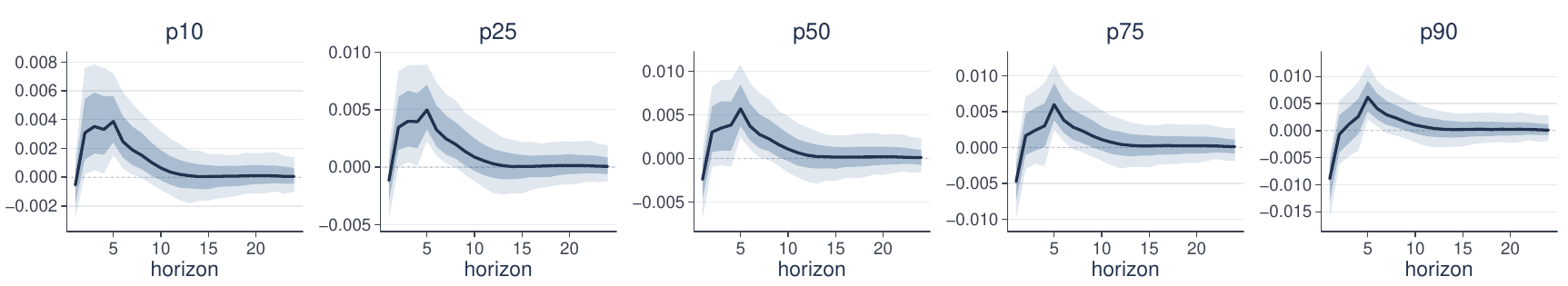}
  \end{subfigure}
  \caption*{\footnotesize \textbf{Notes}: Panels (a) and (b) report the density change $\Delta f(x;h)=f_{\mathrm{shocked}}(x;h)-f_{\mathrm{baseline}}(x;h)$ at selected horizons in response to a three-standard-deviation monetary policy shock, and panels (c) and (d) report the 10th, 25th, 50th, 75th, and 90th percentiles of earnings and consumption. Navy lines are posterior medians, and shaded areas are 68\% and 90\% credible bands.}
  \caption{Density and quantile IRFs to the monetary policy shock}
  \label{fig:emp_mp_dist}
\end{figure}

The posterior median density responses in panels (a) and (b) of \autoref{fig:emp_mp_dist} trace a short-lived redistribution of probability mass in both distributions. On impact, earnings mass moves from the upper part of the distribution toward its center. By quarter four, the direction reverses, with mass shifting from below-average toward above-average earnings. Consumption mass moves on impact from the middle and upper parts of the distribution toward low values. This shift begins to unwind around quarter four and has largely disappeared by quarter twelve.

At the posterior medians, the monetary tightening compresses the earnings distribution on impact: the tenth percentile rises slightly, while the ninetieth and other upper quantiles fall (panel (c) of \autoref{fig:emp_mp_dist}). All reported consumption quantiles also fall on impact (panel (d)). Across most percentiles, both distributions then exhibit a small positive hump between quarters three and six. The cross-quantile location of the consumption decline provides a diagnostic for the mechanism in \citet{kaplan2018monetary}. If transmission were concentrated among hand-to-mouth households and these households sat mainly in the lower part of the consumption distribution, the decline would concentrate in the lower and middle consumption quantiles, whereas our posterior medians place it at and above the median. The credible sets admit both patterns, so neither transmission mechanism is ruled out. Consumption ranks do not identify liquidity status, so this comparison is informative only through the mapping a given model implies.

\subsubsection{Inequality IRFs}
\begin{figure}[!t]
  \centering
  \begin{subfigure}[b]{\textwidth}\centering
    \caption{Fiscal shock: earnings}
    \includegraphics[width=\textwidth, trim=0 0 0 10bp, clip]{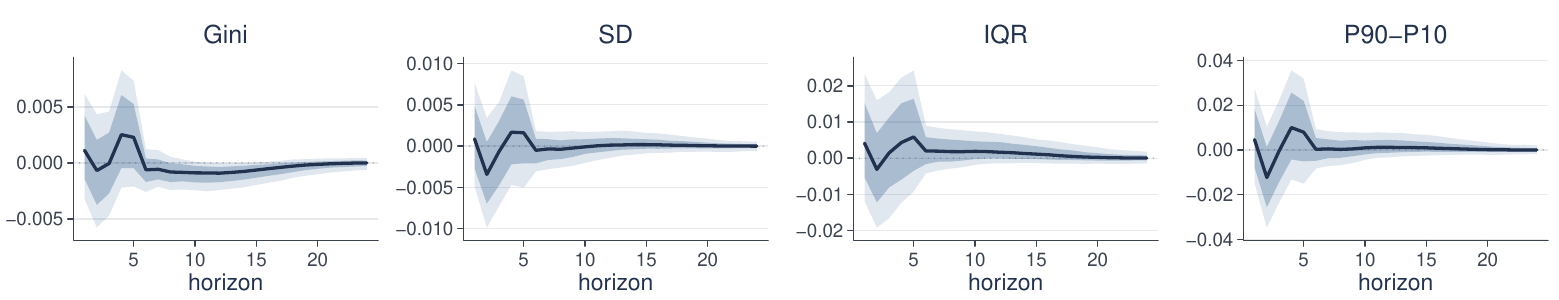}\end{subfigure}\\[0.2em]
  \begin{subfigure}[b]{\textwidth}\centering
    \caption{Fiscal shock: consumption}
    \includegraphics[width=\textwidth, trim=0 0 0 10bp, clip]{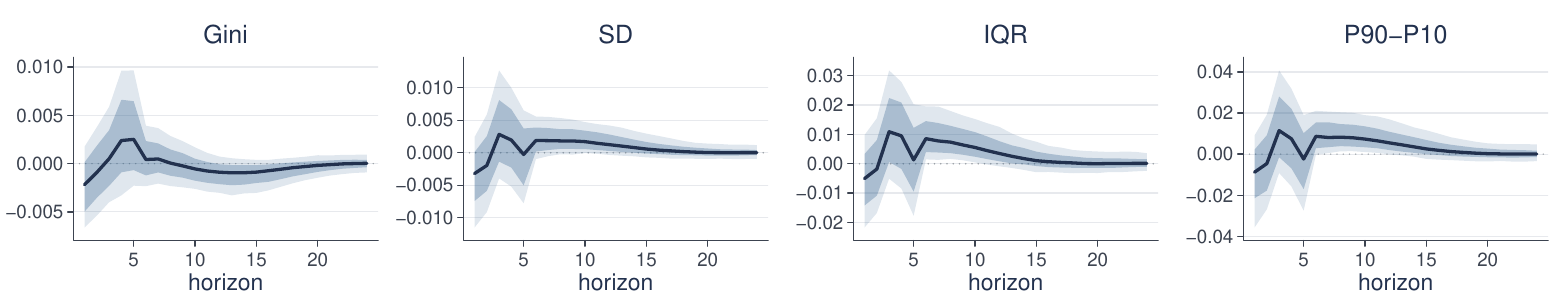}\end{subfigure}\\[0.2em]
  \begin{subfigure}[b]{\textwidth}\centering
    \caption{Monetary-policy shock: earnings}
    \includegraphics[width=\textwidth, trim=0 0 0 10bp, clip]{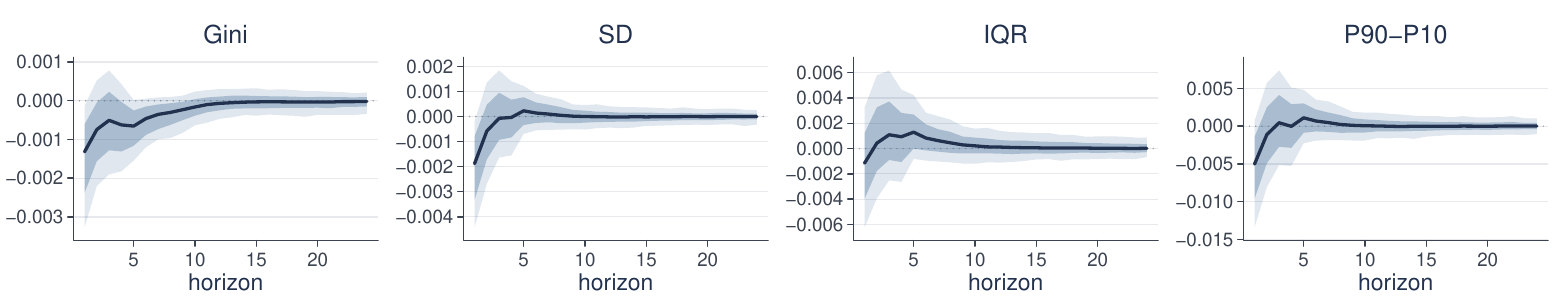}\end{subfigure}\\[0.2em]
  \begin{subfigure}[b]{\textwidth}\centering
    \caption{Monetary-policy shock: consumption}
    \includegraphics[width=\textwidth, trim=0 0 0 10bp, clip]{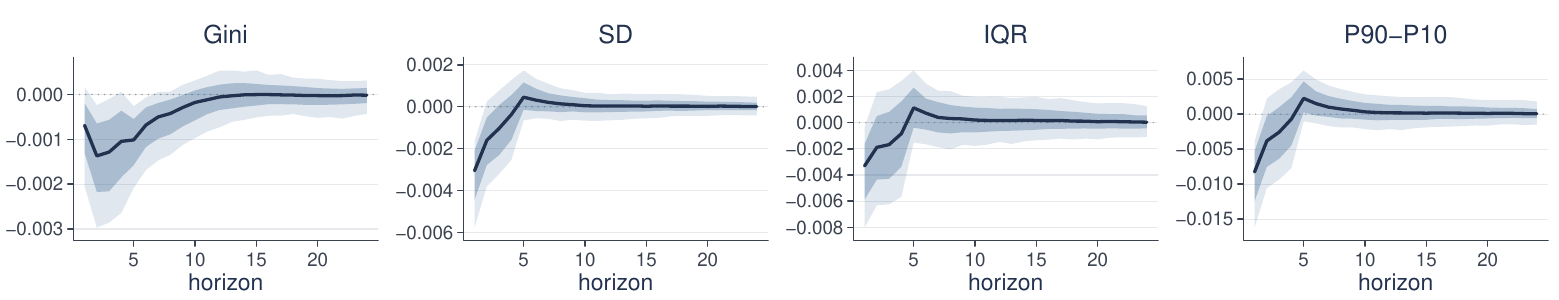}\end{subfigure}
  \caption*{\footnotesize \textbf{Notes}: Panels (a) and (b) report responses to a three-standard-deviation fiscal shock, panels (c) and (d) responses to a three-standard-deviation monetary policy shock. Each panel reports the responses of the Gini coefficient, standard deviation, interquartile range, and P90--P10 spread (90th minus 10th percentile) of earnings or consumption. Navy lines are posterior medians, and shaded areas are 68\% and 90\% credible bands.}
  \caption{Inequality IRFs to the fiscal and monetary policy shocks}
  \label{fig:emp_inequality_main}
\end{figure}

To summarize the distributional responses, \autoref{fig:emp_inequality_main} reports impulse responses of four inequality measures: the Gini coefficient, the standard deviation, the interquartile range, and the P90--P10 spread, each computed from the fitted mixture on the scale used in estimation (IHS-transformed earnings and normalized consumption). Panels (a) and (b) report earnings and consumption responses to the fiscal shock; panels (c) and (d) report the corresponding responses to the monetary policy shock.

The fiscal shock produces little precisely estimated movement in either earnings or consumption inequality (panels (a) and (b)). The posterior medians of all four earnings measures decline in quarter two, rise in quarters four and five, and fade after about six quarters, mirroring the oscillation in the earnings quantiles. For consumption, the posterior median responses rise around quarter three, briefly reverse, and remain slightly positive through about quarter ten. The measures differ in timing and magnitude. Only the rise in consumption dispersion around quarter three is estimated with some precision; the 90\% credible bands include zero at almost all horizons.

The monetary tightening compresses both distributions on impact (panels (c) and (d)). For earnings, the posterior medians of the standard deviation, interquartile range, and P90--P10 spread fall on impact. The 68\% credible bands of the standard deviation and the P90--P10 spread lie below zero. These responses return to zero within about five quarters. The compression in consumption is larger and more precisely estimated. The P90--P10 spread falls by about $0.8$ percentage points of the consumption ratio, and the 68\% credible bands for all four consumption measures lie below zero on impact. The bands of the Gini coefficient and the standard deviation remain below zero during the first three quarters. These responses return to zero around quarter five.

The inequality measures summarize both impact responses as a compression. The quantile responses in \autoref{fig:emp_mp_dist} show that the incidence differs: the tenth earnings percentile rises while the upper earnings percentiles fall, whereas consumption falls across all reported percentiles, with the largest declines in the upper tail. The summary measures alone do not reveal these cross-quantile patterns, which matter for evaluating the transmission mechanisms of heterogeneous-agent models.

Conditional on our model and priors, the earnings responses place an empirical bound on the quantitative strength of the earnings-heterogeneity channel in \citet{auclert2019monetary}. In that channel, the unequal incidence of labor-income changes causes a contractionary shock to widen the earnings distribution. Our posterior median instead shows a transitory compression in the P90--P10 spread. At each horizon, the upper edge of the 90\% credible band provides a posterior bound on widening.

\subsection{Aggregate Effects of a Common Micro Shock}\label{sub:emp_micro_shock}
The common micro shock is an aggregate innovation to the latent factor $f_t$, which captures movements shared by the earnings and consumption distributions beyond the macro block. One structural interpretation comes from heterogeneous-agent theory. In \citet{bayer2019risk}, a rise in idiosyncratic income risk raises precautionary saving and the demand for liquid assets, causing output and the policy rate to fall. We orient our shock in the reverse direction for dispersion and output: a positive innovation compresses both distributions on impact, lowering realized dispersion, while raising output and inflation.

For models with time-varying idiosyncratic risk, these responses provide two empirical diagnostics:
\begin{enumerate}
\item A dynamic multiplier from dispersion to activity, defined as the peak output gap response divided by the impact response of either the P90--P10 spread or the standard deviation.
\item The policy-rate path. Output and inflation rise on impact by construction, but the sharp impact decline in the federal funds rate is not imposed because its factor loading is unrestricted. The rate path provides a diagnostic for how a model's monetary rule responds to distributional shocks.
\end{enumerate}
\autoref{fig:emp_micro_macro} and \autoref{app:micro_inequality} contain the responses needed to construct these diagnostics; \autoref{fig:emp_micro_dist} reports the underlying density and quantile responses.

\subsubsection{Macro, Density and Quantile Responses}
Beyond the impact restrictions described above, the macro, density, and quantile response paths are learned from the posterior. The factor is excluded contemporaneously from the BZP equation, so the common micro shock does not affect identification of the fiscal shock.

\begin{figure}[!htb]
  \centering
  \includegraphics[width=\textwidth, trim=0 3bp 0 10bp, clip]{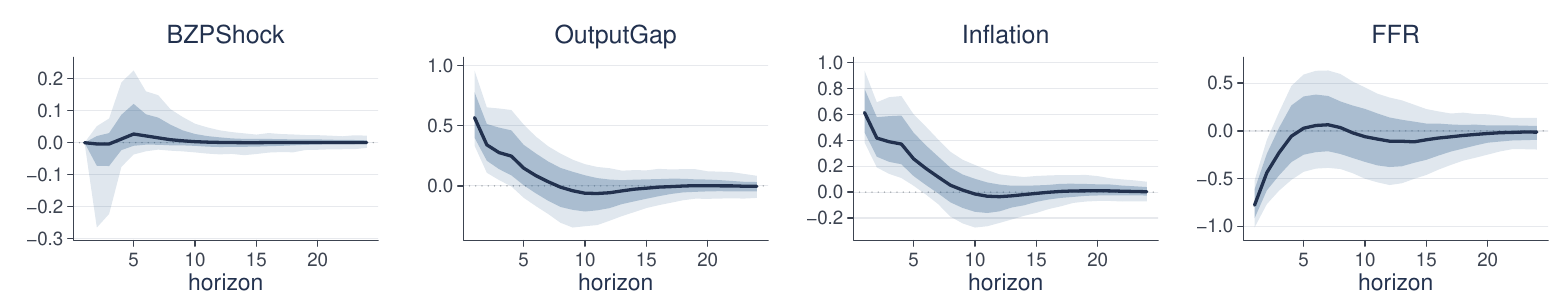}
  \caption*{\footnotesize \textbf{Notes}: Aggregate responses (BZP, the output gap, inflation, and the federal funds rate) to a three-standard-deviation shock to $f_t$. BZP is zero on impact by construction. Navy lines are posterior medians, and shaded areas are 68\% and 90\% credible bands.}
  \caption{Macro IRFs to the micro shock}
  \label{fig:emp_micro_macro}
\end{figure}

The posterior median shows a sharp impact decline in the federal funds rate (\autoref{fig:emp_micro_macro}). The aggregate responses then return toward zero and are generally close to zero after eight to ten quarters.

\begin{figure}[!t]
  \centering
  \begin{subfigure}[b]{\textwidth}
    \centering
    \caption{Density IRF: earnings}
    \label{fig:emp_micro_dens_earn}
    \includegraphics[width=\textwidth, trim=0 0 0 10bp, clip]{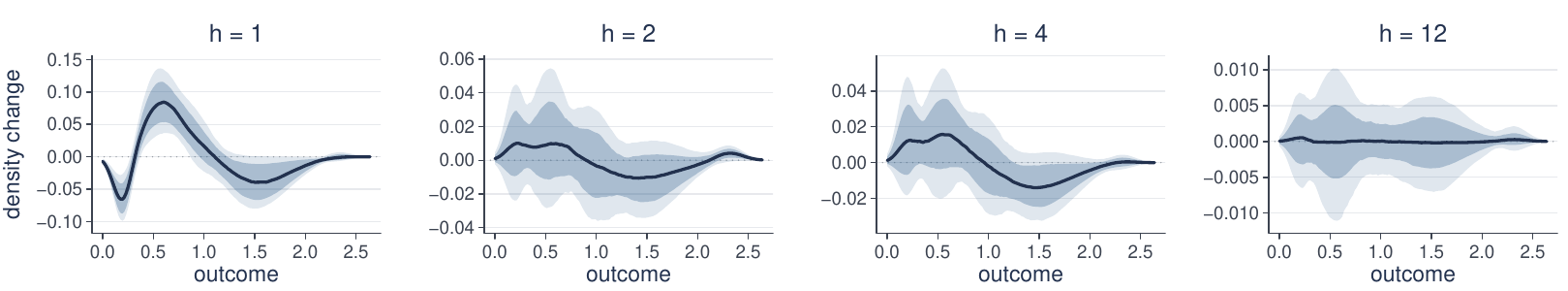}
  \end{subfigure}\\[0.2em]
  \begin{subfigure}[b]{\textwidth}
    \centering
    \caption{Density IRF: consumption}
    \label{fig:emp_micro_dens_cons}
    \includegraphics[width=\textwidth, trim=0 0 0 10bp, clip]{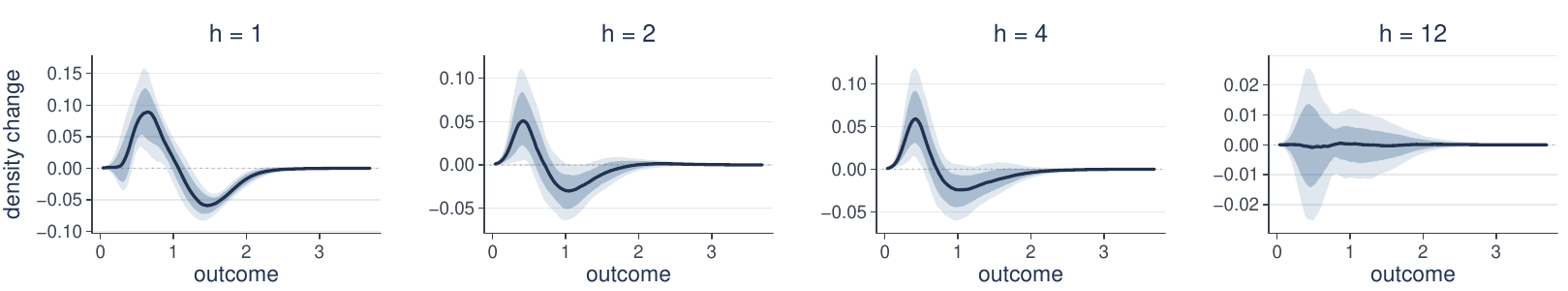}
  \end{subfigure}\\[0.2em]
  \begin{subfigure}[b]{\textwidth}
    \centering
    \caption{Quantile IRF: earnings}
    \label{fig:emp_micro_quant_earn}
    \includegraphics[width=\textwidth, trim=0 0 0 10bp, clip]{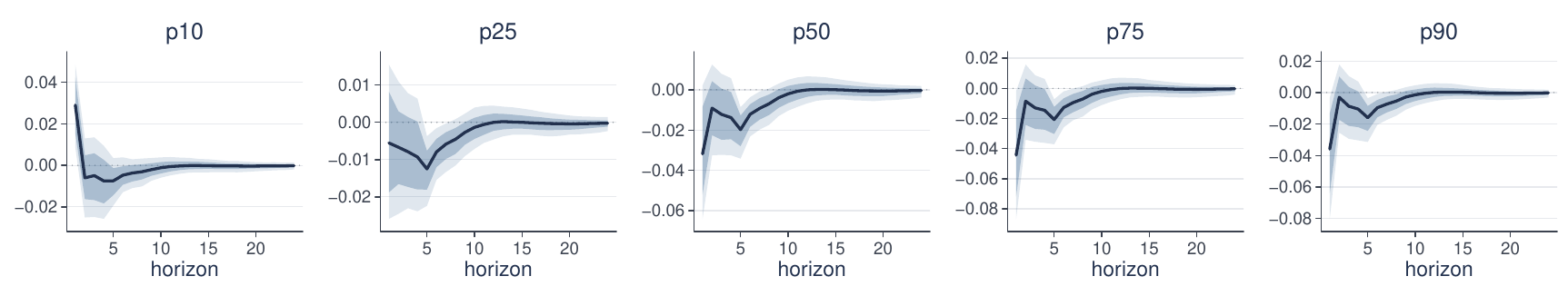}
  \end{subfigure}\\[0.2em]
  \begin{subfigure}[b]{\textwidth}
    \centering
    \caption{Quantile IRF: consumption}
    \label{fig:emp_micro_quant_cons}
    \includegraphics[width=\textwidth, trim=0 0 0 10bp, clip]{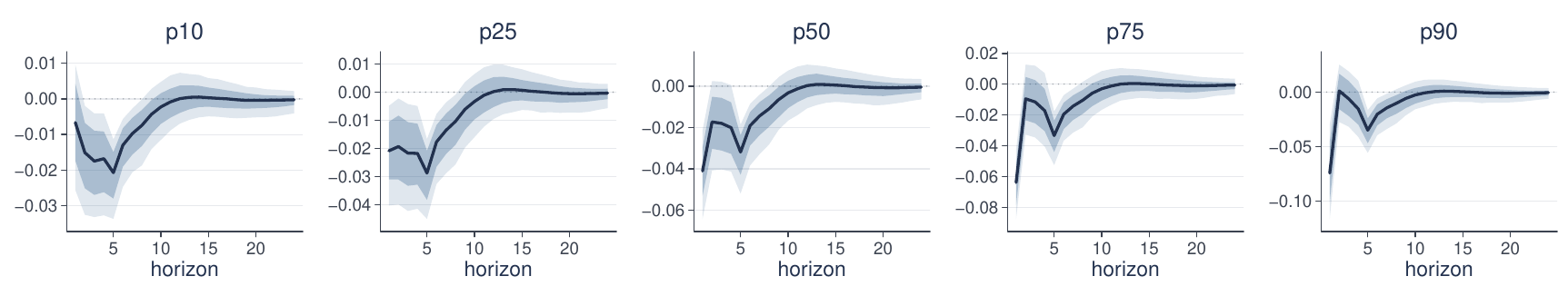}
  \end{subfigure}
  \caption*{\footnotesize \textbf{Notes}: Panels (a) and (b) report density changes $\Delta f(x;h)$ at selected horizons in response to a three-standard-deviation shock to $f_t$, and panels (c) and (d) report the 10th, 25th, 50th, 75th, and 90th percentiles of earnings and consumption. Navy lines are posterior medians, and shaded areas are 68\% and 90\% credible bands.}
  \caption{Density and quantile IRFs to the micro shock}
  \label{fig:emp_micro_dist}
\end{figure}

The density responses in panels (a) and (b) of \autoref{fig:emp_micro_dist} show the impact compression in both distributions as probability mass moves from the tails toward the center. At the posterior medians, the tenth earnings percentile rises on impact, while the median and upper percentiles fall (panel (c)). All reported consumption percentiles fall on impact, with the largest declines in the upper tail (panel (d)). Declines at the median and upper percentiles are relatively precise at short horizons, but posterior uncertainty widens rapidly thereafter.

The inequality responses in \autoref{fig:emp_micro_inequality} summarize the same impact compression. For both earnings and consumption, all four measures fall sharply on impact, as implied by the identifying restrictions on the factor loadings, and then rebound before fading. Posterior uncertainty widens rapidly after impact. Because the initial compression is imposed, the strength and persistence of the rebound are the informative features.

\section{Benchmarking an Estimated HANK Model}\label{sec:hankbench}
We benchmark the JAMM-VAR posterior responses to a monetary policy shock against the estimated HANK model of \citet{bayer2024shocks}. We take the model directly from the authors' public repository and do not alter its economic structure.\footnote{Code and parameter values are those distributed at \url{https://github.com/BASEforHANK/HANK_BusinessCycleAndInequality}. We re-estimate nothing and choose no parameter ourselves. The repository ships a solved steady state and the linear system that governs the response to each shock, so we load the model rather than re-solve it.} Two features make the model well suited to this exercise: monetary policy affects its earnings distribution, and its shock processes were estimated using data that include measures of inequality. Following \autoref{sec:hank_targets}, we compare the model-implied responses with our posterior bands horizon by horizon.

The empirical analysis in \autoref{sec:empirical} implements the first four steps of the protocol. To begin the fifth step, we scale the monetary policy shock in the HANK model so that the policy rate rises on impact by the JAMM-VAR posterior median of about $0.20$ percentage points.

\subsection{Making the HANK Objects Comparable}\label{sec:bblstep5}
Households in the model face idiosyncratic productivity risk, hold liquid and illiquid assets, and pay a cost to move funds between them. Prices and wages are sticky, and a small group of entrepreneurs receives profit income rather than labor income.

Having matched the shock size, we align the remaining model and empirical objects in their variable definitions, transformations, measurement conventions, and response horizons. We construct model earnings $z_{i,t}$ for each productivity state from the model's income function: a productivity-dependent share of the wage bill plus an equal allocation of union rents. We then apply the same inverse hyperbolic sine transformation used for the survey data, $x_{i,t}=\operatorname{asinh}(z_{i,t})$. We exclude the entrepreneurial state because its income consists of profits rather than labor earnings. Finally, we convert the model's annualized quarterly inflation rate into the four-quarter measure used in the empirical analysis by averaging it over the current and previous three quarters.

The nonlinear inverse hyperbolic sine transformation makes the level of $z_{i,t}$ relevant for comparing responses, so we make one level adjustment. The empirical mean of $z_{i,t}$ is $1.328$, whereas the model mean is $0.934$. We therefore multiply model earnings by their ratio, $1.328/0.934=1.42$. This normalization aligns the steady-state means rather than fitting the model's responses.\footnote{Because the inverse hyperbolic sine transformation is nonlinear, the level at which it is applied affects the scale of the transformed responses.}

Before comparing responses, we check whether the model's steady-state earnings distribution resembles the empirical cross section. In units of $x_{i,t}$, the tenth and ninetieth percentiles are $0.35$ and $1.68$ in the survey and $0.54$ and $1.65$ in the model. The model closely matches the ninetieth percentile but places the tenth percentile too high. Its P90--P10 range is therefore $83$ percent of the empirical range. The two distributions are broadly comparable, with the main discrepancy concentrated in the lower tail.

Two limitations restrict the comparison. First, the model has no unemployment, so it cannot reproduce movements into and out of employment. Because the CPS earnings cross sections cover employed individuals, the comparison concerns the intensive margin among the employed on both sides. Empirical earnings responses may nevertheless reflect changes in the composition of the employed, which the model does not capture. Second, the public code does not expose the joint distribution of income and wealth needed to recover the model's consumption distribution. We therefore restrict the comparison to earnings.

\subsection{Diagnosing the Model}\label{sec:bblstep6}
\begin{figure}[t]
\centering
\includegraphics[width=\textwidth]{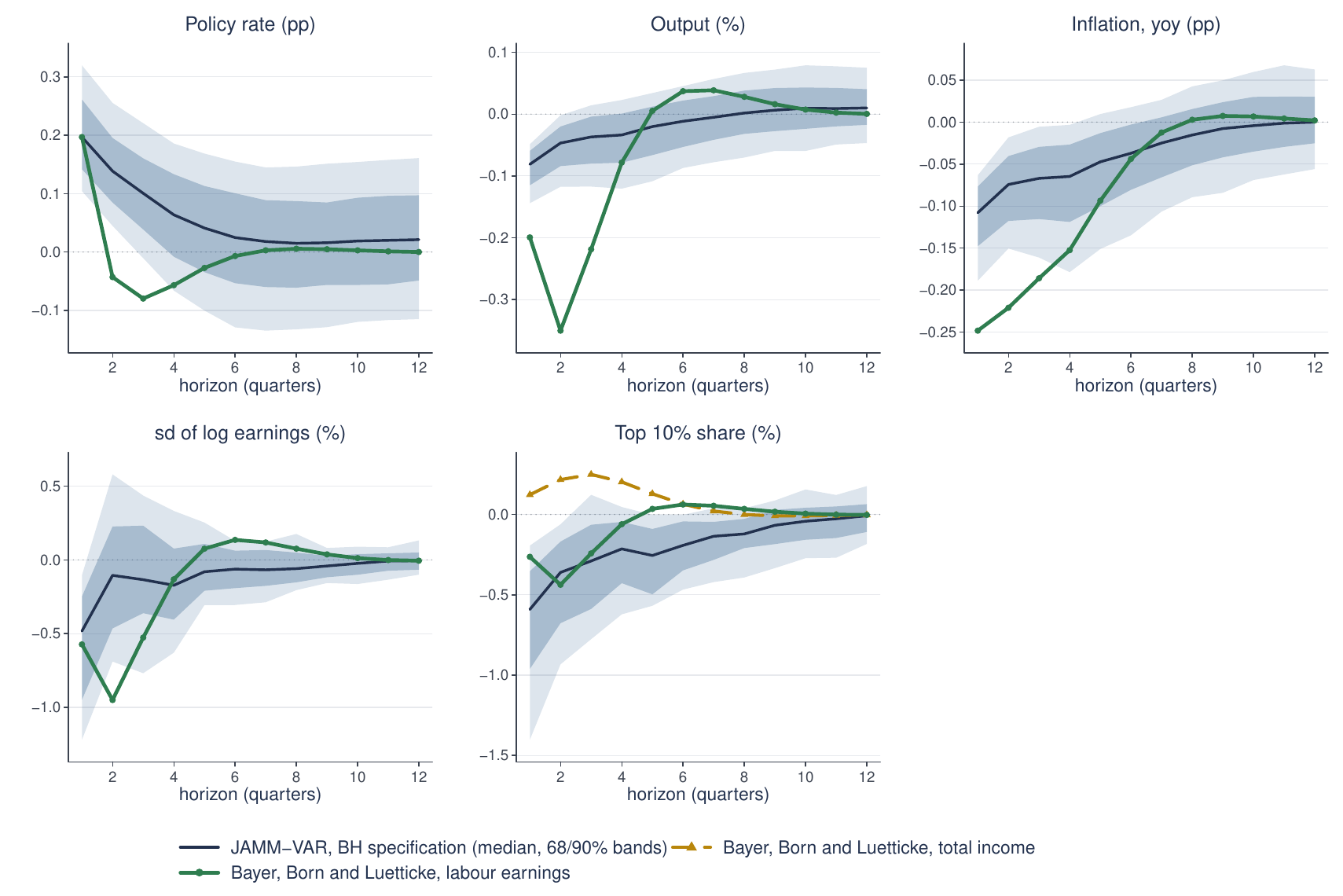}
\caption*{\footnotesize \textbf{Notes}: Navy lines are posterior medians, and shaded areas are 68\% and 90\% credible bands. The HANK model is in dark green, scaled so that the policy rate rises $0.197$ percentage points on impact. The last two panels report the two inequality measures the model itself reports, computed on our side from the fitted mixture. Both the empirical and the model measures are computed on the IHS-transformed earnings scale, $x_{i,t}=\operatorname{asinh}(z_{i,t})$. In the last panel the green line is the top 10\% earnings share, which is the object the survey measures, and the amber line is the share of total income including income from capital.}
\caption{Aggregate and inequality IRFs to the monetary policy shock}
\label{fig:bblmacro}
\end{figure}

\begin{figure}[t]
\centering
\includegraphics[width=\textwidth]{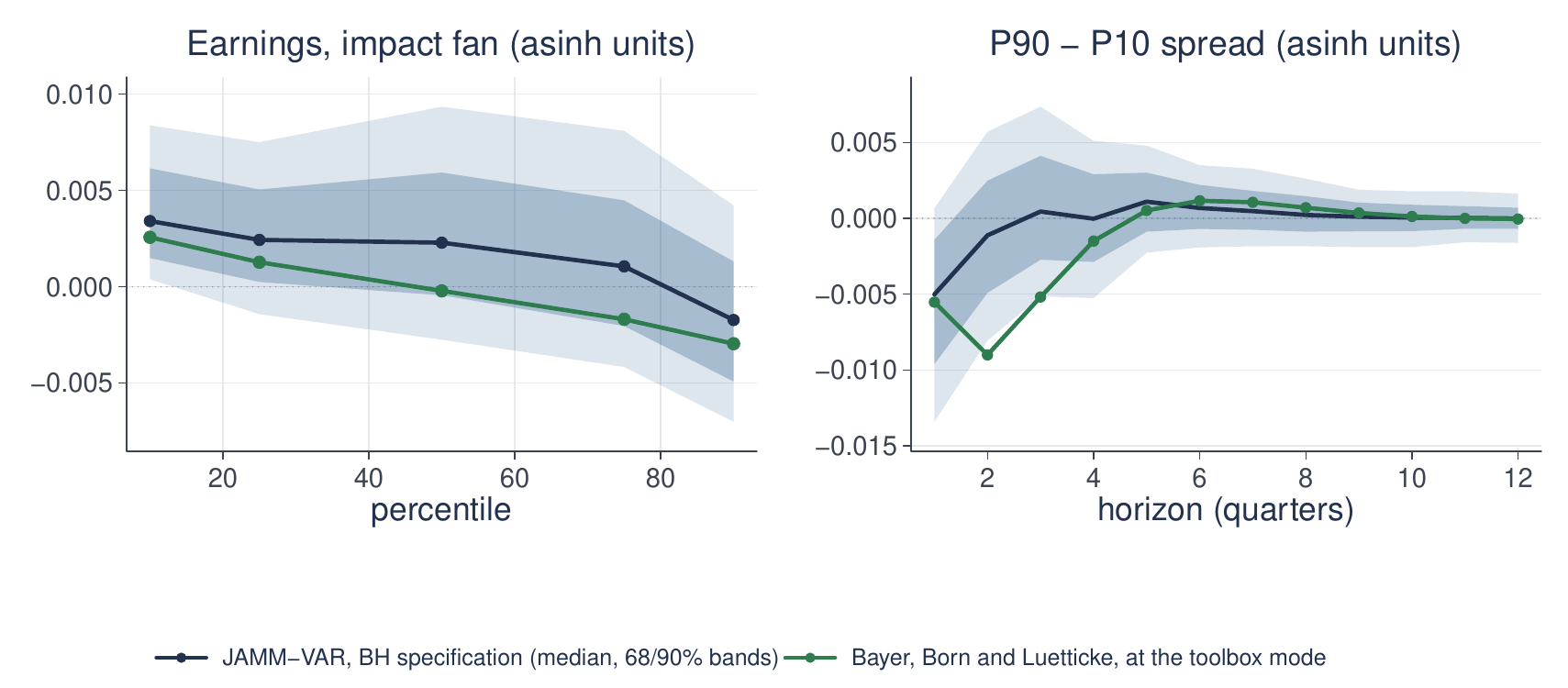}
\caption*{\footnotesize \textbf{Notes}: Earnings quantile responses on impact, and the P90--P10 spread over quarters. Navy lines are posterior medians, and shaded areas are 68\% and 90\% credible bands.}
\caption{Earnings quantile IRFs on impact and the P90--P10 spread}
\label{fig:bblfan}
\end{figure}

\begin{figure}[h!]
\centering
\includegraphics[width=\textwidth]{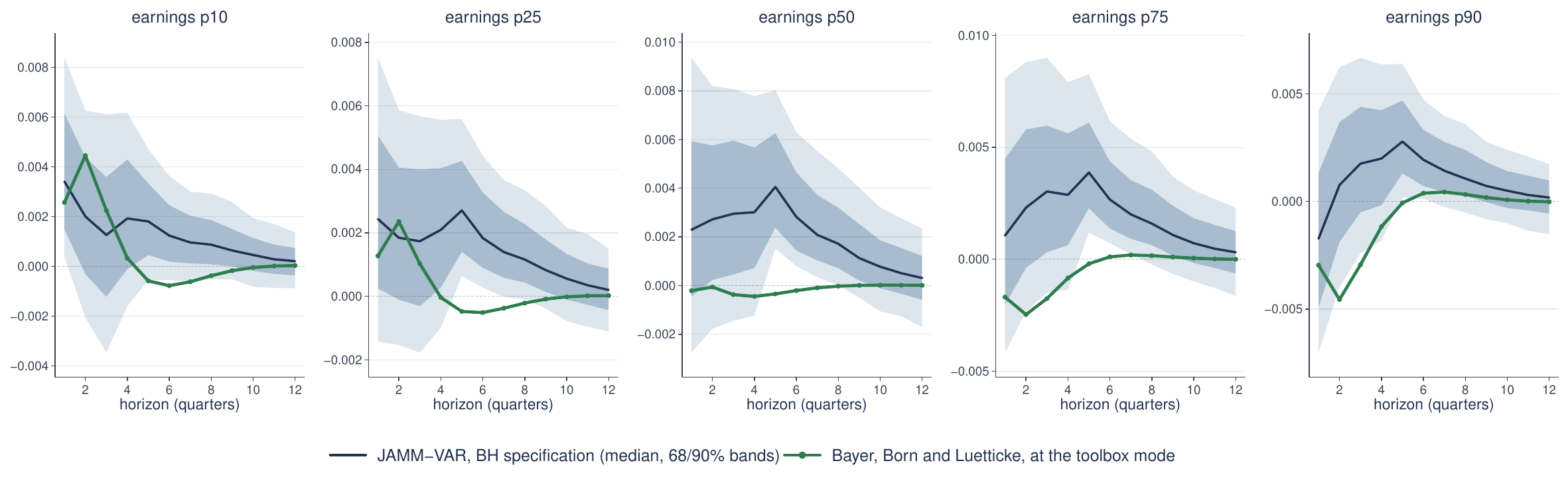}
\caption*{\footnotesize \textbf{Notes}: One panel per reported percentile, over twelve quarters. Navy lines are posterior medians, and shaded areas are 68\% and 90\% credible bands.}
\caption{Earnings quantile IRFs by horizon}
\label{fig:bbldyn}
\end{figure}

The sixth step evaluates the HANK model against the JAMM-VAR posterior of the response objects rather than only their posterior medians. We use the credible bands to assess whether differences in point estimates are large relative to posterior uncertainty. \autoref{fig:bblmacro} compares the aggregate and inequality responses. \autoref{fig:bblfan} compares the earnings-quantile responses on impact and the P90--P10 response across horizons, while \autoref{fig:bbldyn} traces each reported earnings-quantile response over twelve quarters.

On impact, the HANK model reproduces the cross-quantile pattern in the JAMM-VAR earnings responses (\autoref{fig:bblfan}). It raises earnings at lower quantiles and lowers them at upper quantiles, matching the ordering of the posterior medians. The model-implied responses also match the magnitudes. At every reported percentile, the model lies within $0.72$ posterior standard deviations of the posterior median. The model-implied P90--P10 impact response equals $1.08$ times the posterior median. The impact responses are therefore not a dimension along which the model and the evidence disagree.

The model generates this cross-quantile pattern through wage-setting rents. Because wages adjust slowly, a contraction widens the gap between what firms pay and what workers would accept. The resulting rents rise $4.3$ percent on impact and reach $7.0$ percent in the second quarter. They are distributed equally across households rather than in proportion to earnings, so this flat transfer lifts lower earnings quantiles relative to upper quantiles. Thus, the unequal incidence of monetary policy across workers stems from wage setting rather than labor supply. The comparison links the empirical distributional pattern to a specific model mechanism that aggregate responses alone cannot reveal.

The HANK model also reproduces the impact compression in two summary measures of earnings inequality (last two panels of \autoref{fig:bblmacro}). Neither response was directly targeted in the model's estimation or calibration. The model's standard deviation of IHS-transformed earnings falls $0.57$ percent on impact, close to the JAMM-VAR posterior median of $0.48$ percent and within the 68\% credible band of $0.25$ to $0.95$ percent. The top 10\% earnings share falls $0.26$ percent in the model, compared with a posterior median of $0.59$ percent. Both measures therefore imply the same transitory compression reported in \autoref{sub:emp_macro_shocks}, though the model overstates the fall in dispersion and understates the fall in the top share. By contrast, the model's top 10\% share of total income, which includes capital income, rises $0.12$ percent. The model's labor-earnings distribution is therefore broadly consistent with the evidence, but its capital-income channel reverses this response when total income is considered.

The main disagreement concerns strength and persistence: the HANK model's contraction is too large on impact and too short-lived (\autoref{fig:bblmacro}). Output falls about $0.20$ percent on impact, compared with a JAMM-VAR posterior median of $0.08$ percent, while inflation falls $0.25$ percentage points, compared with a posterior median of $0.11$ percentage points. The model-implied aggregate responses leave the credible bands at several horizons, most clearly for output. The policy-rate response reverses within two quarters, so the model's contraction is largely over by the fifth quarter, when the JAMM-VAR posterior medians remain near their troughs. The same lack of persistence appears in the earnings distribution (\autoref{fig:bbldyn}). The JAMM-VAR posterior places the largest earnings responses between the fourth and sixth quarters, whereas the model's quantiles return to their starting values by the fourth quarter because its wage-setting rents dissipate once the policy-rate response reverses.

Since the model does not reproduce the persistence of the aggregate responses, the comparison does not show whether the earnings response would remain too short-lived in a version of the model with more persistent aggregate dynamics. \citet{auclert2020microjumps} show that sticky household expectations can generate hump-shaped aggregate responses while preserving the front-loaded response of consumption to transitory household income shocks.

\subsection{Implications of the Comparison}\label{sec:bblneeds}
The comparison points to one requirement for the model's monetary transmission mechanism and one choice about how to map empirical earnings observables to model-based objects.
\begin{itemize}
  \item The model needs weaker but more persistent monetary transmission. One possibility is a policy rule that generates a more persistent policy-rate response. This would extend both the aggregate contraction and the cross-sectional earnings response, since both currently end when the policy rate reverses. Matching their magnitudes is harder. The degree to which wages lag prices governs both the strength of aggregate transmission and the size of the earnings-distribution response. Weakening this channel to improve the aggregate fit would therefore also attenuate the cross-sectional response. The model needs an additional channel that affects workers unequally without amplifying aggregate transmission.
  \item Because the model abstracts from unemployment, the comparison concerns the intensive margin among employed workers. The CPS earnings cross sections cover employed individuals, so the empirical statistics are already conditional on employment. Compositional changes in the pool of the employed can still move the empirical distribution without a model counterpart, and we interpret discrepancies in light of this difference.
\end{itemize}

\section{Extensions}\label{sec:extensions}

We now highlight two extensions that can be added in a straightforward manner to our model, namely cross sections observed at only some dates and stochastic volatility. Each adds or replaces one block in the sampler of \autoref{sub:posterior} and requires only limited adjustments to the remaining steps. This section states what each extension adds. \autoref{app:extensions} provides the modified samplers.
\subsection{Cross Sections Observed in Some Periods Only}\label{sub:ext_missing}

Cross-sectional surveys often start later than aggregate series, contain gaps, or are fielded less frequently. This mismatch binds in our application: the consumption cross sections begin in 1990, although the aggregate series extend several decades further back. The extension lets the macro block use those earlier aggregate observations while treating the corresponding cross sections as missing. It also accommodates lower-frequency surveys, such as annual income data paired with quarterly national accounts or triennial Survey of Consumer Finances (SCF) data paired with monthly aggregate variables.

Conditional on the macro state, common factors, and parameters, the mixture model defines a population density for each cross section at every date, whether or not the cross section is observed. Missing micro data matter only when their sample quantiles enter an in-sample macro equation through the feedback term in \autoref{eq: VAR}. We therefore augment the posterior with the missing cross sections needed to construct those quantiles and update them in a Metropolis--Hastings imputation block \citep{tanner1987}. Conditional on the completed cross sections, the existing blocks retain their form. Missing cross sections whose quantiles do not enter an in-sample macro equation are integrated out. If all quantile-feedback coefficients are set to zero, no imputation is needed, and the micro likelihood runs only over observed dates.

The posterior therefore yields model-implied cross-sectional densities, with credible bands, at the frequency of the aggregate block, including dates between survey waves. At dates without micro data, the aggregate likelihood informs the common factors through their loadings in the macro equations; the macro state and factors then determine the mixture weights. For example, pairing the triennial SCF with monthly aggregate variables produces a monthly path for the wealth distribution between SCF waves. These paths are model-based interpolations rather than direct survey measurements.

One modeling choice matters. When a survey was fielded but its micro data are unavailable, the latent cross section inherits the survey's design sample size. When no survey was fielded, the latent sample size is a modeling choice. It controls the sampling noise in the latent quantiles and therefore the model itself.

\subsection{Stochastic Volatility}\label{sub:ext_sv}

Structural shock variances change over long macroeconomic samples. The Great Moderation is the leading example. Such variation may matter for the normalization of shocks and the interpretation of magnitude restrictions. Because the structural shocks are orthogonal and $\bm D$ is diagonal, volatility enters equation by equation. Replace each constant variance $d_i$ with $d_{i,t}=\exp(h_{i,t})$ and let the log volatility follow an autoregression. One per-equation block, built on the mixture sampler of \citet{kim1998stochastic}, draws the volatility path and its parameters. This block replaces the structural-variance step, exactly as in VARs with time-varying volatility \citep{primiceri2005time}. Every likelihood term that previously depended on $d_i$ now depends on $d_{i,t}$, so the coefficient conditionals take generalized-least-squares form. The coefficient prior must also be modified to accommodate time-varying volatility. \autoref{app:ext_sv} provides the details.

\section{Conclusions}\label{sec:conclusions}

Aggregate SVARs provide empirical benchmarks for structural macroeconomic models. Our JAMM-VAR extends this role to HANK models by adding several marginal distributions to a Bayesian SVAR. The model retains standard aggregate-shock identification, separately identifies a common micro shock, and estimates aggregate and distributional responses within one posterior. The household observations are repeated cross sections, possibly from separate surveys. Household histories and first-stage density estimates are unnecessary. Our approach jointly estimates all relevant objects, taking into account the estimation uncertainty in the aggregate and distributional blocks jointly, while using the Bayesian and VAR toolkits macroeconomists are familiar with.

The main payoff of our approach is a direct, posterior-based comparison between estimated aggregate and distributional responses and their counterparts in heterogeneous-agent equilibrium models. The HANK simulation shows that this comparison remains informative even when the JAMM-VAR does not nest the equilibrium model. Conditional on matching the policy variable's impact response, the JAMM-VAR recovers the broad propagation of a monetary policy shock across aggregates, earnings quantiles, and consumption quantiles, and the posterior bands generally contain the true paths. In the U.S. application, aggregate data, CPS earnings, and CEX consumption yield a joint posterior of response objects that we compare with HANK-model responses horizon by horizon. This comparison distinguishes features the HANK model reproduces, such as the impact cross-quantile pattern, from dimensions along which its propagation differs, especially strength and persistence. More generally, our approach lets researchers use repeated cross sections to assess and refine heterogeneous-agent models against a coherent joint posterior of response objects rather than isolated point estimates.

\newpage

{\setstretch{1.15}\bibliographystyle{custom}\bibliography{references}}

\begin{thebibliography}{60}
\newcommand{\enquote}[1]{``#1''}
\providecommand{\natexlab}[1]{#1}

\bibitem[{Adams and Barrett(2025)}]{adams2025empirical}
\textsc{Adams JJ, and Barrett P} (2025), \enquote{What Are Empirical Monetary Policy Shocks? Estimating the Term Structure of Policy News,} Imf working paper, International Monetary Fund.

\bibitem[{Adams and Matthes(2026)}]{adams2026ricardian}
\textsc{Adams JJ, and Matthes C} (2026), \enquote{How Ricardian Are We?} Working paper, Federal Reserve Bank of Kansas City.

\bibitem[{Altig \emph{et~al.}(2011)Altig, Christiano, Eichenbaum, and Lind\'{e}}]{altig2011firm}
\textsc{Altig D, Christiano LJ, Eichenbaum M, and Lind\'{e} J} (2011), \enquote{Firm-Specific Capital, Nominal Rigidities and the Business Cycle,} \emph{Review of Economic Dynamics} \textbf{14}(2), 225--247.

\bibitem[{Auclert(2019)}]{auclert2019monetary}
\textsc{Auclert A} (2019), \enquote{Monetary Policy and the Redistribution Channel,} \emph{American Economic Review} \textbf{109}(6), 2333--2367.

\bibitem[{Auclert \emph{et~al.}(2021)Auclert, Bard{\'o}czy, Rognlie, and Straub}]{auclert2021using}
\textsc{Auclert A, Bard{\'o}czy B, Rognlie M, and Straub L} (2021), \enquote{Using the Sequence-Space {J}acobian to Solve and Estimate Heterogeneous-Agent Models,} \emph{Econometrica} \textbf{89}(5), 2375--2408.

\bibitem[{Auclert \emph{et~al.}(2020)Auclert, Rognlie, and Straub}]{auclert2020microjumps}
\textsc{Auclert A, Rognlie M, and Straub L} (2020), \enquote{Micro Jumps, Macro Humps: Monetary Policy and Business Cycles in an Estimated {HANK} Model,} NBER Working Paper 26647, National Bureau of Economic Research.

\bibitem[{Auclert \emph{et~al.}(2024)Auclert, Rognlie, and Straub}]{auclert2024intertemporal}
---{}---{}--- (2024), \enquote{The Intertemporal {K}eynesian Cross,} \emph{Journal of Political Economy} \textbf{132}(12), 4068--4121.

\bibitem[{Baumeister \emph{et~al.}(2026)Baumeister, Frank, Huber, and Koop}]{baumeister2026havar}
\textsc{Baumeister C, Frank P, Huber F, and Koop G} (2026), \enquote{Oil, Inflation Expectations, and Household Characteristics: A Nonlinear Heterogeneous Agent VAR Approach,} Presentation slides.

\bibitem[{Baumeister and Hamilton(2015)}]{baumeister2015sign}
\textsc{Baumeister C, and Hamilton JD} (2015), \enquote{Sign restrictions, structural vector autoregressions, and useful prior information,} \emph{Econometrica} \textbf{83}(5), 1963--1999.

\bibitem[{Baumeister and Hamilton(2018)}]{baumeister2018inference}
---{}---{}--- (2018), \enquote{Inference in structural vector autoregressions when the identifying assumptions are not fully believed: Re-evaluating the role of monetary policy in economic fluctuations,} \emph{Journal of Monetary Economics} \textbf{100}, 48--65.

\bibitem[{Bayer \emph{et~al.}(2024)Bayer, Born, and Luetticke}]{bayer2024shocks}
\textsc{Bayer C, Born B, and Luetticke R} (2024), \enquote{Shocks, Frictions, and Inequality in {US} Business Cycles,} \emph{American Economic Review} \textbf{114}(5), 1211--1247.

\bibitem[{Bayer \emph{et~al.}(2019)Bayer, Luetticke, Pham-Dao, and Tjaden}]{bayer2019risk}
\textsc{Bayer C, Luetticke R, Pham-Dao L, and Tjaden V} (2019), \enquote{Precautionary Savings, Illiquid Assets, and the Aggregate Consequences of Shocks to Household Income Risk,} \emph{Econometrica} \textbf{87}(1), 255--290.

\bibitem[{Ben~Zeev and Pappa(2017)}]{benzeev2017}
\textsc{Ben~Zeev N, and Pappa E} (2017), \enquote{Chronicle of a War Foretold: The Macroeconomic Effects of Anticipated Defence Spending Shocks,} \emph{The Economic Journal} \textbf{127}(603), 1568--1597.

\bibitem[{Canova(1994)}]{canova1994statistical}
\textsc{Canova F} (1994), \enquote{Statistical Inference in Calibrated Models,} \emph{Journal of Applied Econometrics} \textbf{9}(S1), S123--S144.

\bibitem[{Canova(2007)}]{canova2007methods}
---{}---{}--- (2007), \emph{Methods for Applied Macroeconomic Research}, Princeton, NJ: Princeton University Press.

\bibitem[{Canova and Paustian(2011)}]{canova2011business}
\textsc{Canova F, and Paustian M} (2011), \enquote{Business Cycle Measurement with Some Theory,} \emph{Journal of Monetary Economics} \textbf{58}(4), 345--361.

\bibitem[{Carter and Kohn(1994)}]{carter1994}
\textsc{Carter CK, and Kohn R} (1994), \enquote{On {G}ibbs Sampling for State Space Models,} \emph{Biometrika} \textbf{81}(3), 541--553.

\bibitem[{Carvalho \emph{et~al.}(2010)Carvalho, Polson, and Scott}]{carvalho2010}
\textsc{Carvalho CM, Polson NG, and Scott JG} (2010), \enquote{The Horseshoe Estimator for Sparse Signals,} \emph{Biometrika} \textbf{97}(2), 465--480.

\bibitem[{Chang \emph{et~al.}(2024)Chang, Chen, and Schorfheide}]{chang2024heterogeneity}
\textsc{Chang M, Chen X, and Schorfheide F} (2024), \enquote{Heterogeneity and aggregate fluctuations,} \emph{Journal of Political Economy} \textbf{132}(12), 4021--4067.

\bibitem[{Chang and Schorfheide(2026)}]{chang2024monetary}
\textsc{Chang M, and Schorfheide F} (2026), \enquote{On the Effects of Monetary Policy Shocks on Income and Consumption Heterogeneity,} \emph{American Economic Journal: Macroeconomics} Conditionally accepted; NBER Working Paper 32166; PIER Working Paper 24-003.

\bibitem[{Chang \emph{et~al.}(2023)Chang, Park, and Pyun}]{chang2024functional}
\textsc{Chang Y, Park JY, and Pyun D} (2023), \enquote{From Functional Autoregressions to Vector Autoregressions,} Working paper, Indiana University.

\bibitem[{Christiano \emph{et~al.}(2005)Christiano, Eichenbaum, and Evans}]{christiano2005nominal}
\textsc{Christiano LJ, Eichenbaum M, and Evans CL} (2005), \enquote{Nominal Rigidities and the Dynamic Effects of a Shock to Monetary Policy,} \emph{Journal of Political Economy} \textbf{113}(1), 1--45.

\bibitem[{Coibion \emph{et~al.}(2021)Coibion, Gorodnichenko, and Koustas}]{coibion2021consumption}
\textsc{Coibion O, Gorodnichenko Y, and Koustas D} (2021), \enquote{Consumption Inequality and the Frequency of Purchases,} \emph{American Economic Journal: Macroeconomics} \textbf{13}(4), 449--482.

\bibitem[{DeJong \emph{et~al.}(1996)DeJong, Ingram, and Whiteman}]{dejong1996bayesian}
\textsc{DeJong DN, Ingram BF, and Whiteman CH} (1996), \enquote{A {B}ayesian Approach to Calibration,} \emph{Journal of Business \& Economic Statistics} \textbf{14}(1), 1--9.

\bibitem[{Del~Negro and Schorfheide(2004)}]{delnegro2004priors}
\textsc{Del~Negro M, and Schorfheide F} (2004), \enquote{Priors from General Equilibrium Models for {VARs},} \emph{International Economic Review} \textbf{45}(2), 643--673.

\bibitem[{Del~Negro \emph{et~al.}(2007)Del~Negro, Schorfheide, Smets, and Wouters}]{delnegro2007fit}
\textsc{Del~Negro M, Schorfheide F, Smets F, and Wouters R} (2007), \enquote{On the Fit of New {K}eynesian Models,} \emph{Journal of Business \& Economic Statistics} \textbf{25}(2), 123--143.

\bibitem[{Ettmeier \emph{et~al.}(2024)Ettmeier, Kim, and Schorfheide}]{ettmeier2024functional}
\textsc{Ettmeier S, Kim CH, and Schorfheide F} (2024), \enquote{Distributional Effects of Aggregate Shocks: Functional vs. Panel Approaches,} Presentation slides, May 27.

\bibitem[{Fr\"{u}hwirth-Schnatter(1994)}]{fruhwirth1994}
\textsc{Fr\"{u}hwirth-Schnatter S} (1994), \enquote{Data Augmentation and Dynamic Linear Models,} \emph{Journal of Time Series Analysis} \textbf{15}(2), 183--202.

\bibitem[{Fr\"{u}hwirth-Schnatter(2006)}]{FS_book}
---{}---{}--- (2006), \emph{Finite Mixture and Markov Switching Models}, Springer, New York.

\bibitem[{Gal\'{i}(1999)}]{gali1999technology}
\textsc{Gal\'{i} J} (1999), \enquote{Technology, Employment, and the Business Cycle: Do Technology Shocks Explain Aggregate Fluctuations?} \emph{American Economic Review} \textbf{89}(1), 249--271.

\bibitem[{Gelman \emph{et~al.}(2014)Gelman, Hwang, and Vehtari}]{gelman2014waic}
\textsc{Gelman A, Hwang J, and Vehtari A} (2014), \enquote{Understanding Predictive Information Criteria for {B}ayesian Models,} \emph{Statistics and Computing} \textbf{24}(6), 997--1016.

\bibitem[{Geweke and Zhou(1996)}]{geweke1996}
\textsc{Geweke J, and Zhou G} (1996), \enquote{Measuring the Pricing Error of the Arbitrage Pricing Theory,} \emph{Review of Financial Studies} \textbf{9}(2), 557--587.

\bibitem[{Heathcote \emph{et~al.}(2010)Heathcote, Storesletten, and Violante}]{heathcote2010macro}
\textsc{Heathcote J, Storesletten K, and Violante GL} (2010), \enquote{The Macroeconomic Implications of Rising Wage Inequality in the {U}nited {S}tates,} \emph{Journal of Political Economy} \textbf{118}(4), 681--722.

\bibitem[{Kaplan \emph{et~al.}(2018)Kaplan, Moll, and Violante}]{kaplan2018monetary}
\textsc{Kaplan G, Moll B, and Violante GL} (2018), \enquote{Monetary Policy According to {HANK},} \emph{American Economic Review} \textbf{108}(3), 697--743.

\bibitem[{Kaplan and Violante(2018)}]{kaplan2018microeconomic}
\textsc{Kaplan G, and Violante GL} (2018), \enquote{Microeconomic Heterogeneity and Macroeconomic Shocks,} \emph{Journal of Economic Perspectives} \textbf{32}(3), 167--194.

\bibitem[{Kim \emph{et~al.}(1998)Kim, Shephard, and Chib}]{kim1998stochastic}
\textsc{Kim S, Shephard N, and Chib S} (1998), \enquote{Stochastic Volatility: Likelihood Inference and Comparison with {ARCH} Models,} \emph{Review of Economic Studies} \textbf{65}(3), 361--393.

\bibitem[{Koop \emph{et~al.}(2026)Koop, McIntyre, Mitchell, and Wu}]{koop2026pseudo}
\textsc{Koop G, McIntyre S, Mitchell J, and Wu P} (2026), \enquote{Incorporating Micro Data into Macro Models Using Pseudo {VARs},} Working Paper 26-04, Federal Reserve Bank of Cleveland, dOI: 10.26509/frbc-wp-202604.

\bibitem[{Krusell and Smith(1998)}]{krusell1998income}
\textsc{Krusell P, and Smith AA Jr} (1998), \enquote{Income and Wealth Heterogeneity in the Macroeconomy,} \emph{Journal of Political Economy} \textbf{106}(5), 867--896.

\bibitem[{Litterman(1986)}]{litterman1986}
\textsc{Litterman RB} (1986), \enquote{Forecasting with {B}ayesian Vector Autoregressions---Five Years of Experience,} \emph{Journal of Business \& Economic Statistics} \textbf{4}(1), 25--38.

\bibitem[{Liu and Plagborg-M{\o}ller(2023)}]{liu2023full}
\textsc{Liu L, and Plagborg-M{\o}ller M} (2023), \enquote{Full-Information Estimation of Heterogeneous Agent Models Using Macro and Micro Data,} \emph{Quantitative Economics} \textbf{14}(1), 1--35.

\bibitem[{Loria \emph{et~al.}(2022)Loria, Matthes, and Wang}]{loria2022economic}
\textsc{Loria F, Matthes C, and Wang MC} (2022), \enquote{Economic Theories and Macroeconomic Reality,} \emph{Journal of Monetary Economics} \textbf{126}, 105--117.

\bibitem[{Makalic and Schmidt(2016)}]{makalic2016}
\textsc{Makalic E, and Schmidt DF} (2016), \enquote{A Simple Sampler for the Horseshoe Estimator,} \emph{IEEE Signal Processing Letters} \textbf{23}(1), 179--182.

\bibitem[{Marcellino \emph{et~al.}(2025)Marcellino, Renzetti, and Tornese}]{marcellino2025firm}
\textsc{Marcellino M, Renzetti A, and Tornese T} (2025), \enquote{Firm Heterogeneity and Aggregate Fluctuations: A Functional {VAR} Model for Multidimensional Distributions,} Working paper, arXiv:2411.05695.

\bibitem[{Marron and Wand(1992)}]{marron1992exact}
\textsc{Marron JS, and Wand MP} (1992), \enquote{Exact Mean Integrated Squared Error,} \emph{The Annals of Statistics} \textbf{20}(2), 712--736.

\bibitem[{Matthes \emph{et~al.}(2025)Matthes, Nagasaka, and Schwartzman}]{matthes2025missing}
\textsc{Matthes C, Nagasaka N, and Schwartzman F} (2025), \enquote{Estimating the Missing Intercept,} Working paper.

\bibitem[{McKay \emph{et~al.}(2016)McKay, Nakamura, and Steinsson}]{mckay2016power}
\textsc{McKay A, Nakamura E, and Steinsson J} (2016), \enquote{The Power of Forward Guidance Revisited,} \emph{American Economic Review} \textbf{106}(10), 3133--3158.

\bibitem[{Moffitt and Gottschalk(2012)}]{moffitt2012trends}
\textsc{Moffitt RA, and Gottschalk P} (2012), \enquote{Trends in the Transitory Variance of Male Earnings: Methods and Evidence,} \emph{Journal of Human Resources} \textbf{47}(1), 204--236.

\bibitem[{Nagasaka(2026)}]{nagasaka2026identifying}
\textsc{Nagasaka N} (2026), \enquote{Identifying Macro Shocks From Micro Evidence: A Mixed Autoregressive Approach,} Working paper, Indiana University.

\bibitem[{Omori \emph{et~al.}(2007)Omori, Chib, Shephard, and Nakajima}]{omori2007}
\textsc{Omori Y, Chib S, Shephard N, and Nakajima J} (2007), \enquote{Stochastic Volatility with Leverage: Fast and Efficient Likelihood Inference,} \emph{Journal of Econometrics} \textbf{140}(2), 425--449.

\bibitem[{Ottonello and Winberry(2020)}]{ottonello2020financial}
\textsc{Ottonello P, and Winberry T} (2020), \enquote{Financial Heterogeneity and the Investment Channel of Monetary Policy,} \emph{Econometrica} \textbf{88}(6), 2473--2502.

\bibitem[{Parra-Alvarez \emph{et~al.}(2023)Parra-Alvarez, Posch, and Wang}]{parraalvarez2023estimation}
\textsc{Parra-Alvarez JC, Posch O, and Wang MC} (2023), \enquote{Estimation of Heterogeneous Agent Models: A Likelihood Approach,} \emph{Oxford Bulletin of Economics and Statistics} \textbf{85}(2), 304--330.

\bibitem[{Plagborg-M{\o}ller and Wolf(2021)}]{plagborg2021local}
\textsc{Plagborg-M{\o}ller M, and Wolf CK} (2021), \enquote{Local Projections and {VARs} Estimate the Same Impulse Responses,} \emph{Econometrica} \textbf{89}(2), 955--980.

\bibitem[{Polson \emph{et~al.}(2013)Polson, Scott, and Windle}]{polson2013}
\textsc{Polson NG, Scott JG, and Windle J} (2013), \enquote{Bayesian Inference for Logistic Models Using {P}\'{o}lya--{G}amma Latent Variables,} \emph{Journal of the American Statistical Association} \textbf{108}(504), 1339--1349.

\bibitem[{Primiceri(2005)}]{primiceri2005time}
\textsc{Primiceri GE} (2005), \enquote{Time Varying Structural Vector Autoregressions and Monetary Policy,} \emph{Review of Economic Studies} \textbf{72}(3), 821--852.

\bibitem[{Raftery and Lewis(1992)}]{raftery1992}
\textsc{Raftery AE, and Lewis SM} (1992), \enquote{How Many Iterations in the {G}ibbs Sampler?} in \textsc{JM~Bernardo, JO~Berger, AP~Dawid, and AFM Smith} (eds.) \enquote{Bayesian Statistics 4,} 763--773, Oxford: Oxford University Press.

\bibitem[{Smith(1993)}]{smith1993estimating}
\textsc{Smith AA} (1993), \enquote{Estimating Nonlinear Time-Series Models Using Simulated Vector Autoregressions,} \emph{Journal of Applied Econometrics} \textbf{8}(S1), S63--S84.

\bibitem[{Tanner and Wong(1987)}]{tanner1987}
\textsc{Tanner MA, and Wong WH} (1987), \enquote{The Calculation of Posterior Distributions by Data Augmentation,} \emph{Journal of the American Statistical Association} \textbf{82}(398), 528--540.

\bibitem[{Watanabe(2010)}]{watanabe2010}
\textsc{Watanabe S} (2010), \enquote{Asymptotic Equivalence of {B}ayes Cross Validation and Widely Applicable Information Criterion in Singular Learning Theory,} \emph{Journal of Machine Learning Research} \textbf{11}, 3571--3594.

\bibitem[{Winberry(2018)}]{winberry2018method}
\textsc{Winberry T} (2018), \enquote{A Method for Solving and Estimating Heterogeneous Agent Macro Models,} \emph{Quantitative Economics} \textbf{9}(3), 1123--1151.

\bibitem[{Windle \emph{et~al.}(2014)Windle, Polson, and Scott}]{windle2014}
\textsc{Windle J, Polson NG, and Scott JG} (2014), \enquote{Sampling {P}\'{o}lya--{G}amma Random Variates: Alternate and Approximate Techniques,} \emph{arXiv preprint arXiv:1405.0506} .

\end{thebibliography}
\clearpage

\begin{appendices}
\crefalias{section}{appendix}

\begin{center}
{\sffamily\Large\textbf{Online Appendix\\\huge\titletext}}
\end{center}

\setcounter{page}{1} \setcounter{section}{0} \setcounter{equation}{0} \setcounter{footnote}{0}
\renewcommand{\thepage}{A--\arabic{page}}

\renewcommand\thesection{\Alph{section}} \renewcommand\theequation{\Alph{section}.\arabic{equation}} \renewcommand\thefigure{\Alph{section}.\arabic{figure}}
\section{Technical Appendix}\label{app:technical}

This appendix gives the sampler summarized in \autoref{sub:posterior}. It covers the full conditionals, the P\'olya--Gamma augmentation for the log-weight coefficients, the Laplace--Metropolis update of the contemporaneous matrix, the exact Metropolis--Hastings update of the shared factors, the empirical identifying restrictions, and the construction of impulse responses. We retain the notation of \autoref{sec:econometrics}. There are $S$ cross sections. Each is represented by a Gaussian mixture with $G_s$ components and reference component $G_s$. The macro block contains $M$ variables and $P$ lags, and $R$ common factors $\bm f_t$ link the two blocks. We write $\mathcal{N}(\cdot\mid\mu,\sigma^2)$ for a Gaussian density and $\mathcal{N}(\bm\mu,\bm\Sigma)$ for a Gaussian law. $\mathcal{IG}(a,b)$ has density proportional to $x^{-a-1}e^{-b/x}$, and $\mathcal{GIG}(\lambda,\chi,\psi)$ has density proportional to $x^{\lambda-1}e^{-(\chi/x+\psi x)/2}$. $\mathrm{PG}(\cdot)$ denotes the P\'olya--Gamma distribution. The notation $\mathcal{N}_{[\bm l,\bm u]}$ denotes a Gaussian distribution truncated to $[\bm l,\bm u]$.

\paragraph{Augmented likelihood and design matrices.}
Let $z_{it,s}\in\{1,\dots,G_s\}$ denote the latent component label for unit $i$ in cross section $s$ at date $t$. Let $n_{t,s}$ be the number of units and $n_{tg,s}=\sum_i\mathbb{1}(z_{it,s}=g)$. The log-weight design vector is
\begin{equation*}
  \bm X_t=(1,\bm Q_t',\bm Q_{t-1}',\dots,\bm Q_{t-P}',\bm f_t')'\in\mathbb{R}^{K},
  \qquad K=1+(P+1)M+R,
\end{equation*}
so that $\eta_{tg,s}=\bm X_t'\bm b_{g,s}$ for $g=1,\dots,G_s-1$, with reference index $\eta_{t,G_s,s}=0$. Write the macro block as $\bm A_0\bm Q_t=\bm\Phi'\bm m_t+\bm u_t$, where $\bm u_t\sim\mathcal{N}(\bm 0,\bm D)$, $\bm A_0=\bm I_M-\bm W$, $\bm D=\diag(d_1,\dots,d_M)$, and
\begin{equation*}
  \bm m_t=\big(\bm Q_{t-1}',\dots,\bm Q_{t-P}',
  \{\mathcal{Q}_r(\bm y_{t-1,s})\}_{r\in\mathcal{R},s=1,\dots,S}',1,\bm f_t'\big)'
  \in\mathbb{R}^{R_m},\quad R_m=MP+Sn_q+1+R.
\end{equation*}
Stacking the $T$ post-lag dates gives $\bm X=(\bm X_1,\dots,\bm X_T)'$, $\bm M=(\bm m_1,\dots,\bm m_T)'$, and $\tilde{\bm Y}=\bm Q-\bm Q\bm W'$, whose row $t$ equals $(\bm A_0\bm Q_t)'$. One sampler sweep executes the following blocks in order.

\paragraph{Component allocation.}
Conditional on the weights and component parameters, draw labels independently from
\begin{equation*}
  \Pr(z_{it,s}=g\mid\cdot)=
  \frac{w_{tg,s}\mathcal{N}(y_{it,s}\mid\mu_{g,s},\sigma^2_{g,s})}
       {\sum_{h=1}^{G_s}w_{th,s}\mathcal{N}(y_{it,s}\mid\mu_{h,s},\sigma^2_{h,s})},
  \qquad g=1,\dots,G_s,
\end{equation*}
using a log-sum-exp stabilization. These draws determine the counts $n_{tg,s}$ used below.

\paragraph{Component means and variances.}
Because component parameters are time invariant, their sufficient statistics pool all observations assigned to a component. For component $g$ in cross section $s$, let $n_{g,s}$, $\bar y_{g,s}$, and $\mathrm{SS}_{g,s}=\sum(y-\bar y_{g,s})^2$ denote the count, mean, and sum of squared deviations among $\{y_{it,s}:z_{it,s}=g\}$.

The ordering restriction $\mu_{g-1,s}<\mu_{g,s}<\mu_{g+1,s}$ makes the joint conditional of $(\mu_{g,s},\sigma^2_{g,s})$ a Normal--inverse-Gamma density truncated in its mean argument. Drawing $\sigma^2_{g,s}$ from the unrestricted marginal inverse-Gamma and then $\mu_{g,s}$ from a truncated normal would not target this joint, because the probability of the ordering interval depends on $\sigma^2_{g,s}$. We therefore update the pair by two exact Gibbs steps. Given the current mean, the ordering indicator is constant and
\begin{equation*}
  \sigma^2_{g,s}\mid\mu_{g,s},\cdot\sim\mathcal{IG}\Big(a_0+\tfrac{n_{g,s}+1}{2},
  b_0+\tfrac12\big[\mathrm{SS}_{g,s}+n_{g,s}(\bar y_{g,s}-\mu_{g,s})^2
  +\kappa_{0,s}(\mu_{g,s}-m_{0,s})^2\big]\Big).
\end{equation*}
Given the new variance,
\begin{equation*}
  \mu_{g,s}\mid\sigma^2_{g,s},\cdot\sim\mathcal{N}_{[\mu_{g-1,s},\mu_{g+1,s}]}\big(m_{g,s},
  \sigma^2_{g,s}/\kappa_{g,s}\big),\qquad
  \kappa_{g,s}=\kappa_{0,s}+n_{g,s},\quad
  m_{g,s}=\frac{\kappa_{0,s}m_{0,s}+n_{g,s}\bar y_{g,s}}{\kappa_{g,s}},
\end{equation*}
with $\mu_{0,s}=-\infty$ and $\mu_{G_s+1,s}=+\infty$, which enforces $\mu_{1,s}<\dots<\mu_{G_s,s}$. Components are updated in ascending order, so the lower endpoint is the current-sweep value of the preceding mean and the upper endpoint the previous-sweep value of the following mean, a standard systematic scan. The same two conditionals apply to an empty component with $n_{g,s}=0$, for which they reduce to the prior conditionals.

\paragraph{Mean-dispersion hyperparameter.}

The precision $\kappa_{0,s}$ controls shrinkage of the component means toward $m_{0,s}$. Let $x_s=\kappa_{0,s}^{-1}$ denote the corresponding variance ratio, so that $\mu_{g,s}\mid\sigma^2_{g,s},x_s\sim\mathcal{N}(m_{0,s},\sigma^2_{g,s}x_s)$ before ordering. Its hyperprior is the normal-gamma scale mixture
\begin{equation*}
  x_s\mid c_s\sim\mathcal{G}\big(\tfrac12,\tfrac{1}{2c_s}\big),\qquad
  c_s\sim\mathcal{G}(a,b),
\end{equation*}
where $\mathcal{G}(\alpha,\beta)$ denotes the Gamma distribution with shape $\alpha$ and rate $\beta$. Equivalently, $x_s^{1/2}$ is half-normal with variance $c_s$ given $c_s$. With $a=b=1$, the marginal prior of $x_s^{1/2}$ is exponential, so the hyperprior places substantial mass near zero, where the component means are pulled toward $m_{0,s}$, and retains a heavy right tail. The two full conditionals are
\begin{equation*}
  c_s\mid x_s\sim\mathcal{GIG}\big(a-\tfrac12,x_s,2b\big),\qquad
  x_s\mid c_s,\cdot\sim\mathcal{GIG}\Big(\tfrac{1-G_s}{2},
  \textstyle\sum_{g=1}^{G_s}\frac{(\mu_{g,s}-m_{0,s})^2}{\sigma^2_{g,s}},c_s^{-1}\Big).
\end{equation*}
The second follows because the $G_s$ Gaussian kernels of the component means contribute
\begin{equation*}
  x_s^{-G_s/2}\exp\Big\{-\frac{1}{2x_s}\sum_{g=1}^{G_s}\frac{(\mu_{g,s}-m_{0,s})^2}{\sigma^2_{g,s}}\Big\}
\end{equation*}
and the prior contributes $x_s^{-1/2}\exp\{-x_s/(2c_s)\}$. Under the joint-truncation convention of \autoref{sub:priors}, the ordering indicator leaves this conditional unchanged. The draw is exact and uses no cap. Because $x_s$ enters the component updates only through $\kappa_{0,s}=x_s^{-1}$, which is added to counts $n_{g,s}$ in the thousands, this hyperparameter has a negligible effect on the results.

\paragraph{Log-weight coefficients and the P\'olya--Gamma step.}
Conditional on the labels, the multinomial logistic likelihood factorizes across cross sections and is updated one component at a time in one-vs-rest form. For $g\in\{1,\dots,G_s-1\}$, define the log-sum-exp offset over all remaining components, including the reference $\eta_{t,G_s,s}=0$,
\begin{equation*}
  \mathrm{a}_{tg,s}=\log\sum_{h\ne g}e^{\eta_{th,s}},\qquad
  \psi_{tg,s}=\eta_{tg,s}-\mathrm{a}_{tg,s}.
\end{equation*}
Introducing $\omega_{tg,s}\sim\mathrm{PG}(n_{t,s},\psi_{tg,s})$ makes the P\'olya--Gamma likelihood Gaussian in $\bm b_{g,s}$. We draw these weights using the saddlepoint approximation of \citet{windle2014}, whose cost per draw does not increase with $n_{t,s}$. In our applications the saddlepoint draws add about one percent to the total run time. A fast deterministic alternative plugs in the conditional expectation,
\begin{equation*}
  \bar\omega_{tg,s}=\mathbb{E}[\omega_{tg,s}]
  =\frac{n_{t,s}}{2\psi_{tg,s}}\tanh\Big(\frac{\psi_{tg,s}}{2}\Big),
  \qquad
  \lim_{\psi\to0}\bar\omega_{tg,s}=\frac{n_{t,s}}{4},
\end{equation*}
as in an expectation--maximization step. We compared the two options directly in the reduced-form simulation. Across all $1{,}494$ date-by-component mixture weights $w_{tg,s}$, the posterior means differ by at most $0.002$ (on average $0.0004$) and the posterior standard deviations by at most $0.001$. Both differences are within the split-half Monte Carlo noise of a single chain. Kolmogorov--Smirnov statistics for each weight between the two sets of draws are no larger than their within-chain benchmark. The coefficient of variation of $\mathrm{PG}(n_{t,s},\psi)$ is of order $n_{t,s}^{-1/2}$, which explains the agreement. We use the saddlepoint draw throughout. Define the working response $\zeta_{tg,s}=(n_{tg,s}-n_{t,s}/2)/\omega_{tg,s}+\mathrm{a}_{tg,s}$, the weight matrix $\bm\Omega_{g,s}=\diag_t(\omega_{tg,s})$, and the horseshoe prior $\bm b_{g,s}\sim\mathcal{N}(\bm 0,\underline{\bm V}_{g,s})$. The Gaussian conditional is
\begin{equation*}
  \bar{\bm V}_{g,s}=\big(\bm X'\bm\Omega_{g,s}\bm X
  +\underline{\bm V}_{g,s}^{-1}\big)^{-1},\qquad
  \bar{\bm b}_{g,s}=\bar{\bm V}_{g,s}\bm X'\bm\Omega_{g,s}\bm\zeta_{g,s}.
\end{equation*}
Draw $\bm b_{g,s}\sim\mathcal{N}(\bar{\bm b}_{g,s},\bar{\bm V}_{g,s})$. If component $g$ carries identifying restrictions on contemporaneous macro loadings (entries $2,\dots,M+1$ of $\bm X_t$) or factor loadings (the final $R$ entries), draw the restricted subvector from the corresponding truncated multivariate normal.

\paragraph{Horseshoe scale updates.}
Both horseshoe blocks, the log-weight coefficients above and the VAR coefficients below, use the inverse-Gamma augmentation of \citet{makalic2016}. For $\bm b\in\mathbb{R}^{K}$ with local scales $\varphi_l$, global scale $\tau$, and auxiliaries $\nu_l,\xi$,
\begin{equation*}
  \varphi_l^2\sim\mathcal{IG}\Big(1,\tfrac{1}{\nu_l}+\tfrac{b_l^2}{2\tau^2}\Big),
  \nu_l\sim\mathcal{IG}\Big(1,1+\tfrac{1}{\varphi_l^2}\Big),
  \tau^2\sim\mathcal{IG}\Big(\tfrac{K+1}{2},\tfrac{1}{\xi}
  +\tfrac12\textstyle\sum_l\tfrac{b_l^2}{\varphi_l^2}\Big),
  \xi\sim\mathcal{IG}\Big(1,1+\tfrac{1}{\tau^2}\Big),
\end{equation*}
so the prior variance of $b_l$ is $\varphi_l^2\tau^2$. For the log-weight block, $\bm b$ is the coefficient vector $\bm b_{g,s}$ itself and $K$ is its length. For VAR equation $i$, the prior in \autoref{eq:prior_phi} is centered at $\underline{\bm\phi}_i$ and scaled by $d_i$, so the same updates are applied to the standardized deviations $\tilde b_{l,i}=(\phi_{l,i}-\underline{\phi}_{l,i})/\sqrt{d_i}$, with $K$ replaced by the number $K_i$ of free coefficients of equation $i$. Coefficients fixed at zero by a hard restriction are excluded. They carry a degenerate prior, do not enter the conditional of the free block, and their local scales are not updated. Restricted coefficients with sign or magnitude bounds follow the joint-truncation convention of \autoref{sub:priors}, under which the scale conditionals are unchanged.

\paragraph{VAR coefficients.}
Conditional on $\bm W$, $\bm D$, the factors, and lagged quantiles, the structural system separates by equation. For equation $i$, with prior $\bm\phi_i\mid d_i\sim\mathcal{N}(\underline{\bm\phi}_i,d_i\underline{\bm V}_i)$,
\begin{equation*}
  \bar{\bm V}_i=\big(\bm M'\bm M+\underline{\bm V}_i^{-1}\big)^{-1},\qquad
  \bar{\bm\phi}_i=\bar{\bm V}_i\big(\bm M'\tilde{\bm Y}_{\cdot i}
  +\underline{\bm V}_i^{-1}\underline{\bm\phi}_i\big),\qquad
  \bm\phi_i\sim\mathcal{N}\big(\bar{\bm\phi}_i,d_i\bar{\bm V}_i\big).
\end{equation*}
The prior mean $\underline{\bm\phi}_i$ follows the Minnesota convention. The own first lag is centered at $\delta$ and all other coefficients at zero. The prior precision is $\underline{\bm V}_i^{-1}=\diag\big(1/(\tau_i^2\varphi_{l,i}^2)\big)$, with the horseshoe scales updated as described above. Let $\mathcal{F}$ index the $R$ factor coefficients in $\bm\phi_i$, let $\mathcal{F}_0\subseteq\mathcal{F}$ collect those fixed at zero, and let $\mathcal{F}_\pm\subseteq\mathcal{F}$ collect those carrying a sign restriction. Set $\bm\phi_{i,\mathcal{F}_0}=\bm 0$ and draw the remaining coefficients from the Gaussian conditional of $\mathcal{N}(\bar{\bm\phi}_i,d_i\bar{\bm V}_i)$ given $\bm\phi_{i,\mathcal{F}_0}=\bm 0$. If $\mathcal{F}_\pm$ is nonempty, that conditional is truncated to the sign-restricted region and drawn as a truncated multivariate normal. With $R=1$ at most one of the two sets is nonempty. In the HANK simulation with $R=2$, a zero restriction and a sign restriction occur in the same equation, and the two steps are combined as described.

\paragraph{Structural variances.}
Because the coefficient prior in \autoref{eq:prior_phi} is scaled by $d_i$, the Gaussian prior factor contributes $d_i^{-K_i/2}\exp\{-\mathrm{Q}_i/(2d_i)\}$ to the full conditional of $d_i$, where $K_i$ denotes the number of free coefficients of equation $i$ and
$\mathrm{Q}_i=(\bm\phi_i-\underline{\bm\phi}_i)'\underline{\bm V}_i^{-1}(\bm\phi_i-\underline{\bm\phi}_i)$ is the prior quadratic form. With $\mathrm{SSE}_i=\sum_{t}(\tilde Y_{ti}-\bm m_t'\bm\phi_i)^2$ the structural residual sum of squares of equation $i$ and the prior in \autoref{eq:prior_D}, the full conditional is
\begin{equation*}
  d_i\sim\mathcal{IG}\Big(\tfrac{M+2}{2}+\tfrac{T+K_i}{2},
  \tfrac{\hat\sigma^2_i}{2}+\tfrac{\mathrm{SSE}_i+\mathrm{Q}_i}{2}\Big).
\end{equation*}
Coefficients pinned to zero by a hard restriction carry a degenerate prior and are excluded from $K_i$ and $\mathrm{Q}_i$ (under the horseshoe the conditional prior precision of the free block is simply the corresponding diagonal sub-block). Under the joint-truncation convention for restricted coefficients (\autoref{sub:priors}), the truncation contributes no $d_i$-dependent term, so this conditional is unchanged. The prior scale $\hat\sigma^2_i$ is the residual sum of squares of equation $i$ from the reduced-form initialization, a least-squares fit with a unit ridge penalty that stabilizes the initialization, divided by $T$. If magnitude bounds on the variance-scaled impact matrix $\bm A_0^{-1}\bm D^{1/2}$ were imposed, they would also restrict $d_i$. The draw from the inverse-Gamma conditional is then accepted only if the bounds hold. This is an exact independence Metropolis step. The empirical application imposes no such bounds.

\paragraph{Contemporaneous matrix and the sign--magnitude restrictions.}
Write structural equation $i$ as $Q_{ti}=\sum_{j\ne i}W_{ij}Q_{tj}+\bm m_t'\bm\phi_i+u_{ti}$. Let $\bm a=\bm W_{i,-i}\in\mathbb{R}^{M-1}$ collect its free off-diagonal entries and $r_{ti}=Q_{ti}-\bm m_t'\bm\phi_i$ its partial residual. Ignoring the Jacobian, the Gaussian conditional is $\mathcal{N}(\bm m_i^0,\bm V_i^0)$, where
\begin{equation*}
  (\bm V_i^0)^{-1}=d_i^{-1}\bm Q_{-i}'\bm Q_{-i}+\underline{l}_W^{-1}\bm I_{M-1},
  \qquad
  \bm m_i^0=\bm V_i^0d_i^{-1}\bm Q_{-i}'\bm r_i,
\end{equation*}
where $\bm Q_{-i}$ contains the contemporaneous regressors $\{Q_{\cdot j}\}_{j\ne i}$ and $\bm r_i=(r_{1i},\dots,r_{Ti})'$. The exact conditional multiplies this Gaussian kernel by $|\det\bm A_0|^{T}$. Because $\det\bm A_0$ is linear in row $i$,
\begin{equation*}
  \det\bm A_0=c_{ii}-\bm c_i'\bm a,\qquad
  \bm c_i=\det(\bm A_0)[\bm A_0^{-1}]_{-i,i},\qquad
  c_{ii}=\det(\bm A_0)[\bm A_0^{-1}]_{ii},
\end{equation*}
where the cofactors $\bm c_i$ and $c_{ii}$ do not depend on $\bm a$, and $\Delta_0=\det\bm A_0$ at the current $\bm a$. A second-order expansion of $T\log|\det\bm A_0|$ adds $(T/\Delta_0^2)\bm c_i\bm c_i'$ to the precision. The Sherman--Morrison identity yields the $O(M^2)$ update
\begin{equation*}
  \bm V_i^{\ast}=\bm V_i^0-\frac{T/\Delta_0^2}{1+(T/\Delta_0^2)\bm c_i'\bm V_i^0\bm c_i}
  (\bm V_i^0\bm c_i)(\bm V_i^0\bm c_i)',\qquad
  \bm m_i^{\ast}=\bm a+\bm V_i^{\ast}\bm g,\quad
  \bm g=-(\bm V_i^0)^{-1}(\bm a-\bm m_i^0)-\tfrac{T}{\Delta_0}\bm c_i.
\end{equation*}
Propose $\bm a^{\ast}\sim\mathcal{N}_{[\underline{\bm w}_i,\overline{\bm w}_i]}(\bm m_i^{\ast},\bm V_i^{\ast})$. The interval $[\underline{\bm w}_i,\overline{\bm w}_i]$ encodes the restrictions in \autoref{tab:Wbounds}. Sign restrictions set one endpoint to zero, magnitude restrictions set a nonzero bound, and excluded channels are fixed at zero. Accept the proposal using the Metropolis--Hastings log ratio
\begin{equation*}
\begin{aligned}
  \log r={}&\Big[T\log|\Delta^{\ast}|-\tfrac12(\bm a^{\ast}-\bm m_i^0)'(\bm V_i^0)^{-1}
  (\bm a^{\ast}-\bm m_i^0)\Big]\\
  &-\Big[T\log|\Delta_0|-\tfrac12(\bm a-\bm m_i^0)'(\bm V_i^0)^{-1}(\bm a-\bm m_i^0)\Big]
  +\log\frac{q(\bm a\mid\bm a^{\ast})}{q(\bm a^{\ast}\mid\bm a)},
\end{aligned}
\end{equation*}
where $\Delta^{\ast}=c_{ii}-\bm c_i'\bm a^{\ast}$ and $q(\cdot\mid\cdot)$ is the Laplace proposal density, recomputed at $\bm a^{\ast}$ for the reverse move. Because $\bm m_i^{\ast}$ and $\bm V_i^{\ast}$ are rebuilt at the current draw, the proposal is not symmetric, and both directions are evaluated in full. The density $q$ includes the Gaussian kernel, the determinant of $\bm V_i^{\ast}$, and the normalizing constant $\Pr\{\bm a\in[\underline{\bm w}_i,\overline{\bm w}_i]\}$ of the truncated normal. This constant depends on the center and therefore does not cancel. Entries that a hard restriction pins to a fixed value make the proposal degenerate. For those rows $q$ is evaluated on the free block conditional on the pinned entries, which is the density that actually governs the move. The resulting chain is an exact Metropolis--Hastings sampler for the row conditional under the box restrictions on $\bm W$. In our application the normalizing-constant correction is numerically small, but it makes the row-conditional update exact.

The restrictions on impact responses in \autoref{tab:restrictions}, namely the signs of selected entries of $\bm A_0^{-1}$ (and, if imposed, magnitude bounds on $[\bm A_0^{-1}]_{ij}\sqrt{d_j}$), are nonlinear in $\bm W$ and involve all rows at once. They are part of the posterior, so the target of the row-$i$ update is the box-restricted conditional above multiplied by the indicator $\mathbb{1}\{\bm A_0^{-1}\in\mathcal{G}\}$ of the admissible impact set $\mathcal{G}$, evaluated at the matrix formed by the candidate row and the current remaining rows. A proposed row is therefore rejected immediately whenever that matrix violates an impact restriction. Otherwise the ratio above applies. Each row kernel then leaves the correct restricted conditional invariant. Rejecting only after a complete systematic row sweep would not, in general, because the composition of the row kernels is not reversible. The row-level check is switched on once the chain first reaches a state that satisfies the impact restrictions. This occurs during the initial burn-in period from the starting value $\bm W=\bm 0$. Every retained draw is generated by the restricted kernels. The impact restrictions bind at the row level. In the empirical application the acceptance rates of the three free rows are between $0.74$ and $0.78$ for the output gap equation, between $0.89$ and $0.91$ for the inflation equation, and about $0.98$ for the policy-rate equation. The lowest rate is for the output gap equation. Its contemporaneous coefficients determine the impact responses of the output gap to the fiscal and monetary policy shocks.
If $\bm A_0$ is lower triangular, $\det\bm A_0\equiv 1$, the Jacobian disappears, and each row has an exact truncated-Gaussian Gibbs draw. Under impact restrictions that draw is accepted only when the implied $\bm A_0^{-1}$ is admissible, which is an exact independence Metropolis step because the truncated Gaussian is the row conditional under the box restrictions alone. Finally, set $\bm\Sigma=\bm A_0^{-1}\bm D\bm A_0^{-\prime}$.

\paragraph{Shared factors.}
Under the baseline-category multinomial logistic model, the full conditional of $\bm f_t$ is \emph{not} Gaussian. The one-versus-rest P\'olya--Gamma representation used for the log-weight coefficients is valid because it holds the competing component indices fixed while $\bm b_{g,s}$ is updated. The shared factor, by contrast, enters \emph{every} index, so when $\bm f_t$ moves, the offset $\mathrm{a}_{tg,s}$ defined above moves with it and the P\'olya--Gamma kernel ceases to be quadratic in $\bm f_t$. We therefore treat the Gaussian implied by frozen offsets as a \emph{proposal} and add a Metropolis--Hastings correction, so that the factor block targets the exact conditional
\begin{equation*}
  \pi(\bm f_t\mid\cdot)\propto\mathbf 1\{\bm f_t\in\mathcal A_t\}
  \mathcal{N}(\bm f_t\mid\bm m_t^{0},\bm C_t^{0})
  \prod_{s=1}^{S}\prod_{g=1}^{G_s} w_{tg,s}(\bm f_t)^{n_{tg,s}},
\end{equation*}
where $n_{tg,s}$ are the current component counts and $\mathcal A_t$ is the narrative-restriction region at date $t$. Here $\bm r_t^{\mathrm{var}}=\bm A_0\bm Q_t-\bm\Phi'\bm m_t+\bm\Lambda_q\bm f_t$ is the structural macro residual with the current factor contribution added back. It is the residual of the macro equations evaluated without the factor term. The Gaussian factor of the target is the product of the prior and the macro likelihood. This gives $\mathcal{N}(\bm f_t\mid\bm 0,\bm I_R)\mathcal{N}(\bm r_t^{\mathrm{var}}\mid\bm\Lambda_q\bm f_t,\bm D)\propto\mathcal{N}(\bm f_t\mid\bm m_t^0,\bm C_t^0)$ with
\begin{equation*}
  \bm C_t^0=\big(\bm I_R+\bm\Lambda_q'\bm D^{-1}\bm\Lambda_q\big)^{-1},\qquad
  \bm m_t^0=\bm C_t^0\bm\Lambda_q'\bm D^{-1}\bm r_t^{\mathrm{var}}.
\end{equation*}
The same residual enters the proposal below, and the multinomial logistic weights $w_{tg,s}(\bm f_t)$ are evaluated at the candidate factor value with the log-weight coefficients held fixed.

Let $\bm\Lambda_q$ ($M\times R$) contain the VAR factor loadings and $\bm\Lambda_{w}^{(s)}$ ($(G_s-1)\times R$) the log-weight factor loadings for cross section $s$, and let $\hat{\bm\Omega}_t^{(s)}(\bm f_t)=\diag_g(\hat\omega_{tg,s}(\bm f_t))$ with $\hat\omega_{tg,s}(\bm f_t)=\tfrac{n_{t,s}}{2\psi_{tg,s}(\bm f_t)}\tanh\big(\tfrac{\psi_{tg,s}(\bm f_t)}{2}\big)$, the conditional mean of the P\'olya--Gamma variable evaluated at the one-versus-rest log odds $\psi_{tg,s}(\bm f_t)$ implied by the factor value at which the proposal is centered. The working residuals are formed with the same $\hat\omega_{tg,s}(\bm f_t)$ and the offsets $\mathrm{a}_{tg,s}(\bm f_t)$. Entry $g$ of $\bm r_t^{\mathrm{w},(s)}(\bm f_t)$ is
\begin{equation*}
  r^{\mathrm{w}}_{tg,s}(\bm f_t)=\frac{n_{tg,s}-n_{t,s}/2}{\hat\omega_{tg,s}(\bm f_t)}+\mathrm{a}_{tg,s}(\bm f_t)-\tilde{\bm X}_t'\tilde{\bm b}_{g,s},
\end{equation*}
where $\tilde{\bm X}_t$ and $\tilde{\bm b}_{g,s}$ are $\bm X_t$ and $\bm b_{g,s}$ without the factor entries, so that $\tilde{\bm X}_t'\tilde{\bm b}_{g,s}+\bm\lambda_{g,s}'\bm f_t=\eta_{tg,s}$. It is the P\'olya--Gamma working response for component $g$ net of the non-factor part of its index, and it depends on $\bm f_t$ through $\hat\omega_{tg,s}(\bm f_t)$ and $\mathrm{a}_{tg,s}(\bm f_t)$. Add the factor contributions back to the macro residual $\bm r_t^{\mathrm{var}}$ and the log-weight working residuals $\bm r_t^{\mathrm{w},(s)}$. With prior $\bm f_t\sim\mathcal{N}(\bm 0,\bm I_R)$, the proposal centered at the current value $\bm f_t$, $q(\cdot\mid\bm f_t)=\mathcal{N}(\bm m_t(\bm f_t),\bm C_t(\bm f_t))$, is
\begin{align*}
  \bm C_t(\bm f_t)^{-1}&=\bm I_R+\bm\Lambda_q'\bm D^{-1}\bm\Lambda_q
  +\sum_{s=1}^{S}\bm\Lambda_{w}^{(s)\prime}\hat{\bm\Omega}_t^{(s)}(\bm f_t)\bm\Lambda_{w}^{(s)},\\
  \bm m_t(\bm f_t)&=\bm C_t(\bm f_t)\Big[\bm\Lambda_q'\bm D^{-1}\bm r_t^{\mathrm{var}}
  +\sum_{s=1}^{S}\bm\Lambda_{w}^{(s)\prime}\hat{\bm\Omega}_t^{(s)}(\bm f_t)
  \bm r_t^{\mathrm{w},(s)}(\bm f_t)\Big],
\end{align*}
truncated to $\mathcal A_t$ at the dates carrying a narrative restriction. When $R>1$, the proposal at such a date is untruncated. A candidate outside $\mathcal A_t$ is rejected. This defines the same kernel. The factor path starts inside $\mathcal A_t$. The proposal is accepted with probability
\begin{equation*}
  \min\Big\{1,
  \frac{\pi(\bm f_t^{\star}\mid\cdot)q(\bm f_t\mid\bm f_t^{\star})}
       {\pi(\bm f_t\mid\cdot)q(\bm f_t^{\star}\mid\bm f_t)}\Big\}.
\end{equation*}
The proposal is a deterministic function of the value it is centered on. Its mean, covariance, and truncation constant depend on $\bm f_t$ through the offsets and curvature terms $\hat\omega_{tg,s}(\bm f_t)$. Both proposal densities therefore enter the ratio. Each density uses its own covariance and determinant. We compute the reverse density by rebuilding the proposal at $\bm f_t^{\star}$. The step is a standard Metropolis--Hastings transition for $\pi$ and leaves it invariant. We evaluate the multinomial logistic probabilities in $\pi$ exactly with a log-sum-exp stabilization. The P\'olya--Gamma draws of the coefficient block are auxiliaries for that block only. They do not enter the factor update. Using them in place of $\hat\omega_{tg,s}(\bm f_t)$ would make the proposal depend on random quantities generated from an earlier state. The resulting kernel would not be exactly invariant. Acceptance rates are between $0.92$ and $0.95$ across chains in the empirical application and about $0.98$ in the HANK simulation. Relative to drawing $\bm f_t$ directly from the displayed Gaussian, the correction leaves the empirical results essentially unchanged. The posterior median factor path correlates at $0.98$ with its uncorrected counterpart, and the macro impulse-response surface at $0.998$. The approximation is nevertheless not innocuous in general. The uncorrected draw inflates the posterior dispersion of $\bm f_t$ by roughly $15\%$ in the calibration underlying our simulation design. In the HANK simulation, where two factors enter, the first half of the burn-in period uses the uncorrected draw as an initialization device. Every retained draw is generated by the exact kernel.

\paragraph{Identifying restrictions in the empirical application.}
\autoref{tab:Wbounds} lists the contemporaneous restrictions on the off-diagonal elements $\bm W$ of $\bm A_0=\bm I_M-\bm W$ for $\bm Q_t=(\text{BZP},OG,\pi,R)'$. Reading equation $i$ as $Q_{ti}=\sum_{j\ne i}W_{ij}Q_{tj}+\dots$, the restrictions define a small Baumeister--Hamilton structural macro model. The policy-rate equation responds contemporaneously to the output gap and inflation, with lower bounds calibrated using the priors in \citet{baumeister2018inference}. Output does not respond within the quarter to the policy rate ($W_{2,4}=0$). Inflation falls with the contemporaneous policy rate. The fiscal instrument is contemporaneously exogenous ($\bm W_{1,\cdot}=\bm 0$).

\begin{table}[H]
  \centering
  \caption{Sign and magnitude restrictions on $\bm W$ in the four-variable
  application, $\bm Q_t=(\text{BZP},OG,\pi,R)'$. Unreported bounds are
  $\pm\infty$.}
  \label{tab:Wbounds}
  \begin{threeparttable}
  \begin{tabular}{lll}
    \toprule
    Equation (row $i$) & Regressor (col $j$) & Restriction \\
    \midrule
    BZP $(i=1)$        & all $j\ne 1$ & $W_{1j}=0$ \quad(exogeneity) \\
    Output gap $(i=2)$ & inflation $(j=3)$ & $W_{2,3}\ge 0.5$ \\
    Output gap $(i=2)$ & FFR $(j=4)$       & $W_{2,4}=0$ \\
    Inflation $(i=3)$  & FFR $(j=4)$       & $W_{3,4}\le -0.5$ \\
    FFR $(i=4)$        & output gap $(j=2)$ & $W_{4,2}\ge 0.1$ \\
    FFR $(i=4)$        & inflation $(j=3)$  & $W_{4,3}\ge 0.1$ \\
    \bottomrule
  \end{tabular}
  \end{threeparttable}
\end{table}

The remaining restrictions are the following.
\begin{enumerate}[label=(\roman*), nosep]
\item impact signs for $\bm A_0^{-1}$: a fiscal shock raises the output gap, while a monetary policy shock lowers the output gap and inflation and raises the federal funds rate, imposed through the row-level rejection described above;
\item restrictions on contemporaneous log-weight loadings: positive output gap or inflation innovations shift mass toward higher-mean components, and positive policy-rate innovations shift mass toward lower-mean components, with magnitude floors of $0.02$ and $0.08$;
\item an inequality-compression pattern on the factor loadings that identifies $f_t$: a positive micro shock moves mass from both tails toward the middle components, with tail and middle magnitude floors of $0.05$ and $0.08$;
\item six narrative sign restrictions on $f_t$. We impose $f_t\ge0$ at the 1993 Omnibus Budget Reconciliation Act in 1993Q3 (signed August 1993) and at the minimum-wage increases to \$4.75 in 1996Q4, \$5.15 in 1997Q3, and \$5.85 in 2007Q3. We impose $f_t\le0$ at the 2001 Bush rebates in 2001Q3 and the 2003 Jobs and Growth Tax Relief Reconciliation Act in 2003Q2.
\item the VAR factor-loading pattern $(0,+,+,\cdot)$: the BZP loading is fixed at zero, the output gap and inflation loadings are positive, and the policy-rate loading is unrestricted.
\end{enumerate}

\paragraph{Impulse response computation.}
We compute impulse responses for each retained draw by forward simulation from the sample mean of the state. Horizon $1$ denotes the impact period throughout, in the text, in the tables, and in every figure. The initial state sets the aggregate lags to their sample means, the quantile regressors to the quantiles of the pooled lagged cross sections, and the factors to zero. The structural impact matrix is $\bm P=\bm A_0^{-1}\bm D^{1/2}$, whose column $\ell$ is the impact response to a one-standard-deviation structural innovation $u_{t\ell}$. Partition $\bm\Phi'=[\bm A_1,\dots,\bm A_P,\bm\alpha,\bm c,\bm\Lambda_q]$ conformably with $\bm m_t$. The reduced-form companion matrix has first block row $\bm A_0^{-1}[\bm A_1,\dots,\bm A_P]$ and identity blocks below. It involves only the aggregate-lag block, not the quantile, intercept, or factor coefficients. We exclude draws whose companion matrix has spectral radius at least one. This screen concerns the linear aggregate-lag dynamics. It is not a stability condition for the full nonlinear recursion with quantile feedback, whose paths we simulate directly. A shock of $\delta$ standard deviations to structural innovation $\ell$ changes the impact vector by $\delta\bm P_{\cdot\ell}$, and we report $\delta=3$. For a factor shock, we set the factor to $\delta$ on the shocked path and to zero on the baseline path at horizon $1$. At later horizons both paths set the factor to its unconditional mean of zero, because $\bm f_t$ is i.i.d., so the perturbation equals $\delta$ at impact and zero thereafter. We propagate baseline and shocked paths through the nonlinear recursion. At horizon $h$, we compute the macro vector $\bm Q_h$ from the regressor vector $\bm m_h$ (plus the impact term at $h=1$), evaluate the mixture weights at $\bm Q_h$, its lags, and the factor through the multinomial logistic map, and obtain the cross-sectional quantiles $\mathcal{Q}_r$ of $\sum_g w_{hg,s}\mathcal{N}(\cdot\mid\mu_{g,s},\sigma^2_{g,s})$ from $10{,}000$ simulated draws. These horizon-$h$ quantiles enter the regressor vector $\bm m_{h+1}$ of the next period, matching the one-period lag in \autoref{eq: VAR}. They never enter $\bm m_h$ contemporaneously. During estimation, the feedback term in \autoref{eq: VAR} uses the empirical quantiles of the observed cross sections. In the impulse-response simulation, the quantiles are simulated from the fitted mixture along both paths. Macro impulse responses are differences between the two macro paths, rescaled by the standard deviations of $\bm Q$ when the data are standardized. Density responses equal $f^{\mathrm{shock}}_h(x)-f^{\mathrm{base}}_h(x)$. Quantile responses are differences in $\mathcal{Q}_r$, computed for the reported figures by inverting the mixture distribution function. Inequality responses are differences in the Gini coefficient, standard deviation, interquartile range, and P90--P10 gap, evaluated in closed form (Gini coefficient and standard deviation) or by inversion of the mixture distribution function (interquartile range and P90--P10 gap).

\section{Validation Appendix}\label{app:validation}
This appendix reports the parameterization of the controlled DGP of \autoref{sec:validation}, the associated recovery results, and the fit of the micro block in the HANK exercise.

\subsection{Controlled DGP: Parameterization}\label{app:controlled_dgp}
The aggregate block is the VAR(1)
\begin{equation*}
  \bm Q_t = \begin{pmatrix} 0.6 & 0 \\ 0.1 & 0.5 \end{pmatrix}\bm Q_{t-1}
          + \bm\alpha \widetilde{\bm m}_{t-1} + \bm\Lambda_q f_t + \bm u_t,
  \qquad
  \bm u_t \sim \mathcal{N}\left(\bm 0,\begin{pmatrix} 0.10 & 0.02 \\ 0.02 & 0.10 \end{pmatrix}\right),
\end{equation*}
initialized at $\bm Q_0 = \bm 0$ with $T = 250$. The vector
$\widetilde{\bm m}_{t-1} = (\widehat m^{(1)}_{t-1}, \widehat m^{(2)}_{t-1})'$ contains the medians of the two cross sections. These medians feed back into the macro block with $\bm \alpha = 0.3\bm I_2$. The common factor $f_t \sim \mathcal{N}(0,1)$ enters the macro block with loadings $\bm\Lambda_q = (0.10, -0.06)'$.

At each date, we draw $n_{t,s} = 500$ observations for each cross section. For unit $i$ and cross section $j \in \{1,2\}$, we first draw the component assignment from the time-varying weights and then draw the outcome from that component,
\begin{equation*}
  z^{(j)}_{it} \sim \text{Categorical}\big(\bm w^{(j)}_{t}\big),
  \qquad
  y^{(j)}_{it} \big| \{z^{(j)}_{it} = k\} \sim
  \mathcal{N}\big(\mu^{(j)}_{k}, \sigma^{2(j)}_{k}\big),
\end{equation*}
using $G = 3$ components with fixed locations and scales
\begin{equation*}
  \begin{aligned}
    \mu^{(1)} &= (-2, 0, 2)', &\qquad \sigma^{2(1)} &= (0.4, 0.3, 0.5)', \\
    \mu^{(2)} &= (-1, 0, 2)', &\qquad \sigma^{2(2)} &= (0.5, 0.4, 0.3)'.
  \end{aligned}
\end{equation*}
The mixture weights depend on a constant, the contemporaneous and lagged macro state, and the common factor, so that $\bm x_t = (1, \bm Q_t', \bm Q_{t-1}', f_t)'$. The factor therefore shifts both blocks, while all time variation in the cross-sectional densities comes from the weights $\bm w^{(j)}_t$. The factor's identifying sign restrictions are a positive macro loading on $Q_1$, a negative loading on $Q_2$, and signed mixture-weight loadings. The structural macro shock is the first shock of the recursive orthogonalization of $\bm u_t$ with $Q_{1t}$ ordered first, that is, the first column of the lower-triangular Cholesky factor of the innovation covariance, $(0.316, 0.063)'$ per standard deviation. The estimated model imposes the same recursive identification through a lower-triangular $\bm A_0$. Reported macro-shock responses are to a three-standard-deviation shock, and factor-shock responses to a two-standard-deviation innovation in $f_t$.

\subsection{Controlled DGP: Recovery Results}\label{app:controlled_results}
\begin{figure}[!htb]
  \centering
  \includegraphics[width=\textwidth]{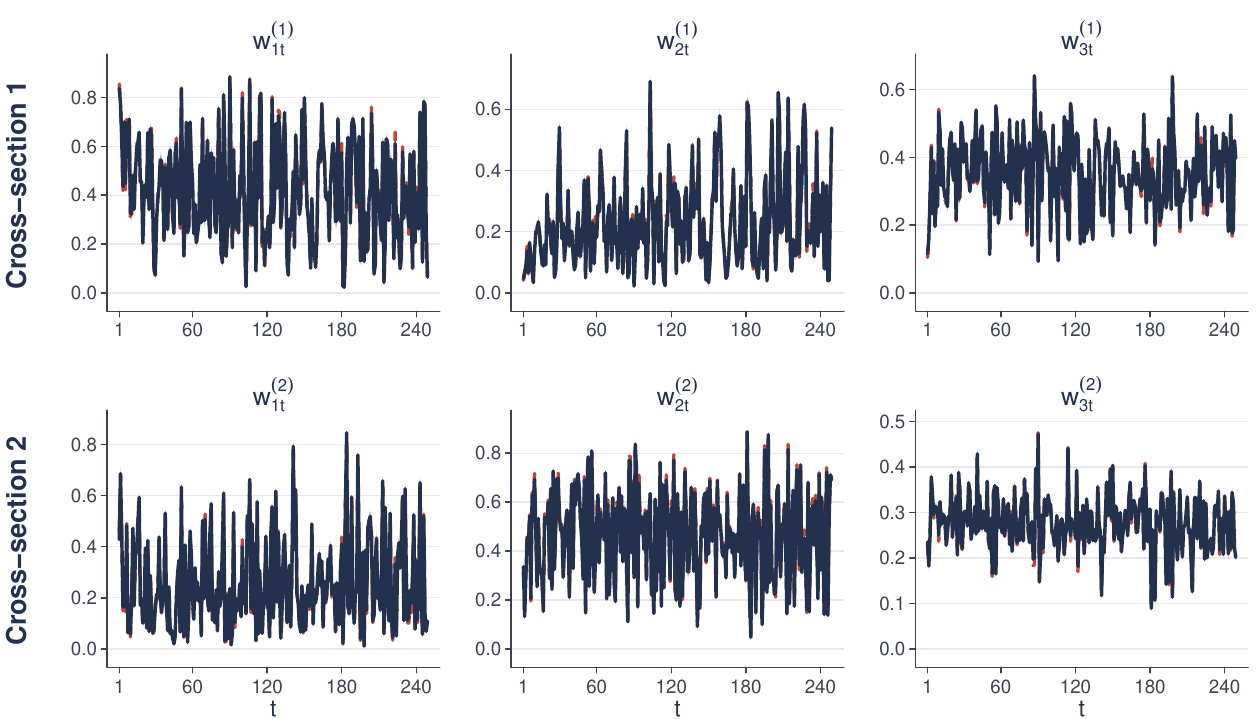}
  \caption*{\footnotesize \textbf{Notes}: Navy lines are posterior median weights, shaded areas are 68\% and 90\% credible bands, and dashed red lines are the true weights, for every component and both cross sections.}
  \caption{Recovery of the mixture weights}
  \label{fig:rf_weights}
\end{figure}

\begin{figure}[!htb]
  \centering
  \includegraphics[width=\textwidth]{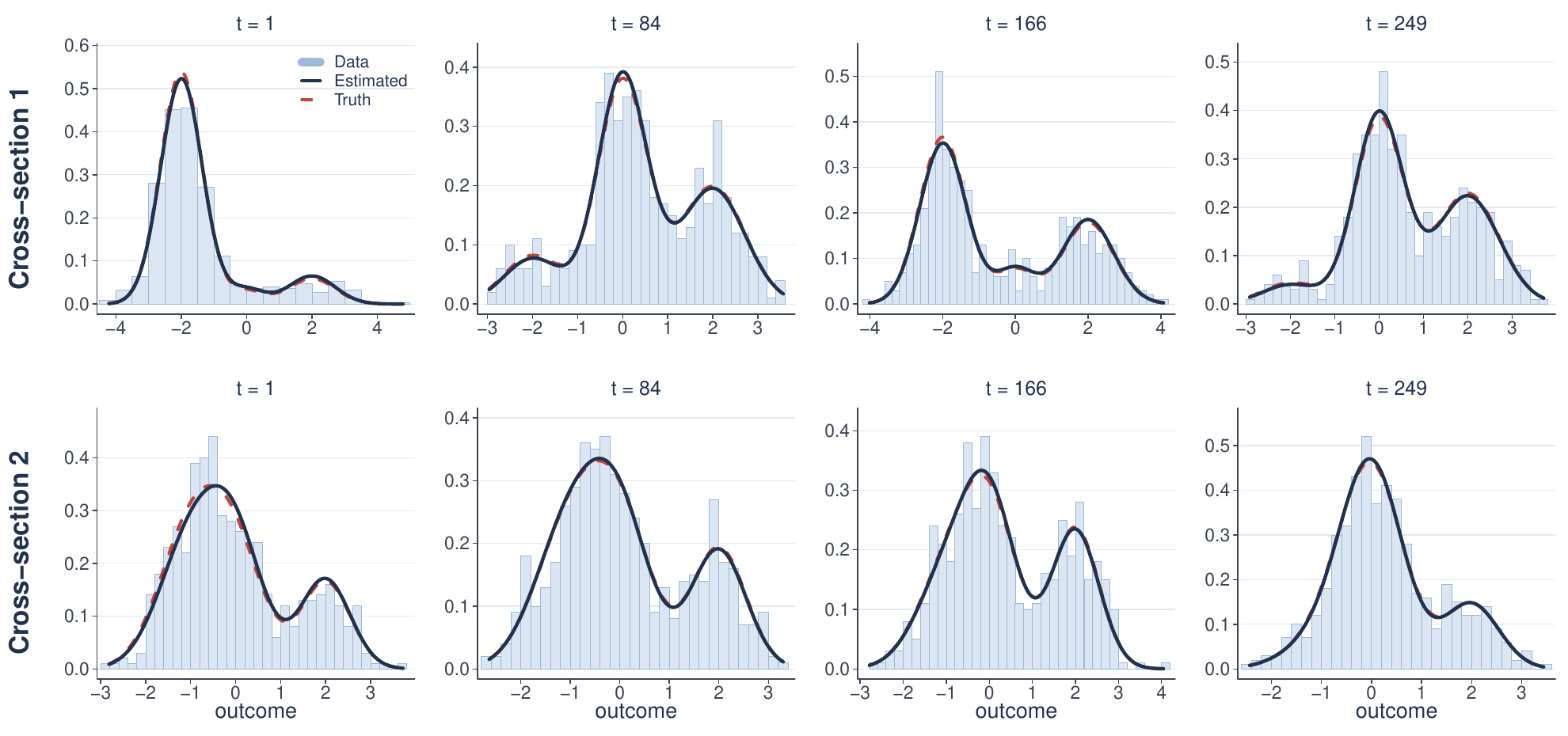}
  \caption*{\footnotesize \textbf{Notes}: Rows correspond to the two cross sections ($j=1$ top, $j=2$ bottom) and columns to $t\in\{1,84,166,249\}$. Light-blue histograms show the simulated data, solid dark lines the posterior mixture densities, and dashed red lines the true densities.}
  \caption{Recovery of the cross-sectional densities}
  \label{fig:rf_densities}
\end{figure}

Because the DGP is a Gaussian mixture, the true weights and densities are known. \autoref{fig:rf_weights} compares the estimated and true mixture weights, and \autoref{fig:rf_densities} the estimated and true densities at selected dates. True values are in red.

The posterior median closely tracks the true mixture weights. The credible bands are barely visible because each date contains $n_{t,s}=500$ observations and the posterior variance of the log-weight coefficients declines linearly with the per-date sample size.

The fitted and true cross-sectional densities are nearly indistinguishable at every date shown. Accurate recovery of component locations, scales, and weights underlies the dynamic recovery below.

\begin{figure}[!htb]
  \centering
  \includegraphics[width=\textwidth]{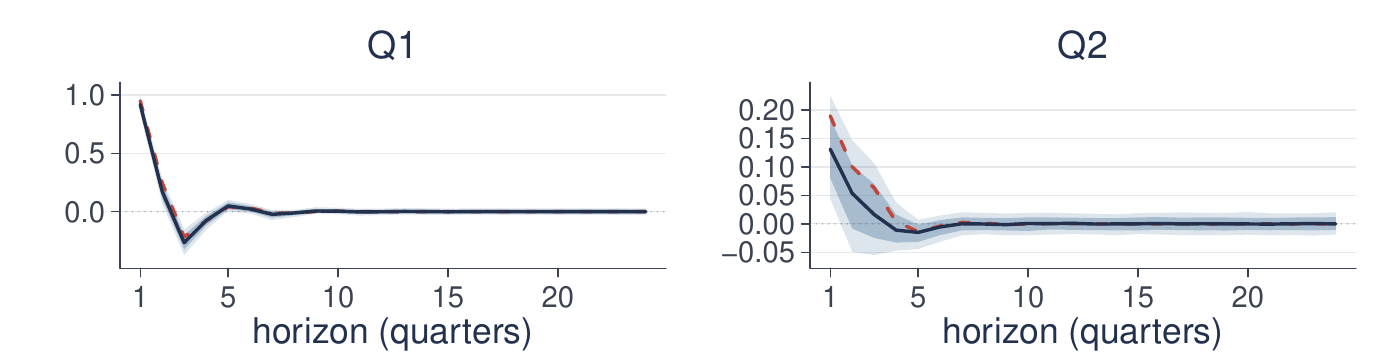}
  \caption*{\footnotesize \textbf{Notes}: Aggregate responses to the structural macro shock. Navy lines are posterior medians, shaded areas are 68\% and 90\% credible bands, and dashed red lines are the truth.}
  \caption{Macro IRFs to the macro shock}
  \label{fig:rf_macro_macro}
\end{figure}

We turn to the structural impulse responses, beginning with a macro shock, an innovation to the first equation of the macro VAR. \autoref{fig:rf_macro_macro} reports the aggregate responses. A three-standard-deviation shock immediately raises both $Q_{1t}$ and $Q_{2t}$. The second response reflects the off-diagonal coefficient and the contemporaneous error correlation. The posterior median for $Q_{1t}$ closely tracks the true response. The 90\% credible band contains the true path at every horizon. The true path leaves the 68\% band only in the second quarter. The median for $Q_{2t}$ understates the transmission on impact, where the truth lies outside the 68\% band, but the 90\% band contains the true path at every horizon.

\begin{figure}[!t]
  \centering
  \begin{subfigure}[b]{\textwidth}
    \centering
    \caption{Distributional responses}
    \label{fig:rf_dist_macro}
    \includegraphics[width=\textwidth]{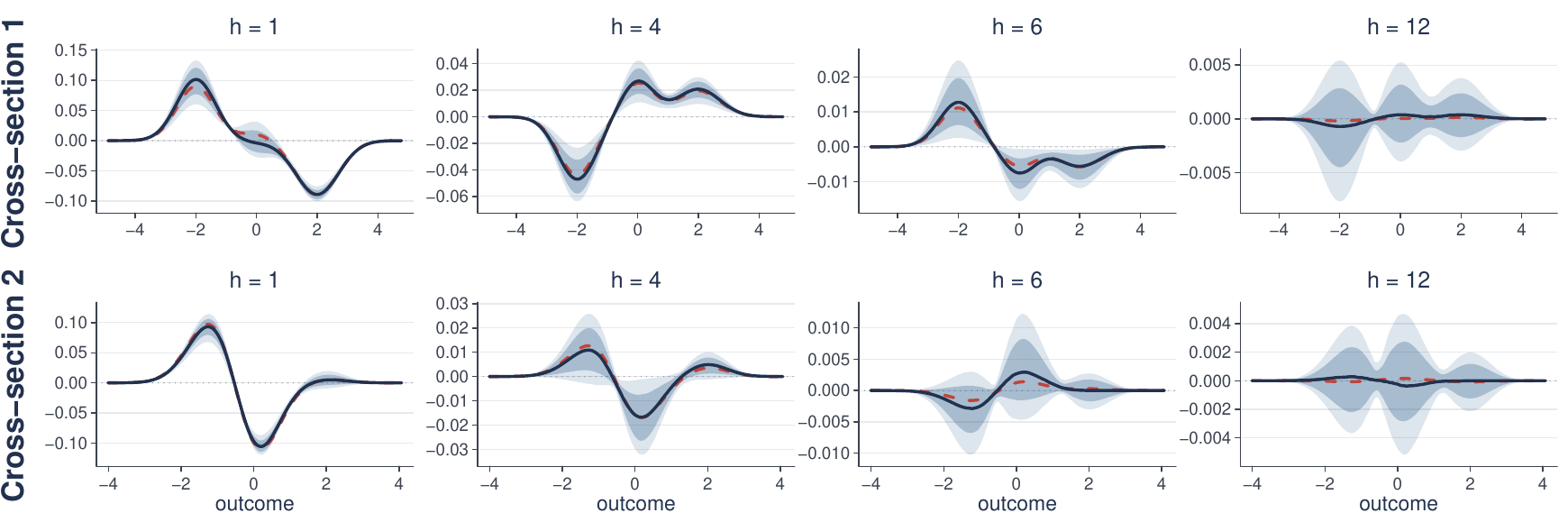}
  \end{subfigure}\\[0.2em]
  \begin{subfigure}[b]{\textwidth}
    \centering
    \caption{Quantile responses}
    \label{fig:rf_quant_macro}
    \includegraphics[width=\textwidth]{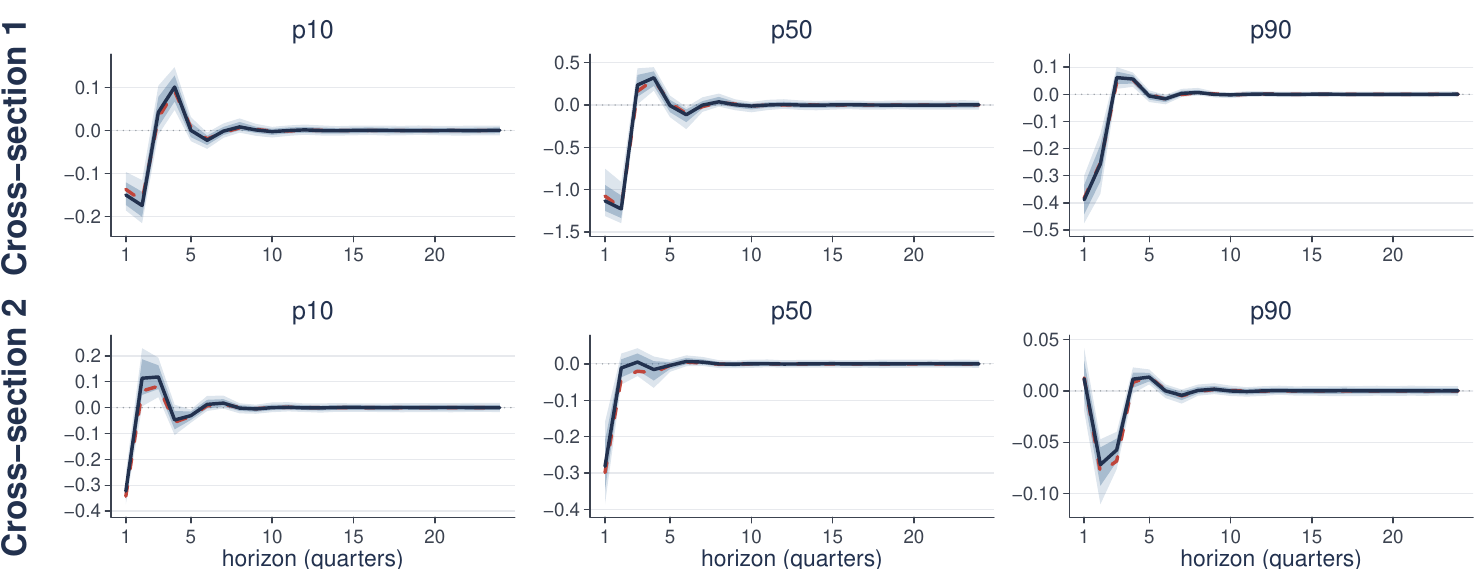}
  \end{subfigure}
  \caption*{\footnotesize \textbf{Notes}: Panel (a) reports the distributional response of each cross section, $\Delta f^{(j)}(x;h)=f^{(j)}_{\text{shocked}}(x;h)-f^{(j)}_{\text{baseline}}(x;h)$, at $h\in\{1,4,6,12\}$. Panel (b) reports responses of the 10th, 50th, and 90th percentiles. Navy lines are posterior medians, shaded areas are 68\% and 90\% credible bands, and dashed red lines are the truth.}
  \caption{Density and quantile IRFs to the macro shock}
  \label{fig:rf_dist_quant_macro}
\end{figure}

The macro shock also reshapes the cross-sectional densities because the contemporaneous and lagged macro state enters the mixture-weight equations. Following \citet{chang2024heterogeneity}, \autoref{fig:rf_dist_macro} summarizes this propagation through density responses. The model recovers the shape and location of the true responses. The posterior median slightly overstates the impact peak in the first cross section ($0.10$ against $0.09$ in density units), and the 90\% bands contain the true response at every point of the support at all four horizons shown, even though the bands are tight because each date contains $500$ observations. The quantile responses in \autoref{fig:rf_quant_macro} lead to the same conclusion. The 90\% bands contain the true response at every percentile--horizon pair, and the posterior median tracks the truth with errors below $0.01$ at almost every pair. The largest deviation is at the impact response of the first cross section's median.

\begin{figure}[!htb]
  \centering
  \includegraphics[width=\textwidth]{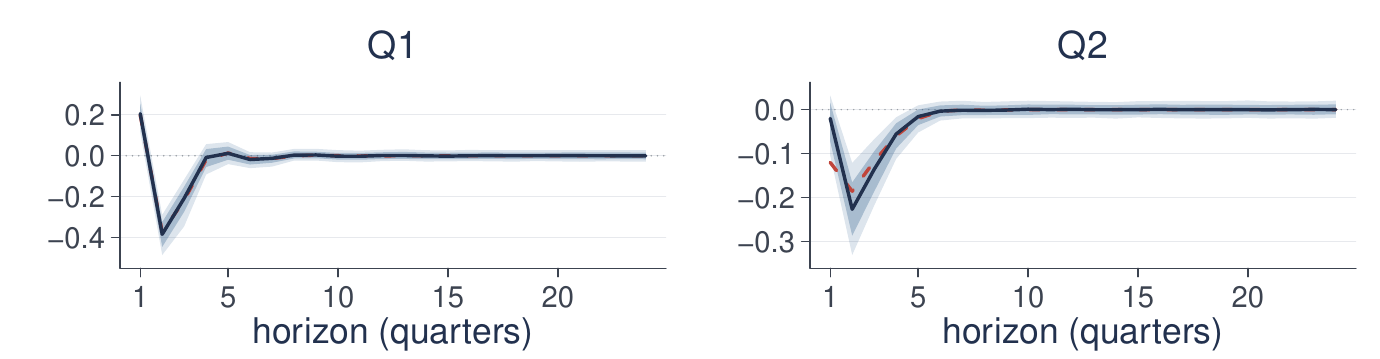}
  \caption*{\footnotesize \textbf{Notes}: Aggregate responses to the common-factor shock. Navy lines are posterior medians, shaded areas are 68\% and 90\% credible bands, and dashed red lines are the truth.}
  \caption{Macro IRFs to the common-factor shock}
  \label{fig:rf_macro_micro}
\end{figure}

\begin{figure}[!t]
  \centering
  \begin{subfigure}[b]{\textwidth}
    \centering
    \caption{Distributional responses}
    \label{fig:rf_dist_micro}
    \includegraphics[width=\textwidth]{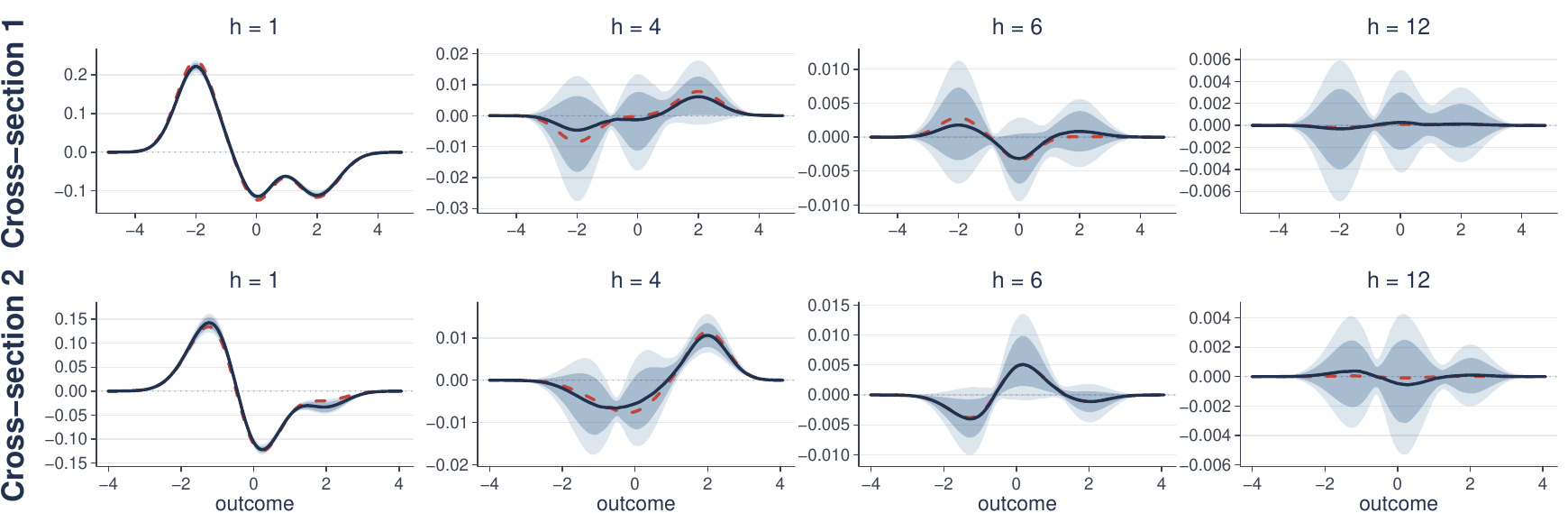}
  \end{subfigure}\\[0.2em]
  \begin{subfigure}[b]{\textwidth}
    \centering
    \caption{Quantile responses}
    \label{fig:rf_quant_micro}
    \includegraphics[width=\textwidth]{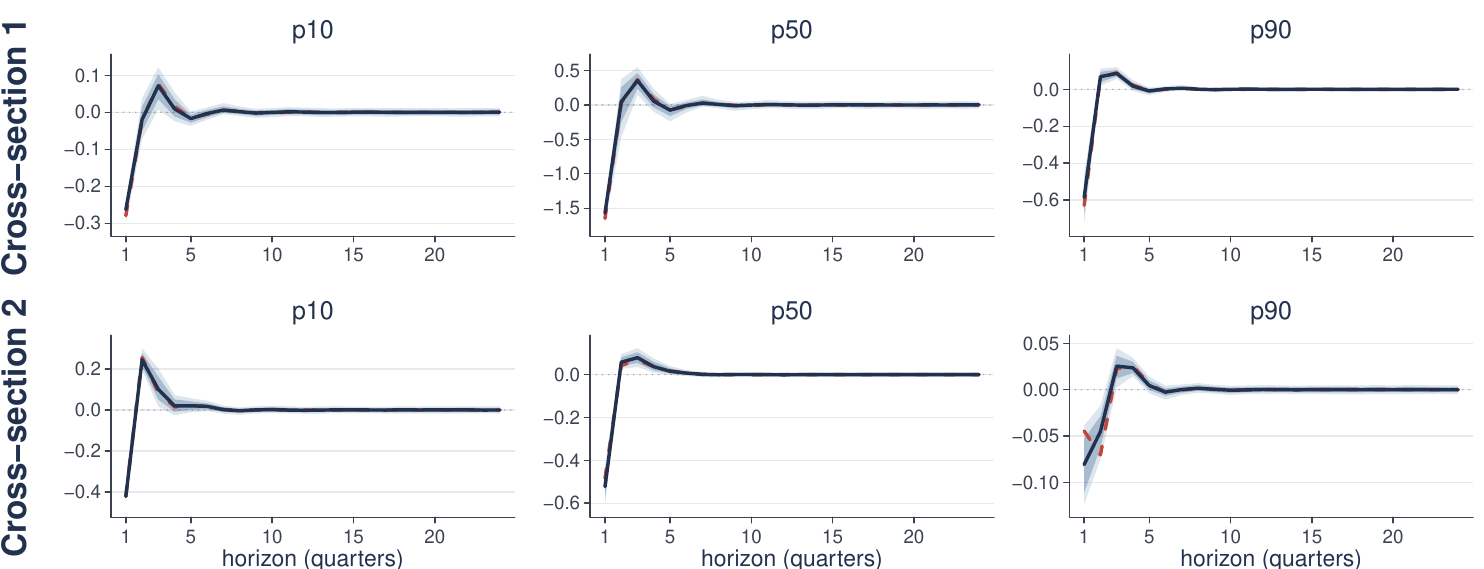}
  \end{subfigure}
  \caption*{\footnotesize \textbf{Notes}: See the notes to \autoref{fig:rf_dist_quant_macro}.}
  \caption{Density and quantile IRFs to the common-factor shock}
  \label{fig:rf_dist_quant_micro}
\end{figure}

We next shock the common factor $f_t$. Because the factor enters both the macro block and the mixture weights, this exercise tests recovery of joint macro--micro propagation. \autoref{fig:rf_macro_micro} reports the aggregate responses, and \autoref{fig:rf_dist_quant_micro} the density and quantile responses. The shock sharply reshapes both cross-sectional distributions on impact, reallocating mass across mixture components, before the effects decay. The aggregate responses peak in the second quarter and return to zero within roughly six quarters. The response of $Q_{1t}$ is positive on impact and then swings negative. Across aggregate, density, and quantile responses, posterior medians track the truth closely. The 90\% credible bands contain the true aggregate responses at every horizon except the impact response of $Q_{2t}$, the true quantile responses at every percentile--horizon pair, and the true density responses at every point of the support and horizon shown except a small part of the support at impact for the first cross section, where the posterior median slightly understates the peak of the immediate reallocation of mass ($0.22$ against $0.24$ in density units). The remaining discrepancy at impact reflects the weak identification of the common factor from two cross sections and the aggregate state, which \autoref{sub:rf_dgp} discusses.

\FloatBarrier

\subsection{HANK DGP: Calibration}\label{app:hank_calibration}
Idiosyncratic productivity follows an AR(1) in logs with quarterly persistence $0.966$ and cross-sectional standard deviation $0.5$, discretized on a $101$-state Rouwenhorst grid with fixed endpoints and matched mean and variance. The asset grid has $500$ points on $[0,150]$. The elasticity of intertemporal substitution and the Frisch elasticity are both $0.5$. The markup is $1.2$, and the New Keynesian Phillips curve has slope $0.1$. The Taylor-rule inflation coefficient is $1.5$. Real bonds are fixed at $5.6$ times quarterly output. The discount factor ($0.979$) and the disutility of labor clear the asset and labor markets at a steady-state quarterly real rate of $0.5$ percent.

Aggregate assets equal the fixed bond supply, so the simulated series is constant up to the numerical precision of the solution (its standard deviation is of order $10^{-11}$). We keep it in the observation vector. The aggregate series are standardized before estimation, so this series becomes a unit-variance numerical-noise series, its least-squares residual variance $\hat\sigma^2_i$ is that of the standardized noise, and its equation is fitted like any other. Both factors are excluded from this equation (\autoref{sec:validation}). Its lags enter the other equations and the log-weight indices as pure-noise regressors, which the horseshoe prior shrinks toward zero. Its own response is numerically zero and is shown on a symmetric scale in \autoref{fig:hank_irfs}.

\subsection{HANK DGP: Fit of the Micro Block}\label{app:hank_fit}
\begin{figure}[htbp]
  \centering
  \begin{subfigure}[b]{\textwidth}
    \centering
    \caption{Consumption}
    \includegraphics[width=\textwidth, trim=0 0 0 5bp, clip]{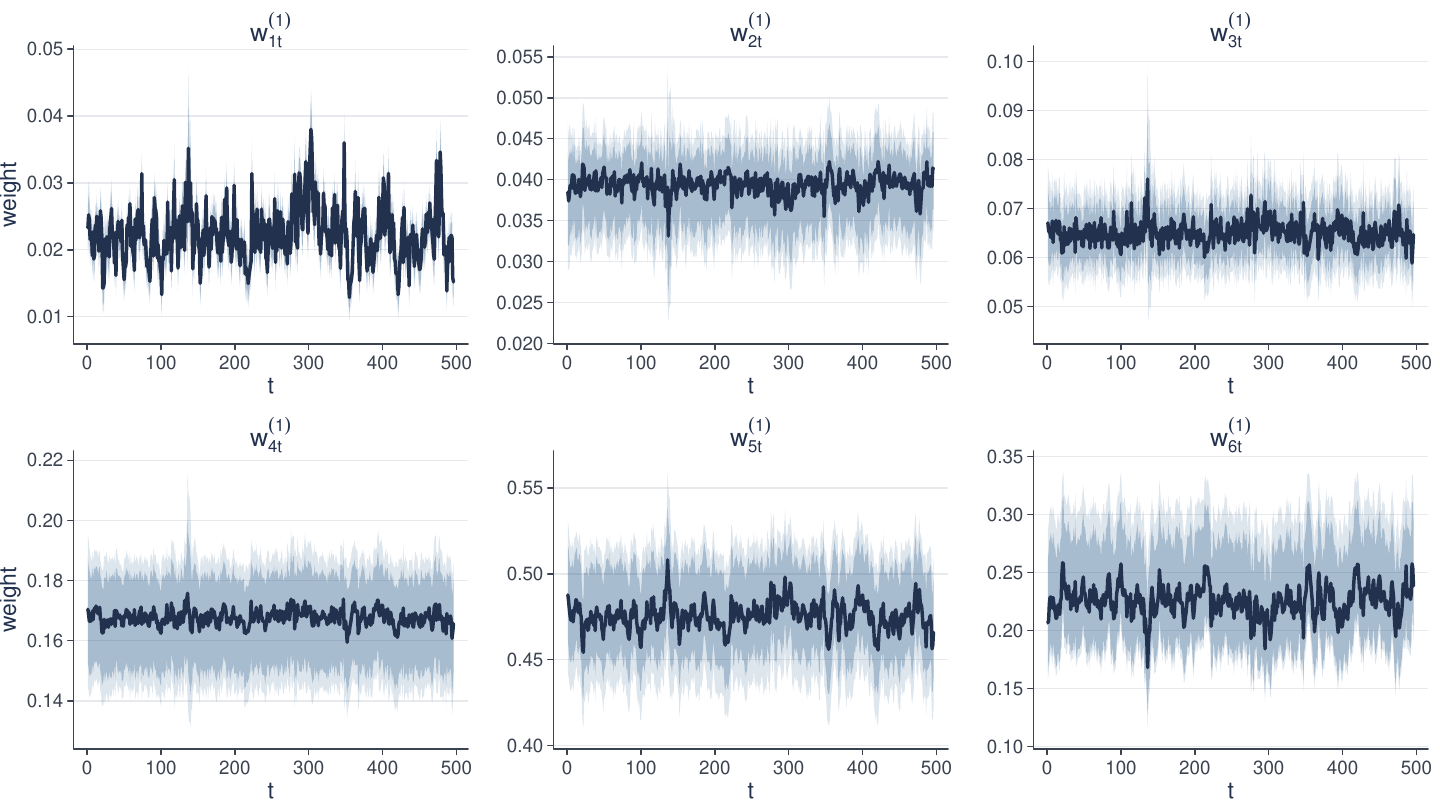}
  \end{subfigure}\\[0.4em]
  \begin{subfigure}[b]{\textwidth}
    \centering
    \caption{Earnings}
    \includegraphics[width=\textwidth, trim=0 0 0 5bp, clip]{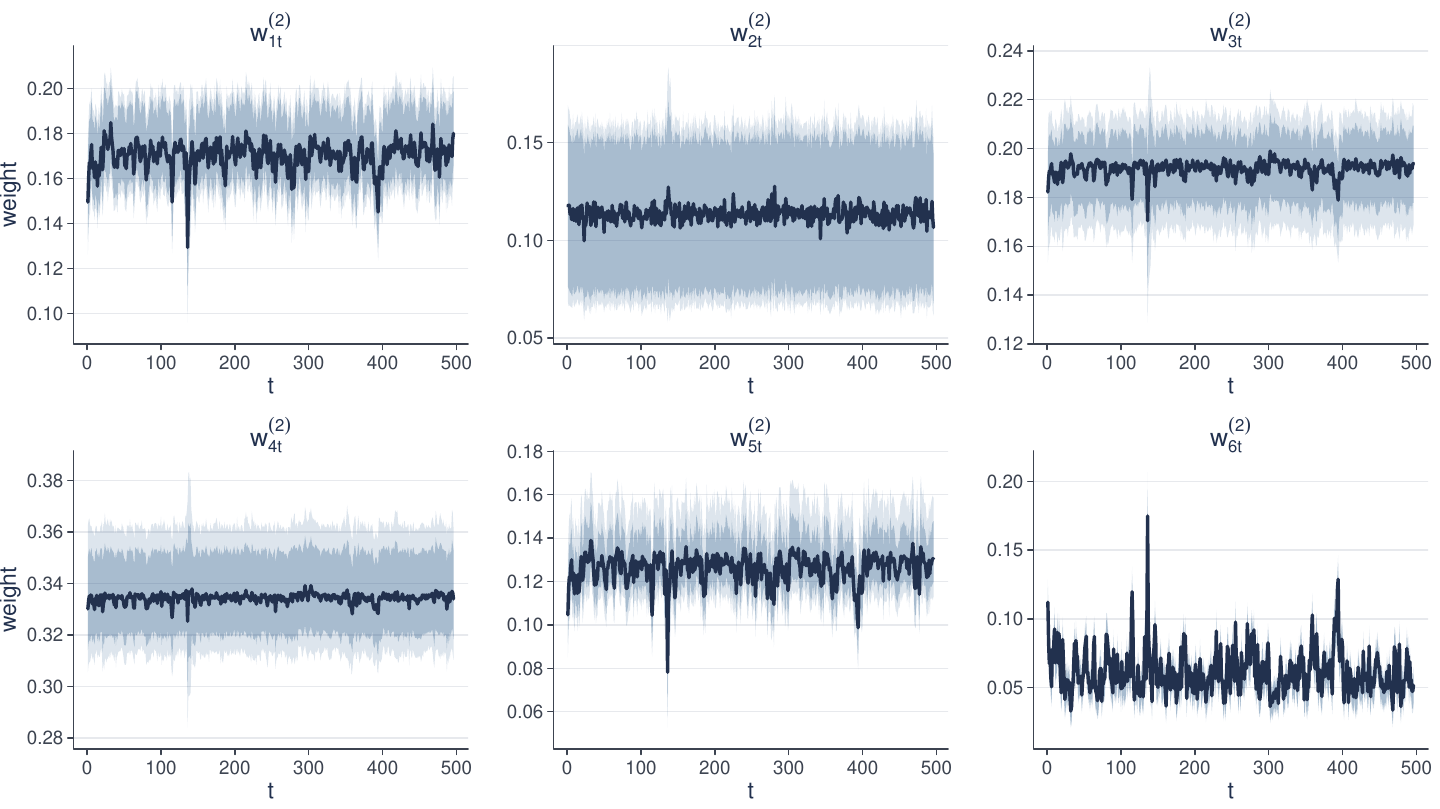}
  \end{subfigure}
  \caption*{\footnotesize \textbf{Notes}: Navy lines are posterior median weights $w_{tg,s}$, and shaded areas are 68\% and 90\% credible bands. Panel (a) reports consumption ($j=1$) and panel (b) earnings ($j=2$). Each panel uses its own vertical scale.}
  \caption{HANK simulation: mixture weights}
  \label{fig:hank_weights}
\end{figure}

\autoref{fig:hank_weights} reports the estimated consumption and earnings weights. Some weights remain nearly constant. Others absorb higher-frequency or episodic changes in distributional shape. Posterior precision varies across components.

\begin{figure}[htbp]
  \centering
  \includegraphics[width=\textwidth, trim=0 0 0 0, clip]{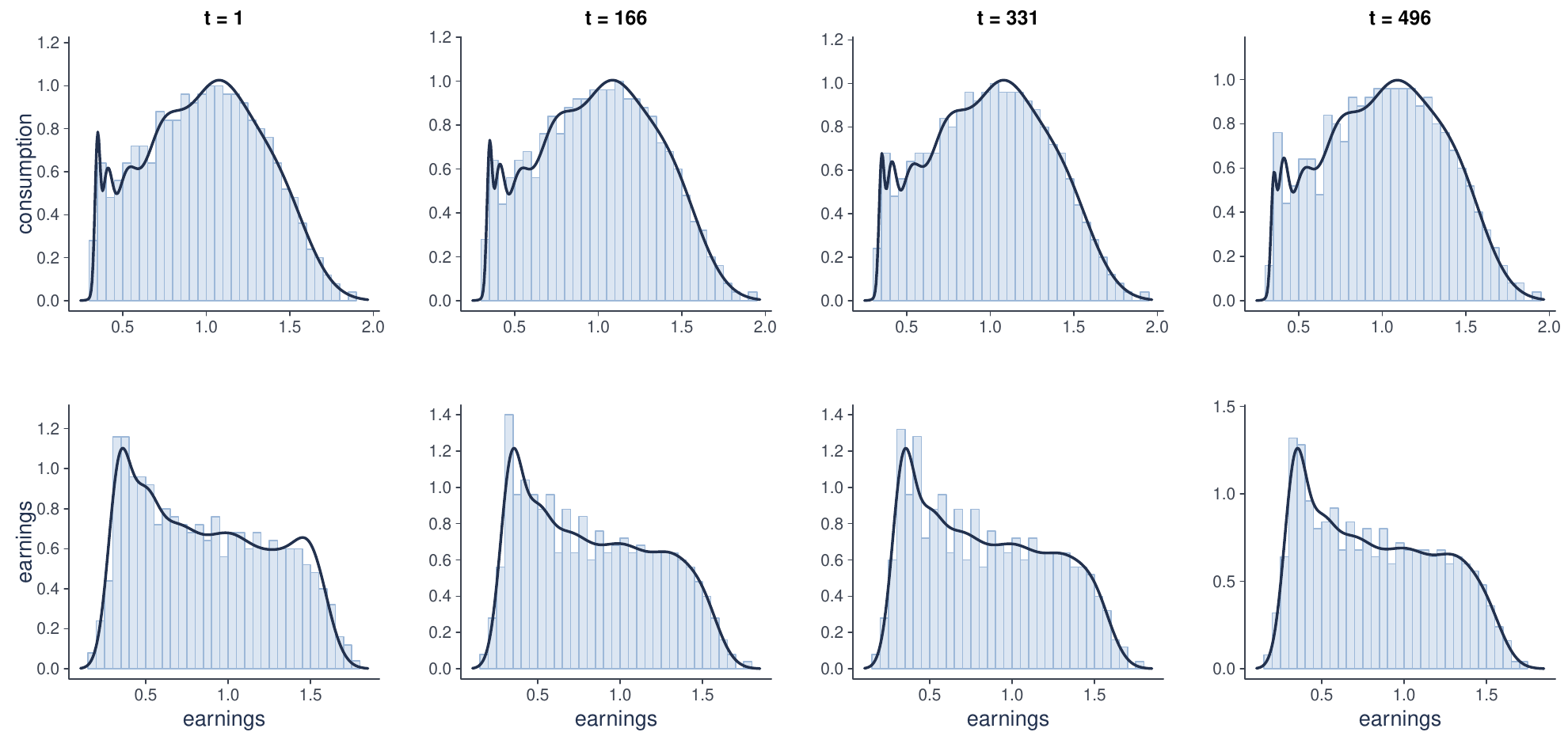}
  \caption*{\footnotesize \textbf{Notes}: Fitted mixture densities (solid) overlaid on the cross-section histograms at selected periods of the HANK simulation.}
  \caption{HANK simulation: fitted cross-sectional densities}
  \label{fig:hank_densities}
\end{figure}

\autoref{fig:hank_densities} overlays the fitted mixture densities on the HANK histograms at selected dates. The six-component mixtures reproduce the broad support, modes, skewness, and tails of both consumption and earnings. The fit is not exact at every bin, especially near the lower support of consumption, but the micro block approximates the evolving marginal distributions well despite the misspecification.

\FloatBarrier

\section{Empirical Appendix}\label{app:empirical}
This appendix collects the empirical results that support \autoref{sec:empirical}. It reports the choice of the number of mixture components, the fitted distributions, MCMC diagnostics, and the inequality responses to the micro shock.

\subsection{Model Selection}\label{app:model_fit_selection}
We choose the number of mixture components in each cross section by the Widely Applicable Information Criterion (WAIC) \citep{watanabe2010,gelman2014waic}, computed from posterior draws of the micro likelihood. \autoref{tab:waic} compares $G_s\in\{4,5,6,8,10\}$. The grid uses a common number of components for both cross sections. WAIC falls sharply between four and six components and is flat over six, eight, and ten. The differences within this plateau are small, and their ranking varies across runs with different seeds, which suggests they are within the Monte Carlo error of the criterion. Six components attain the lowest value in the reported run and are the most parsimonious choice on the plateau. The pointwise unit is one of the $500$ quantile points at a given date, so WAIC is a relative fit criterion under the fixed-grid representation rather than a measure of out-of-sample fit for individual households. The log pointwise predictive density pools earnings and consumption over the $T=67$ post-lag dates.

\begin{table}[htbp]
  \centering
  \caption{WAIC selection of the number of mixture components}
  \label{tab:waic}
  \begin{threeparttable}
  \begin{tabular}{crrrr}
    \toprule
    $G_s$ & lppd & $p_{\text{WAIC}}$ & WAIC & $\Delta$WAIC \\
    \midrule
    $4$ & $-44{,}002.2$ & $89.8$ & $88{,}183.9$ & $172.2$ \\
    $5$ & $-43{,}897.5$ & $115.2$ & $88{,}025.3$ & $13.6$ \\
    $\mathbf{6}$ & $\mathbf{-43{,}878.5}$ & $\mathbf{127.3}$ & $\mathbf{88{,}011.7}$ & $\mathbf{0.0}$ \\
    $8$ & $-43{,}903.3$ & $134.4$ & $88{,}075.5$ & $63.8$ \\
    $10$ & $-43{,}868.8$ & $144.3$ & $88{,}026.3$ & $14.6$ \\
    \bottomrule
  \end{tabular}
  \begin{tablenotes}[flushleft]
    \footnotesize
    \item \textit{Notes}: $\mathrm{lppd}$ is the log pointwise predictive density, $p_{\text{WAIC}}$ is the effective number of parameters, $\mathrm{WAIC}=-2(\mathrm{lppd}-p_{\text{WAIC}})$, and $\Delta$WAIC is measured relative to the minimum. Lower values indicate better fit under the fixed-grid representation. The bold row marks the selected specification. It has the lowest WAIC in the reported grid and run.
  \end{tablenotes}
  \end{threeparttable}
\end{table}

\begin{figure}[htbp]
  \centering
  \includegraphics[width=\textwidth, trim=0 0 0 0, clip]{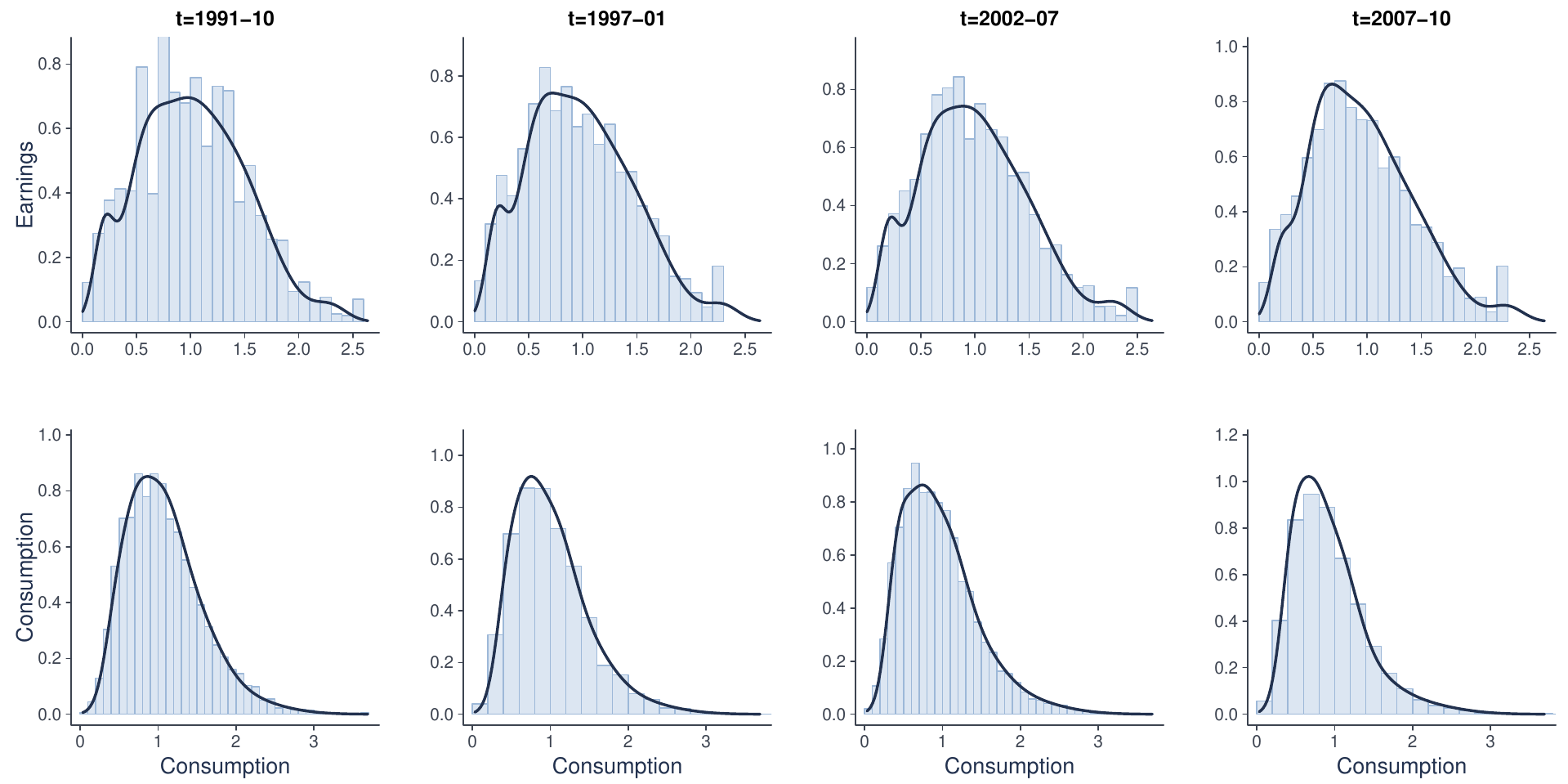}
  \caption*{\footnotesize \textbf{Notes}: Solid lines are fitted mixture densities. Histograms show earnings and consumption at selected quarters.}
  \caption{Fitted cross-sectional densities over time}
  \label{fig:emp_densities_ot}
\end{figure}

\autoref{fig:emp_densities_ot} overlays the fitted mixture densities on the earnings and consumption histograms at selected quarters. The consumption fit is close throughout. The earnings histograms are more irregular. The six-component mixture smooths over some local bumps but matches the location, spread, and skewness of the earnings distribution at every date shown.

\begin{figure}[htbp]
  \centering
  \begin{subfigure}[b]{\textwidth}
    \centering
    \caption{Earnings}
    \includegraphics[width=\textwidth, trim=0 0 0 10bp, clip]{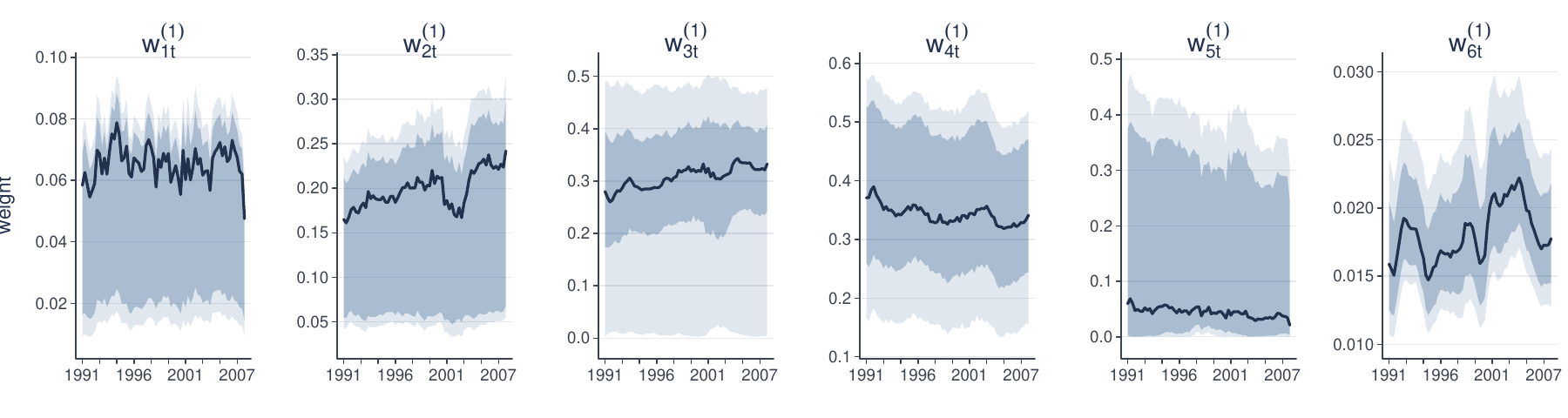}
  \end{subfigure}\\[0.4em]
  \begin{subfigure}[b]{\textwidth}
    \centering
    \caption{Consumption}
    \includegraphics[width=\textwidth, trim=0 0 0 10bp, clip]{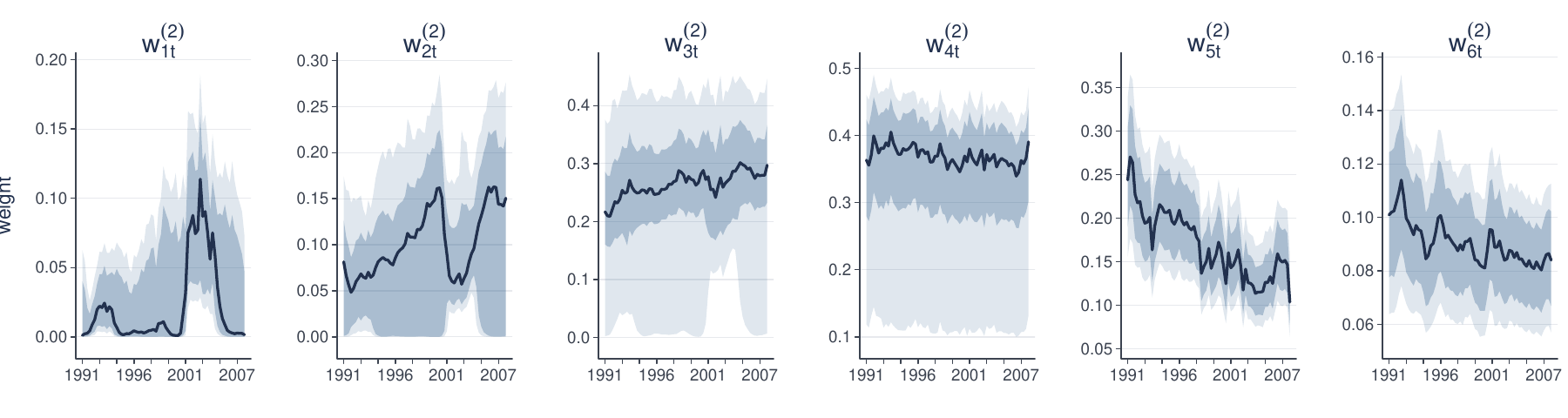}
  \end{subfigure}
  \caption*{\footnotesize \textbf{Notes}: Navy lines are posterior median weights $w_{tg,s}$, and shaded areas are 68\% and 90\% credible bands. Panel (a) reports earnings and panel (b) consumption.}
  \caption{Mixture weights}
  \label{fig:emp_weights}
\end{figure}

\autoref{fig:emp_weights} reports the estimated mixture weights over time. The weights show both persistent and higher-frequency reallocations of probability mass. For earnings, the weight on the second component drifts up over the sample and the weight on the fifth component drifts down. For consumption, the weights on the second and third components rise while those on the fifth and sixth components fall. The bands for the middle components are wide. Adjacent components overlap, so the data do not pin down how mass is split between them. The mixture density is what enters the impulse responses, and it is much less sensitive to this split than the individual weights are.

\FloatBarrier

\subsection{MCMC Diagnostics}\label{app:mcmc_diagnostics}
The empirical application runs ten independent chains from different random seeds. Each chain has $15{,}000$ draws. We discard the first $5{,}000$ and keep every twentieth of the remaining $10{,}000$, so each chain contributes $500$ retained draws. All ten chains target the same posterior. Two chains per shock are used to compute impulse responses, so the pooled posterior for each shock rests on about $1{,}000$ retained draws after dropping the draws with an explosive companion matrix, $0.9$ percent of all retained draws and between $0.6$ and $1.4$ percent per chain.

\autoref{tab:mcmc_diag} follows the convergence appendix of \citet{primiceri2005time} and reports, for each block of quantities, the 20th-order sample autocorrelation of the retained draws, the inefficiency factor, and the diagnostic of \citet{raftery1992}. The inefficiency factor is the ratio of the number of retained draws to the effective sample size, so a value of one means the draws are as informative as independent draws. The Raftery--Lewis statistic is the number of retained draws needed to estimate the 2.5th percentile of the posterior with an accuracy of $0.025$ at a $95\%$ probability. All statistics are computed chain by chain, averaged over chains, and then summarized across the quantities in a block by their median and maximum. Panel~A covers the model parameters. Panel~B covers the objects that enter the impulse responses, namely the fitted mixture densities on a grid of $25$ points and the fitted 10th, 50th, and 90th percentiles, both at four dates. Panel~C covers the macro, quantile, and density impulse responses for the fiscal, monetary policy, and micro shocks, the macro responses at all $24$ horizons and the quantile and density responses at horizons $1$, $2$, $4$, $8$, $12$, and $24$. For the impulse responses only the two chains assigned to each shock are compared.

Three patterns stand out. First, the median diagnostics for the VAR coefficients, structural variances, and factor path are close to standard convergence thresholds. The free entries of $\bm W$ and the factor loadings mix more slowly. Their median inefficiency factors range from $7$ to $11$. Second, the individual mixture parameters mix slowly, with median inefficiency factors between $24$ and $44$. The ordered means prevent literal label switching. The problem instead reflects weak separation among overlapping middle components and different allocations of mass across chains. The fitted distributions mix much better, with median inefficiency factors of at most three. Third, the impulse-response blocks have median inefficiency factors of at most two, and their 20th-order autocorrelations are close to zero. The Metropolis--Hastings acceptance rates range from $0.74$ to $0.78$, from $0.89$ to $0.91$, and about $0.98$ for the three free rows of $\bm A_0$. They range from $0.92$ to $0.95$ for the factor block.

\begin{table}[htbp]
  \centering
  \caption{MCMC convergence diagnostics for the U.S. application}
  \label{tab:mcmc_diag}
  \footnotesize\setlength{\tabcolsep}{2.7pt}
    \begin{threeparttable}
  \begin{tabular}{@{}lrrrrrrr@{}}
    \toprule
    & & \multicolumn{2}{c}{Autocorr.\ (20)} & \multicolumn{2}{c}{Ineff.\ factor} & \multicolumn{2}{c}{Raftery--Lewis} \\
    \cmidrule(lr){3-4}\cmidrule(lr){5-6}\cmidrule(lr){7-8}
    Block & \# & Med.\ & Max & Med.\ & Max & Med.\ & Max \\
    \midrule
    \multicolumn{8}{@{}l}{\textit{Panel A. Parameters (10 chains, 5{,}000 retained draws)}}\\[0.2em]
    VAR coefficients (intercepts, lags, quantile feedback) & 108 & -0.00 & 0.19 & 1.1 & 25.1 & 164 & 534 \\
    Factor loadings in the macro block & 3 & 0.05 & 0.13 & 6.8 & 14.5 & 190 & 374 \\
    Contemporaneous matrix $\bm W$ (free entries) & 8 & 0.06 & 0.34 & 11.1 & 54.2 & 267 & 938 \\
    Structural variances $d_i$ & 4 & 0.02 & 0.09 & 5.7 & 9.2 & 174 & 215 \\
    Factor path $F_{\text{mic},t}$ & 67 & 0.02 & 0.12 & 3.6 & 9.2 & 204 & 272 \\
    Log-weight coefficients $\bm b_{g,s}$ & 220 & 0.01 & 0.30 & 1.5 & 48.3 & 175 & 705 \\
    Mixture means $\mu_{g,s}$ & 12 & 0.39 & 0.67 & 41.7 & 144.2 & 611 & 1527 \\
    Mixture variances $\sigma^2_{g,s}$ & 12 & 0.24 & 0.47 & 23.9 & 84.0 & 447 & 715 \\
    Mixture weights $w_{tg,s}$ & 804 & 0.34 & 0.70 & 43.9 & 150.9 & 504 & 1628 \\
    \addlinespace
    \multicolumn{8}{@{}l}{\textit{Panel B. Fitted cross-sectional distributions (10 chains, 5{,}000 retained draws)}}\\[0.2em]
    Fitted earnings density (25 points, 4 dates) & 100 & 0.04 & 0.30 & 3.0 & 30.4 & 171 & 341 \\
    Fitted earnings quantiles (P10, P50, P90, 4 dates) & 12 & 0.02 & 0.05 & 1.6 & 3.2 & 222 & 283 \\
    Fitted consumption density (25 points, 4 dates) & 100 & 0.02 & 0.13 & 2.4 & 7.9 & 170 & 242 \\
    Fitted consumption quantiles (P10, P50, P90, 4 dates) & 12 & 0.02 & 0.04 & 1.9 & 2.9 & 234 & 313 \\
    \addlinespace
    \multicolumn{8}{@{}l}{\textit{Panel C. Impulse responses (2 chains per shock, about 1{,}000 retained draws)}}\\[0.2em]
    Fiscal shock, macro IRFs & 96 & -0.01 & 0.05 & 1.0 & 1.7 & 152 & 344 \\
    Fiscal shock, quantile IRFs & 60 & -0.01 & 0.07 & 1.3 & 4.9 & 264 & 1634 \\
    Fiscal shock, density IRFs & 240 & 0.01 & 0.08 & 1.3 & 6.0 & 165 & 250 \\
    Monetary-policy shock, macro IRFs & 95 & -0.00 & 0.18 & 1.3 & 35.6 & 180 & 319 \\
    Monetary-policy shock, quantile IRFs & 60 & 0.00 & 0.08 & 1.2 & 8.1 & 318 & 2035 \\
    Monetary-policy shock, density IRFs & 240 & 0.01 & 0.13 & 1.5 & 17.3 & 164 & 296 \\
    Micro shock, macro IRFs & 95 & 0.00 & 0.17 & 1.2 & 23.5 & 178 & 333 \\
    Micro shock, quantile IRFs & 60 & 0.01 & 0.12 & 1.6 & 6.9 & 252 & 1024 \\
    Micro shock, density IRFs & 240 & 0.01 & 0.20 & 1.8 & 19.2 & 164 & 488 \\
    \bottomrule
  \end{tabular}
  \begin{tablenotes}[flushleft]
    \footnotesize
    \item \textit{Notes}: \# is the number of scalar quantities in the block. Autocorr.\ (20) is the sample autocorrelation of the retained draws at lag $20$. The inefficiency factor is the number of retained draws divided by the effective sample size, estimated from the spectral density at frequency zero. Raftery--Lewis is the number of retained draws required to estimate the 2.5th percentile with accuracy $0.025$ at probability $0.95$; each chain has $500$ retained draws and the ten chains together have $5{,}000$. All statistics are computed on the thinned draws, chain by chain, averaged over chains, and summarized by their median and maximum across the quantities in the block. Quantities with zero posterior variance (entries fixed by identifying restrictions) are excluded. Panel~B evaluates the fitted mixtures at four dates spread over the sample. Panel~C uses all $24$ horizons for the macro responses and horizons $1$, $2$, $4$, $8$, $12$, and $24$ for the quantile and density responses (the latter on a grid of $20$ points), and, for the quantile responses, the 10th, 25th, 50th, 75th, and 90th percentiles of both cross sections.
  \end{tablenotes}
  \end{threeparttable}
\end{table}
\FloatBarrier

\subsection{Inequality Responses to the Micro Shock}\label{app:micro_inequality}
\begin{figure}[htbp]
  \centering
  \begin{subfigure}[b]{\textwidth}\centering
    \caption{Earnings}
    \includegraphics[width=\textwidth, trim=0 0 0 10bp, clip]{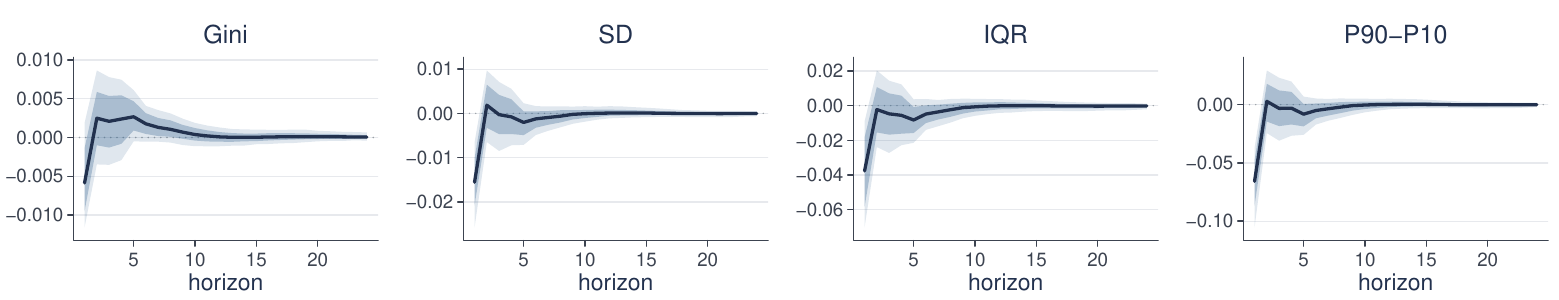}\end{subfigure}\\[0.2em]
  \begin{subfigure}[b]{\textwidth}\centering
    \caption{Consumption}
    \includegraphics[width=\textwidth, trim=0 0 0 10bp, clip]{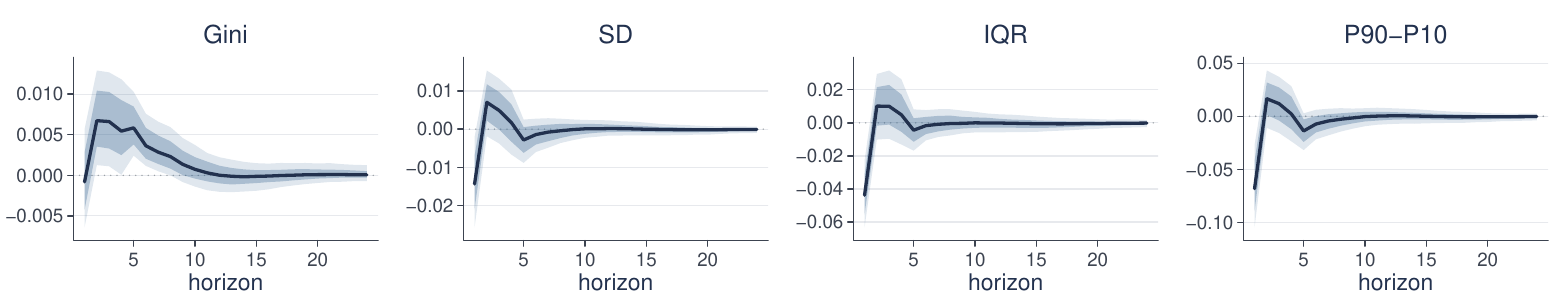}\end{subfigure}
  \caption*{\footnotesize \textbf{Notes}: Panels report responses of the Gini coefficient, standard deviation, interquartile range, and P90--P10 spread for earnings (a) and consumption (b) after a three-standard-deviation micro shock. Navy lines are posterior medians, and shaded areas are 68\% and 90\% credible bands.}
  \caption{Inequality IRFs to the micro shock}
  \label{fig:emp_micro_inequality}
\end{figure}

\autoref{fig:emp_micro_inequality} reports the inequality responses to the micro shock discussed in \autoref{sub:emp_micro_shock}. The identifying restrictions on the factor loadings induce the impact compression. They move mass from both tails toward the middle components. The inequality measures themselves are not restricted. All four measures fall on impact for both distributions, although the impact decline of the consumption Gini coefficient is small and imprecise. The earnings P90--P10 spread falls by about $0.07$, and the consumption spread falls by a similar amount. The compression is short-lived. The spread measures return to zero within about five quarters. The Gini coefficients overshoot. The earnings Gini turns slightly positive after two quarters. The consumption Gini rises to about $0.007$ around quarters two to five and fades by quarter fifteen. The bands widen quickly after impact. As in the main text, the rebound rather than the impact compression is the informative part of the response.

\FloatBarrier

\section{Extensions Appendix}\label{app:extensions}

This appendix details the two extensions of \autoref{sec:extensions}. Both preserve the structure of the sampler in \autoref{app:technical}. The missing-data extension adds one Metropolis--Hastings block. The stochastic-volatility extension replaces the structural-variance step with one per-equation Gibbs block. We use the step numbering of \autoref{sub:posterior} and the notation of \autoref{sec:econometrics}.

\subsection{Cross Sections Observed in Some Periods Only}\label{app:ext_missing}

\paragraph{Setup and ignorability.}
Cross section $s$ is observed at dates $t\in\mathcal{O}_s\subseteq\{1,\dots,T\}$ and missing at $t\in\mathcal{M}_s$, while the aggregates $\bm Q_t$ are observed throughout. A deterministic survey calendar, such as annual waves or the triennial SCF, is ignorable by design. A stochastic calendar may be omitted from the likelihood only under a missing-at-random condition stated relative to observed information. The probability that a survey is fielded at $t$ may depend on the observed history, the aggregates, survey-design variables, and the other observed cross sections at that date. It may not depend on the unobserved outcomes or on $\bm f_t$ after conditioning on those observables. In addition, the parameters of the observation process must be a priori distinct from the economic parameters. Missingness may depend on the unobserved outcomes, for example when a survey is suspended during a crisis \emph{because} the distribution moved. The calendar is then informative. Its likelihood must be modeled, and the ratio of the two missingness likelihoods multiplies the acceptance probability below.

\paragraph{Latent sample sizes.}
At an imputed date, the latent cross section $\bm y^{\text{mis}}_{t,s}$ has a fixed size $n^\star_{t,s}$. This size is part of the model, not a tuning parameter. An empirical quantile computed from $n^\star_{t,s}$ draws varies around the population quantile of the mixture with sampling variance of order $1/n^\star_{t,s}$, and that quantile enters the conditional mean of $\bm Q_{t+1}$ through \autoref{eq: VAR}. Different sizes therefore define different observed-data likelihoods. Two cases arise. When a survey was fielded but its micro data are unavailable, for example because of confidential files or an inaccessible extract, the design sample size is the appropriate value. When no survey was fielded, $n^\star_{t,s}$ is a reference-sample assumption. The average observed size is a reasonable calibration, but results should be reported for several plausible values. Under the fixed-grid representation of the empirical application, the imputed cross section takes the place of the grid, so $n^\star_{t,s}$ equals the grid size of $500$ points and no further compression is applied. All empirical-Bayes prior hyperparameters, including the mixture prior centers and the residual-variance scales $\hat\sigma^2_i$, are computed once from observed data and held fixed. They are never recomputed from imputations.

\paragraph{Active dates.}
A missing pair $(t,s)$ is \emph{active} if its quantiles enter at least one in-sample macro equation with coefficients not fixed at zero. Let $\mathcal{J}_{t,s}$ collect the affected macro dates. In the baseline specification $\mathcal{J}_{t,s}=\{t+1\}$, and with $L_q$ quantile lags it is $\{t+1,\dots,t+L_q\}$ intersected with the sample. Missing dates that are not active are integrated out. Their micro density integrates to one, so they are omitted from every block. In particular, a missing cross section at the final sample date is integrated out, and if $\bm\alpha_{r,s}=\bm 0$ for all $r$, the extension reduces to the baseline sampler with all micro sums restricted to $\mathcal{O}_s$. Pre-sample cross sections needed to initialize lagged quantiles are conditioned on as initial data or assigned an explicit initial distribution. They require a separate initial-condition specification and are not covered by the imputation block.

\paragraph{New block: exact Metropolis--Hastings imputation.}
Augment the posterior with the active latent cross sections and their component labels \citep{tanner1987}. Holding everything else fixed, the full conditional of an active $(t,s)$ is
\begin{equation}
  \pi\big(\bm y^{\text{mis}}_{t,s},\bm z^{\text{mis}}_{t,s}\mid\cdot\big)\propto
  \Bigg\{\prod_{i=1}^{n^\star_{t,s}}
  w_{t,z_{it,s},s}\mathcal{N}\big(y_{it,s}\mid\mu_{z_{it,s},s},\sigma^2_{z_{it,s},s}\big)\Bigg\}
  \prod_{\tau\in\mathcal{J}_{t,s}}\mathcal{N}_M\big(\bm u_\tau\mid\bm 0,\bm D\big),
  \label{eq:ext_mis_conditional}
\end{equation}
where $\bm u_\tau$ is the structural residual of \autoref{eq: VAR}, which depends on $\bm y^{\text{mis}}_{t,s}$ through its empirical quantiles. Propose observations and labels \emph{jointly} from the current mixture. Draw each label from $\text{Categorical}(w_{t1,s},\dots,w_{tG_s,s})$ and the observation from the corresponding Gaussian component. The proposal density is exactly the braced term in \autoref{eq:ext_mis_conditional}, so the entire micro density cancels from the Metropolis--Hastings ratio, and the log acceptance probability is
\begin{equation}
  \log a_{t,s}=\min\Bigg\{0,
  -\frac{1}{2}\sum_{\tau\in\mathcal{J}_{t,s}}
  \Big[(\bm u^{\star}_\tau)'\bm D^{-1}\bm u^{\star}_\tau
  -\bm u_\tau'\bm D^{-1}\bm u_\tau\Big]\Bigg\},
  \label{eq:ext_mis_accept}
\end{equation}
with candidate residuals available without rebuilding the design matrix,
\begin{equation}
  \bm u^{\star}_{t+1}=\bm u_{t+1}
  -\sum_{r\in\mathcal{R}}\bm\alpha_{r,s}
  \Big[\mathcal{Q}_r\big(\bm y^{\star}_{t,s}\big)
  -\mathcal{Q}_r\big(\bm y^{\text{mis}}_{t,s}\big)\Big].
  \label{eq:ext_mis_residual}
\end{equation}
With $L_q>1$ quantile lags, the same correction applies at every $\tau\in\mathcal{J}_{t,s}$ with the coefficients of lag $\tau-t$ in place of $\bm\alpha_{r,s}$. The same sample-quantile convention, including any interpolation rule, applies to observed and imputed cross sections. This block introduces no approximation. Several cross sections missing at the same date can be proposed jointly, with the corrections in \autoref{eq:ext_mis_residual} summed over $s$. Proposing one $(t,s)$ block at a time typically yields higher acceptance rates.

The relevant diagnostics for this block are date-level acceptance rates and effective sample sizes of the imputed quantiles.

\paragraph{Existing blocks.}
Conditional on the completed cross sections, every baseline conditional retains its form. The blocks simply run over completed observations, counts, and quantiles. The imputed observations are latent variables of the augmented posterior, so treating them as data in the conditional updates is valid. Their apparent information is exactly offset when they are marginalized out. Step 1 may redraw the imputed labels. The extra draw is valid and improves mixing. Steps 4--6 use the completed quantile regressors. Step 7 is the baseline exact Metropolis--Hastings factor update with completed counts $n_{tg,s}$. At a date at which every missing cross section is integrated out, the multinomial logistic term in the acceptance ratio is constant, the proposal is accepted with probability one, and the factor conditional combines the macro information with the standard-normal prior.

\paragraph{Validity.}
Under fixed latent sample sizes and an ignorable calendar, the sweep leaves the augmented posterior invariant. The imputation block is a Metropolis--Hastings transition targeting \autoref{eq:ext_mis_conditional}. Conditional on the completed cross sections, Steps 1--5 are exact Gibbs draws, and Steps 6 and 7 retain their exact Metropolis--Hastings corrections. Marginalizing the retained draws over the imputations yields observed-data inference relative to the chosen latent sample sizes. Sensitivity to $n^\star_{t,s}$ is therefore sensitivity to the model, not to the algorithm.

\subsection{Stochastic Volatility}\label{app:ext_sv}

\paragraph{Model change.}
We replace each constant structural variance $d_i$ with $d_{i,t}=\exp(h_{i,t})$, where the log volatility follows
\begin{equation}
  h_{i,t}=\mu_{h,i}+\phi_{h,i}(h_{i,t-1}-\mu_{h,i})+\nu_{i,t},
  \qquad \nu_{i,t}\sim\mathcal{N}\big(0,\sigma^2_{h,i}\big),
  \label{eq:ext_sv_ar}
\end{equation}
with a random walk ($\phi_{h,i}=1$, $\mu_{h,i}$ dropped) as the common special case. Because the structural shocks are mutually orthogonal and $\bm D$ is diagonal, volatility is added equation by equation.

\paragraph{New block.}
Conditional on the coefficients drawn in Steps 4 and 6, the structural residual $u_{i,t}$ of equation $i$ is observed, and $\log u_{i,t}^2=h_{i,t}+\log\varepsilon_{i,t}^2$ with $\varepsilon_{i,t}\sim\mathcal{N}(0,1)$. Following \citet{kim1998stochastic}, we approximate the $\log\chi^2_1$ distribution of $\log\varepsilon^2_{i,t}$ by a seven-component normal mixture, or the ten-component refinement of \citet{omori2007}. Conditional on the mixture indicators, \autoref{eq:ext_sv_ar} is then a linear Gaussian state-space model. The per-equation block draws the indicators from their multinomial conditionals, the path $h_{i,1:T}$ by forward-filtering backward-sampling \citep{carter1994,fruhwirth1994}, and $(\mu_{h,i},\phi_{h,i},\sigma^2_{h,i})$ from Gaussian and inverse-Gamma conditionals. We use the priors $\mu_{h,i}\sim\mathcal{N}(0,10)$, $\phi_{h,i}\sim\mathcal{N}(0.9,0.1^2)$ truncated to $(-1,1)$, and $\sigma^2_{h,i}\sim\mathcal{IG}(3,0.03)$, so that $\mu_{h,i}$ has a Gaussian conditional and $\sigma^2_{h,i}$ an inverse-Gamma conditional. The initial state is $h_{i,0}\sim\mathcal{N}\big(\mu_{h,i},\sigma^2_{h,i}/(1-\phi_{h,i}^2)\big)$, the stationary distribution, which initializes the forward filter. Because its variance depends on $\phi_{h,i}$, the initial-state density contributes the factor $(1-\phi_{h,i}^2)^{1/2}\exp\{-(1-\phi_{h,i}^2)(h_{i,0}-\mu_{h,i})^2/(2\sigma^2_{h,i})\}$ to the conditional of $\phi_{h,i}$, which is therefore not Gaussian. As in \citet{kim1998stochastic}, we draw a candidate from the Gaussian conditional implied by the transition equations alone, truncated to $(-1,1)$, and accept it with a Metropolis--Hastings probability equal to the ratio of this initial-state factor at the candidate and the current value. In the random-walk case, $\mu_{h,i}$ is dropped, $\phi_{h,i}=1$, and $h_{i,0}\sim\mathcal{N}(\log\hat\sigma^2_i,10)$. This block replaces Step 5. Every likelihood term that carried $d_i$ now carries $d_{i,t}$, so the coefficient full conditionals take generalized-least-squares form, exactly as in VARs with time-varying volatility \citep{primiceri2005time}. The coefficient prior must also be modified, as described below. The Laplace--Metropolis row update of $\bm A_0$ is unchanged in form, with the quadratic forms weighted by $d_{i,\tau}^{-1}$. Under stochastic volatility the variance-scaled impact matrix $\bm A_0^{-1}\bm D_t^{1/2}$ is date specific. The sign and zero restrictions on $\bm A_0$, on $\bm A_0^{-1}$, and on the loadings do not involve $\bm D_t$ and are unchanged. Magnitude restrictions on $[\bm A_0^{-1}]_{ij}\sqrt{d_{j,t}}$ and the reporting of shocks in standard-deviation units require a reference volatility. We evaluate both at the reference variance $\exp(\mu_{h,j})$, the variance at the steady-state log volatility and the median of the stationary distribution of $d_{j,t}$. Its unconditional mean is $\exp\big(\mu_{h,j}+\sigma^2_{h,j}/(2(1-\phi_{h,j}^2))\big)$. Under a random walk we use the sample average of $d_{j,t}$. Both conventions reduce to the constant-volatility convention as $\sigma^2_{h,j}\to0$. A shock of $\delta$ reference standard deviations is then a fixed perturbation in structural units, and its responses do not depend on the volatility state at the conditioning date.

\paragraph{Prior re-anchoring.}
The baseline coefficient prior \autoref{eq:prior_phi} is scaled by $d_i$, which is why the full conditional of $d_i$ carries the terms $K_i/2$ and $\mathrm{Q}_i/2$ in \autoref{app:technical}. With time-varying $d_{i,t}$ there is no single scale, so we fix the prior covariance of $\bm\phi_i$ at $\hat\sigma^2_i\underline{\bm V}_i$, using the equation-wise least-squares residual variance $\hat\sigma^2_i$ that already centers the variance prior \autoref{eq:prior_D}. This modifies the prior while leaving the algorithm unchanged. The volatility block replaces the inverse-Gamma step wholesale, and no analogue of the $K_i$ and $\mathrm{Q}_i$ terms appears because the coefficient prior no longer depends on the volatilities.

\paragraph{Validity.}
Conditional on the mixture indicators, the volatility path is an exact draw from a linear Gaussian model, and the remaining blocks retain their form, with the $\bm A_0$ row update, the factor update, and the $\phi_{h,i}$ update keeping their Metropolis--Hastings corrections. The only approximation is the finite normal mixture for $\log\chi^2_1$. Its accuracy is quantified in \citet{kim1998stochastic} and \citet{omori2007}. Exactness can be restored by reweighting the posterior draws with the importance ratio between the true and approximating measurement densities.

\end{appendices}

\end{document}